\documentclass[10pt,a4paper,reqno]{amsart}
\usepackage{acronym}
\usepackage{amsfonts}
\usepackage{amsmath}
\usepackage{amssymb}
\usepackage{amsthm}
\usepackage{bm}
\usepackage{braket}
\usepackage{cite}
\usepackage{color}
\usepackage{geometry}
\usepackage{graphicx}
\usepackage{hyperref}
\usepackage{mathrsfs}
\usepackage{mathtools}
\usepackage{setspace}

\usepackage{booktabs}
\usepackage{threeparttable}
\usepackage{diagbox}
\usepackage{multirow}
\usepackage{float}
\usepackage{array}
\usepackage{graphicx}

\usepackage{stmaryrd}
\usepackage{subcaption}
\usepackage{tikz}

\usepackage{epsfig}
\usepackage{setspace}
\usepackage{epstopdf}
\usepackage{booktabs}
\usepackage{threeparttable}
\usepackage{diagbox}

\usepackage{bbm, dsfont}

\newgeometry{top=2cm,bottom=2cm,outer=1.5cm,inner=1.5cm}
\newcommand{\bse}{\begin{subequations}}
\newcommand{\ese}{\end{subequations}}

\numberwithin{equation}{section}
\DeclareSymbolFont{largesymbols}{OMX}{yhex}{m}{n}
\DeclareMathAccent{\Widehat}{\mathord}{largesymbols}{"62}

\title[The improved backward compatible physics-informed neural networks for reducing error accumulation and applications in data-driven higher-order rogue waves]{The improved backward compatible physics-informed neural networks for reducing error accumulation and applications in data-driven higher-order rogue waves}
\author{Shuning Lin}
\address[SL]{School of Mathematical Sciences, Shanghai Key Laboratory of Pure Mathematics and Mathematical Practice, and Shanghai Key Laboratory of Trustworthy Computing \\
East China Normal University \\ Shanghai 200241 \\ China}
\author{Yong Chen$^*$}
\address[YC]{School of Mathematical Sciences, Shanghai Key Laboratory of Pure Mathematics and Mathematical Practice, and Shanghai Key Laboratory of Trustworthy Computing \\
East China Normal University \\ Shanghai 200241 \\ China}
\address[YC]{College of Mathematics and Systems Science \\ Shandong University of Science and Technology \\ Qingdao 266590 \\ China}
\email{ychen@sei.ecnu.edu.cn}

\begin{document}

\begin{abstract}

Due to the dynamic characteristics of instantaneity and steepness, employing domain decomposition techniques for simulating rogue wave solutions is highly appropriate. Wherein, the backward compatible PINN (bc-PINN) is a temporally sequential scheme to solve PDEs over successive time segments while satisfying all previously obtained solutions. In this work, we propose improvements to the original bc-PINN algorithm in two aspects based on the characteristics of error propagation. One is to modify the loss term for ensuring backward compatibility by selecting the earliest learned solution for each sub-domain as pseudo reference solution. The other is to adopt the concatenation of solutions obtained from individual subnetworks as the final form of the predicted solution. The improved backward compatible PINN (Ibc-PINN) is applied to study data-driven higher-order rogue waves for the nonlinear Schr\"{o}dinger (NLS) equation and the AB system to demonstrate the effectiveness and advantages. Transfer learning and initial condition guided learning (ICGL) techniques are also utilized to accelerate the training. Moreover, the error analysis is conducted on each sub-domain and it turns out that the slowdown of Ibc-PINN in error accumulation speed can yield greater advantages in accuracy. In short, numerical results fully indicate that Ibc-PINN significantly outperforms bc-PINN in terms of accuracy and stability without sacrificing efficiency.


\end{abstract}
\maketitle

\section{Introduction}

Rogues waves (also known as monster waves, killer waves, giant waves and abnormal waves) have always been an important research direction in the field of nonlinear science. Due to the complexity of the rogue wave phenomenon, it is currently difficult to provide a precise definition. Usually, in oceanography, people refer to a type of abnormally steep wave that suddenly appears on the sea level as the rogue wave. This type of wave has the characteristic of 'coming without a shadow, going without a trace'. The concept of rogue waves was first proposed by British scientist Draper in 1964, who described a huge, sudden appearance of monster waves that overturned ships in \cite{freak}. The rogue wave is a very tall and strong solitary wave, and its wave height is usually more than twice that of the highest wave around it. Due to the sudden onset and immense power, it can sometimes cause serious damage to ships and offshore structures. Naturally, rogue wave phenomena are not confined to the field of oceanography. In 2007, Solli et al. \cite{opticalrogue} utilized a new real-time detection technique to observe abnormally steep, massive, and rare optical rogue waves for the first time in nonlinear optical systems. Afterwards, rogue wave phenomena have emerged in numerous research fields, such as Bose Einstein condensation \cite{matterrogue}, plasma physics \cite{plasma}, fluid mechanics \cite{fluid}, meteorology\cite{atmosphere} and finance \cite{financial}.

Scientists aspire to elucidate the ancient and enigmatic phenomenon of rogue waves through the application of nonlinear models. Research has revealed that the rational solution for the Nonlinear Schr\"{o}dinger (NLS) equation can aptly describe rogue waves. This solutions, first discovered by the British physicist Peregrine in 1983 \cite{Peregrine}, is consequently referred to as the Peregrine soliton. In terms of mathematical expression, the first-order rational solution for the NLS equation is composed of a second-order rational polynomial and an exponential function. Physically, it describes a localized, steep wave in both spatial and temporal dimensions, with an amplitude three times higher than the background field. The evolution of this solution precisely aligns with the two fundamental characteristics of rogue waves: suddenness and steepness. Up to this point, utilizing existing soliton theories to investigate various rogue wave solutions has become one of the focal issues capturing the attention of scientists across diverse fields worldwide. Commonly employed research methods include the Hirota bilinear method \cite{bilinearrogue}, the KP hierarchy reduction method \cite{KProgue}, the generalized Darboux transformation \cite{gDTrogue}, the inverse scattering transformation \cite{ISTrogue}, among others.

With the rapid development of computer technology and the explosive growth of available data, methods for solving partial differential equations have expanded from traditional numerical approaches to data-driven methods. The physics-informed neural network (PINN) method \cite{PINN} has demonstrated extraordinary prospects among various data-driven techniques for solving PDEs. The main idea of PINN is to incorporate information from physical equations into the training of neural networks, rather than relying solely on given data. This enables it to provide accurate predictions even in situations where data is scarce or corrupted by noise. In response to various application scenarios and precision requirements, numerous variations and extensions of PINNs have emerged, such as fPINN \cite{fPINN} for solving fractional differential equations, hPINN \cite{hPINN} for inverse design, B-PINN \cite{BPINN} to solve both forward and inverse nonlinear problems involving PDEs and noisy data, NN-aPC \cite{NNaPC} for solving stochastic PDEs and so on. In the field of integrable systems, algorithms based on excellent properties including conserved quantities \cite{LSNJCP}, Miura transformation \cite{LSNMiura} and symmetries \cite{PINNsymmetries}. Moving forward, there is anticipation for the synergistic integration of deep learning methods with other integrable theories, such as Darboux transformation, Riemann-Hilbert method and Lax pair.

Despite promising prospects, PINN has encountered difficulties in accurately and efficiently tackling challenges associated with large domains and multiscale problems. This motivates us to explore additional techniques for enhancing the accuracy. Multiple reasons drive the integration of region decomposition technology into PINNs. Firstly, it can decompose the scale of the original problem, dividing a complex algorithm into equivalent smaller parts according to a certain decomposition method, and then solving them one by one. Secondly, locality can be highlighted. The experience shows that neural networks tend to approximate smooth and continuous functions. This may lead to poor local prediction performance in complex situations, such as approaching steep or discontinuous solutions. Furthermore, in terms of the algorithm design, we can combine traditional numerical algorithms with neural network methods. Finally, parallel computing can also be used to improve efficiency. The cPINN method \cite{cPINN} performs domain decomposition in space and the conserved quantities like fluxes, are preserved by enforcing their continuity in the strong form at the common interfaces of neighboring sub-domains. The extended PINN (XPINN) \cite{XPINN} is a space-time domain decomposition based deep learning framework and is an extension of the cPINN method. XPINN replaces the continuity of flux on the interface of adjacent regions with the more general condition of residual continuity, which can be extended to solve any type of PDEs. On this basis, a parallel algorithm for cPINNs and XPINNs is also proposed \cite{ParallelcPINNXPINN}. With regard to innovation in the algorithm design, FBPINN \cite{FBPINN} is an overlapping domain decomposition framework inspired by classical finite element methods, and hp-VPINNs \cite{hpVPINN} are designed by combining variational principles. In addition to manual decomposition of the computational domain, Stiller et al. put forward an adaptive domain decomposition method called Gated PINN \cite{GatedPINN}. This study suggests that the decomposition of the computational domain can be learned by utilizing the mixture of expert approach \cite{mixtureexpert}. Specifically, the decomposition method of time domain is also a hot research topic due to the particularity of time variables compared to spatial ones. Two time-adaptive approaches, adaptive sampling in time and adaptive time marching strategy, are introduced in \cite{AdaptivePINN}. To ensure backward compatibility of the solution, the bc-PINN scheme \cite{bcPINN} re-trains the neural network over successive time segments while satisfying the solution for all previous time segments. Given that the existing PINN formula cannot respect the inherent spatio-temporal causal structure of the evolution of physical systems, Wang et al. \cite{causalPINN} put forth a general casual training algorithm for explicitly respecting physical causality during model training. In contrast to previously mentioned methods using hard constraints, causal PINN adjusts the initiation of training in subsequent time domains based on the magnitude of weights, employing a soft constraint approach. Moreover, the idea of ensemble learning is adopted in \cite{ensemblelearning}, and the computational region can be automatically expanded without artificially dividing the time domain.

In recent years, the use of deep learning methods to simulate rogue wave solutions has attracted widespread attention and sparked a new research trend. Since Chen proposed the concept of integrable deep learning \cite{LJ1,LJ2}, his team has conducted extensive research in the field of data-driven rogue waves with the aid of the PINN algorithm and its variants. The dynamic behaviors of the rogue wave solution \cite{rouge} for the nonlinear Schr\"{o}dinger equation and the rogue periodic wave solution \cite{periodicrouge} for the Chen-Lee-Liu equation have been reproduced for the first time. Miao and Chen \cite{MZW2022} also studied the data-driven resonance rogue wave solution for the (2+1)-dimensional KP equation, which is the first practice of deep learning methods in high-dimensional integrable systems. Later, Pu et al. also successfully simulated vector rogue waves for the Manakov \cite{PJCchaos} and Yajima-Oikawa systems \cite{PJCYO}. In addition, other scholars have also made rich achievements and progress in this area of research \cite{Yanrogue1,Yanrogue2,Yanrogue3,Libiaorogue,DSIrogue}.

Considering the complexity of the dynamics of rogue wave solutions (instantaneity and steepness), it is a good application scenario for using domain decomposition techniques. In this paper, in order to capture the complex dynamic behavior of higher-order rogue waves, we treat the integrable equation considered here as a time evolution model and adopt a time-phased training approach. The bc-PINN method draws inspiration from the traditional numerical format and employs a training approach that advances layer by layer in the temporal domain, while satisfying all obtained solutions of the previous time periods to ensure backward compatibility of the solution. For algorithms that perform time piecewise training to solve PDEs, error propagation is an unavoidable problem. Therefore, based on the characteristics of error propagation, we propose improvements to the original bc-PINN algorithm in the design of the loss function and the final form of the predicted solution. Then the improved algorithm is applied to study the data-driven high-order rogue waves for the NLS equation \cite{Peregrine} and the AB system \cite{Pedlosky}, while analyzing the effectiveness of these improvements in reducing error accumulation by calculating the errors in each sub-domain.

The structure of the article is as follows. Sec. \ref{Methodology} gives a brief review about the original PINN and bc-PINN algorithms, and puts forward the improved bc-PINN (Ibc-PINN) for reducing error accumulation. In Sec. \ref{experiments}, the Ibc-PINN method is utilized to learn abundant dynamic behaviors of the first-order and higher-order rogue waves for the NLS equation and the AB system. Then we conduct analysis of error accumulation to disclose the necessity of proposing an Ibc-PINN method and analyze the errors across the entire spatiotemporal region to explore the role of each aspect of improvements in Sec. \ref{analysis}. In addition, numerical experiments are carried out to discuss the impact of changes in network structure on both algorithm stability and accuracy. Finally, the conclusion and expectation are given in the last section.

\section{Methodology}\label{Methodology}
\subsection{A brief review of PINN and bc-PINN}
\quad

Consider the general form of a $N+1$-dimensional partial differential equation (PDE)
\begin{align}\label{PDE}
f\left(\boldsymbol{x},t ; \frac{\partial u}{\partial x_1}, \ldots, \frac{\partial u}{\partial x_N}, \frac{\partial u}{\partial t} ; \frac{\partial^2 u}{\partial x_1^2}, \ldots, \frac{\partial^2 u}{\partial x_1 \partial x_N}, \frac{\partial^2 u}{\partial x_1 \partial t} ; \ldots \right)=0, \quad \boldsymbol{x}=\left(x_1, \cdots, x_N\right) \in \Omega, \quad t \in [T_0, T],
\end{align}
where $\Omega$ is a subset of $\mathbb{R}^N$, and $f$ is a nonlinear function of the solution $u(\boldsymbol{x},t)$ and partial derivatives of space coordinate $\boldsymbol{x}$ and time coordinate $t$. Then the initial and boundary conditions are given as follows
\begin{align}
\begin{gathered}
u(\boldsymbol{x},T_0)=u_0(\boldsymbol{x}),\quad \boldsymbol{x} \in \Omega,\\
u(\boldsymbol{x},t)=\mathcal{B}(\boldsymbol{x},t),\quad  (\boldsymbol{x},t) \in \Gamma \times [T_0,T],
\end{gathered}
\end{align}
and $\Gamma$ denotes the boundary of spatial region $\Omega$.

In this paper, our emphasis lies on the forward problems of partial differential equations where PINN encounters challenges in achieving precise solutions.

$\bullet$ \textbf{PINN:}

The physics-informed neural network (PINN) method is an effective tool in solving forward problems of PDEs, i.e. the initial-boundary value problem considered here.

Construct a feedforward neural network with a depth of $L$, which consists of one input layer, $L-1$ hidden layers and one output layer. The $l$th ($l=0,1,\cdots,L$) layer has $N_l$ neurons, which represents that it transmits $N_l$-dimensional output vector $\mathbf{x}^l$ to the ($l+1$)th layer as the input. The connection between layers is achieved by the following affine transformation $\mathcal{A}$ and activation function $\sigma(\cdot)$:
\begin{align}\label{eq2.3}
\mathbf{x}^l=\sigma(\mathcal{A}_l(\mathbf{x}^{l-1}))=\sigma(\mathbf{w}^{l} \mathbf{x}^{l-1}+\mathbf{b}^{l}),	
\end{align}
where $\mathbf{w}^{l}\in \mathbb{R}^{N_{l} \times N_{l-1}}$ and $\mathbf{b}^{l}\in \mathbb{R}^{N_{l}}$ denote the weight matrix and bias vector of the $l$th layer, respectively. Thus, the relation between input $\mathbf{x}^0$ and output $u(\mathbf{x}^0,\boldsymbol{\Theta})$ is given by
\begin{align}\label{eq2.4}
u(\mathbf{x}^0,\boldsymbol{\Theta})=(\mathcal{A}_L \circ \sigma \circ \mathcal{A}_{L-1} \circ \cdots \circ \sigma \circ \mathcal{A}_1)(\mathbf{x}^0),
\end{align}
and here $\boldsymbol{\Theta}=\left\{\mathbf{w}^{l}, \mathbf{b}^{l}\right\}_{l=1}^{L}$ represents the trainable parameters of PINN. For the $N+1$-dimensional PDE mentioned in \eqref{PDE}, the input $\mathbf{x}^0$ is the combination of spatial and temporal coordinates $(\boldsymbol{x},t)$.

Assume we can obtain the training data, including the initial dataset $\{\boldsymbol{x}_k^i, T_0, u_k^i\}_{i=1}^{N_i}$ and boundary dataset $\{\boldsymbol{x}_k^b, t_k^b, u_b^i\}_{i=1}^{N_b}$ via simple random sampling method, and the set of collocation points $\{\boldsymbol{x}_k^r, t_k^r \}$ of the PDE residual
\begin{align}
R:=f\left(\boldsymbol{x},t ; \frac{\partial \hat{u}}{\partial x_1}, \ldots, \frac{\partial \hat{u}}{\partial x_N}, \frac{\partial \hat{u}}{\partial t} ; \frac{\partial^2 \hat{u}}{\partial x_1^2}, \ldots, \frac{\partial^2 \hat{u}}{\partial x_1 \partial x_N}, \frac{\partial^2 \hat{u}}{\partial x_1 \partial t} ; \ldots \right),
\end{align}
by Latin hypercube sampling approach \cite{Latin}. Then the loss function that reflects the initial-boundary conditions and the PDE residual is defined as follows
\begin{align}
\begin{gathered}
\operatorname{MSE}=w_i \operatorname{MSE}_I\left(\boldsymbol{x}_k^i, T_0\right)+w_b \operatorname{MSE}_B\left(\boldsymbol{x}_k^b, t_k^b\right)+w_r \operatorname{MSE}_R\left(\boldsymbol{x}_k^r, t_k^r\right)\\
\boldsymbol{x}_k^i \in \Omega, \quad\left(\boldsymbol{x}_k^b, t_k^b\right) \in \Gamma \times\left[T_0, T\right], \quad\left(\boldsymbol{x}_k^r, t_k^r\right) \in \Omega \times\left[T_0, T\right],
\end{gathered}		
\end{align}
where
\begin{align}\label{PINN-MSE}
\begin{gathered}
 \operatorname{MSE}_I\left(\boldsymbol{x}_k^i, T_0\right)= \frac{1}{N_i} \sum_{k=1}^{N_i} \left|\hat{u}\left(\boldsymbol{x}_k^i, T_0, \boldsymbol{\Theta}\right)-u_k^i\right|^2,\\
 \operatorname{MSE}_B\left(\boldsymbol{x}_k^b, t_k^b\right)	= \frac{1}{N_b} \sum_{k=1}^{N_b} \left|\hat{u}\left(\boldsymbol{x}_k^b, t_k^b, \boldsymbol{\Theta}\right)-u_k^b\right|^2,\\
 \operatorname{MSE}_R\left(\boldsymbol{x}_k^r, t_k^r\right)=\frac{1}{N_r} \sum_{k=1}^{N_r} \left|R\left(\boldsymbol{x}_k^r, t_k^r, \boldsymbol{\Theta}\right)\right|^2.
\end{gathered}	
\end{align}
It measures the difference between the predicted values and the true values of each iteration. Here we use $\hat{u}$ to represent the predicted solution and its partial derivatives of each order with respect to spatial and the temporal variables  can be derived by automatic differentiation \cite{AD}. The neural network is trained to update weights and biases, and they need to be properly initialized before training. The bias vector is usually initialized to 0 and weight matrices can be initialized by some effective methods, like Xavier initialization \cite{Xavier}, He initialization \cite{He} and so on. Then the trainable parameters of the neural network are iteratively updated to minimize the value of the loss function.  There are many commonly used optimization algorithms, such as stochastic gradient descent (SGD), Adam and L-BFGS \cite{LBFGS}.

$\bullet$ \textbf{bc-PINN:}

The backward compatible PINN (bc-PINN) is proposed for application scenarios where the accuracy of the PINN method significantly decreases, such as strong nonlinearity and high-order time-varying partial differential equations. This scheme uses a single neural network to sequentially solve PDE over successive time segments by retraining the same neural network, while satisfying all previously obtained solutions. It is henceforth referred as backward compatible PINN since this method ensures backward compatibility of the solution.

\begin{figure}[htbp]
\centering
\includegraphics[width=12cm,height=8cm]{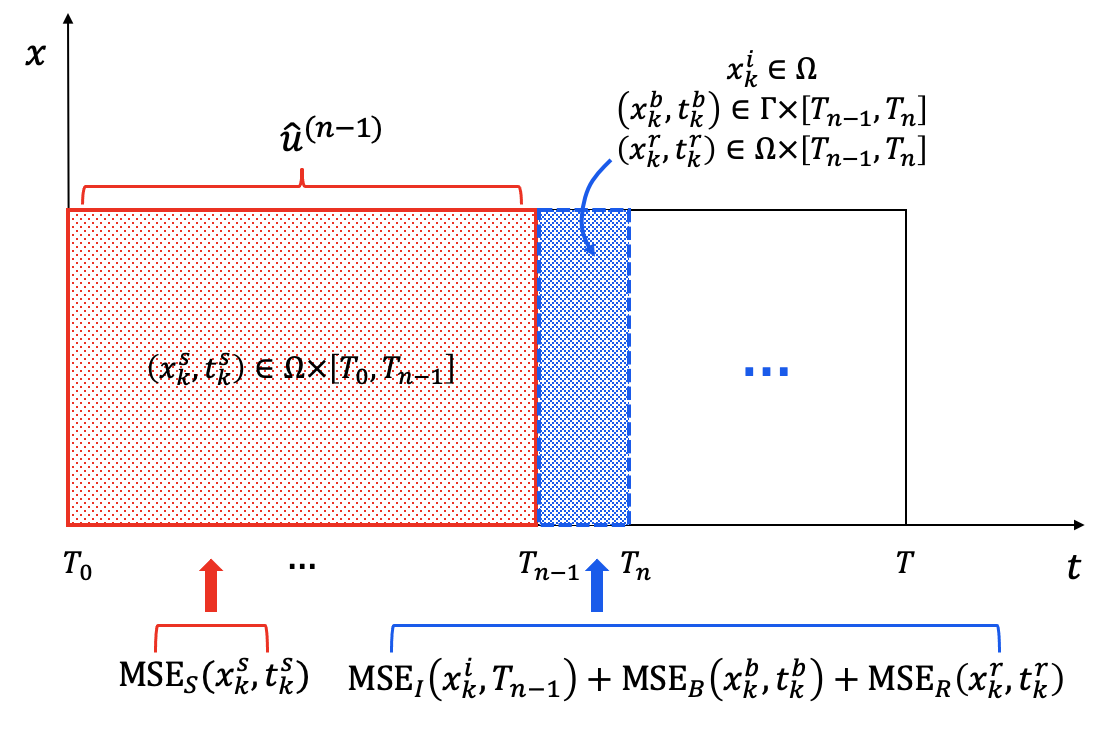}
\caption{(Color online)  Illustration of the bc-PINN for the one-dimensional scenario.}
\label{fig2-1}
\end{figure}

The sequential scheme of bc-PINN is shown in Fig. \ref{fig2-1} and the specific steps are as follows.

Firstly, the time domain $[T_0,T]$ is divided into $n_{\max}$ segments: 
\begin{align}
\left[T_0, T_1\right],\left[T_1, T_2\right], \ldots,\left[T_{n-1}, T_n\right], \cdots,\left[T_{n_{\max }-1}, T_{n_{\max }}=T\right]	
\end{align}
and then we represent the $n$th time interval as $\Delta T_n=\left[T_{n-1}, T_n\right]$ and obtain $n_{\max }$ sub-domains $\Omega \times \Delta T_n$. Therefore, the  neural network needs to be repeatedly trained for $n_{\max }$ times, and we record the solution obtained each time as $\hat{u}^{(n)} \triangleq \hat{u}^{(n)}(\boldsymbol{\Theta}_n^*), (n=1, \ldots, n_{\max })$.

For the first sub-domain $\Omega \times [T_0, T_1]$, the PINN method is applied to minimize the following loss function
\begin{align}
\begin{gathered}
\operatorname{MSE}_{\Delta T_1}=w_i \operatorname{MSE}_I\left(\boldsymbol{x}_k^i, T_0\right)+w_b \operatorname{MSE}_B\left(\boldsymbol{x}_k^b, t_k^b\right)+w_r \operatorname{MSE}_R\left(\boldsymbol{x}_k^r, t_k^r\right) \\
\boldsymbol{x}_k^i \in \Omega, \quad\left(\boldsymbol{x}_k^b, t_k^b\right) \in \Gamma \times\left[T_0, T_1\right], \quad\left(\boldsymbol{x}_k^r, t_k^r\right) \in \Omega \times\left[T_0, T_1\right]
\end{gathered}	
\end{align}
to obtain the solution $\hat{u}^{(1)} \triangleq \hat{u}^{(1)}(\boldsymbol{\Theta}_1^*)$ of PDE, where $\boldsymbol{\Theta}_1^*= \arg \min \operatorname{MSE}_{\Delta T_1}(\boldsymbol{\Theta}_1)$. The trained neural network is named as subnet-$1$. $\operatorname{MSE}_I\left(\boldsymbol{x}_k^i, T_0\right), \operatorname{MSE}_B\left(\boldsymbol{x}_k^b, t_k^b\right)$ and $\operatorname{MSE}_R\left(\boldsymbol{x}_k^r, t_k^r\right)$ represent the errors of initial condition, boundary condition and PDE constraints respectively, the calculation formulae of which are similar to \eqref{PINN-MSE} except for replacing $\hat{u}$ with $\hat{u}^{(1)}$ and $\boldsymbol{\Theta}$ with $\boldsymbol{\Theta}_1$.

For all subsequent sub-domains $\Omega \times \Delta T_n, (n=2, \ldots, n_{\max })$, a new loss function is proposed, which differs in that it is enforced to satisfy the solution obtained from previous training
\begin{align}
\begin{gathered}
\operatorname{MSE}_{\Delta T_n}=w_i \operatorname{MSE}_I\left(\boldsymbol{x}_k^i, T_{n-1}\right)+w_b \operatorname{MSE}_B\left(\boldsymbol{x}_k^b, t_k^b\right)+w_r \operatorname{MSE}_R\left(\boldsymbol{x}_k^r, t_k^r\right) 
+w_s \operatorname{MSE}_S\left(\boldsymbol{x}_k^s, t_k^s\right),  \\
\boldsymbol{x}_k^i \in \Omega, \quad\left(\boldsymbol{x}_k^b, t_k^b\right) \in \Gamma \times\left[T_{n-1}, T_n\right] \\
\left(\boldsymbol{x}_k^r, t_k^r\right) \in \Omega \times\left[T_{n-1}, T_n\right], \quad\left(\boldsymbol{x}_k^s, t_k^s\right) \in \Omega \times\left[T_0, T_{n-1}\right]	
\end{gathered}
\end{align}
where
\begin{align}
	\operatorname{MSE}_S\left(\boldsymbol{x}_k^s, t_k^s\right)=\frac{1}{N_s} \sum_{k=1}^{N_s} \left|\hat{u}^{(n)} (\boldsymbol{x}_k^s, t_k^s, \boldsymbol{\Theta}_{n})-\hat{u}^{(n-1)} (\boldsymbol{x}_k^s, t_k^s)\right|^2,
\end{align}
\begin{align}
\hat{u}^{(n-1)}\triangleq \hat{u}^{(n-1)}(\boldsymbol{\Theta}_{n-1}^*), \quad \boldsymbol{\Theta}_{n-1}^*= \arg \min \operatorname{MSE}_{\Delta T_{n-1}}(\boldsymbol{\Theta}_{n-1}).
\end{align}
Specifically, the solution $\hat{u}^{(n)}$ obtained by the $n$th trained network (subnet-$n$) is constrained to satisfy the predicted values of $\hat{u}^{(n-1)}$ obtained from the previous training at all previous time segments $[T_0, T_{n-1}]$. $\operatorname{MSE}_S\left(\boldsymbol{x}_k^s, t_k^s\right)$ measures the departure between current predicted solution and the one from the previous training, and the introduction of this term ensures backward compatibility.

Finally, the result of bc-PINN is the predicted solution of the neural network trained for the last time in the entire spatiotemporal region, i.e., 
\begin{align}\label{bcPINNsolution}
\hat{u}(\boldsymbol{x}, t)=\hat{u}^{(n_{\max})}(\boldsymbol{x}, t),\quad	(\boldsymbol{x}, t) \in \Omega \times [T_0,T]
\end{align}
where $\hat{u}^{(n_{\max})}$ is obtained by training subnet-$n_{\max}$.

\subsection{The improved bc-PINN (Ibc-PINN) for reducing error accumulation}
\quad

For schemes like bc-PINN that perform time piecewise training to sequentially solve PDEs, error accumulation is inevitable. Here we propose an improved bc-PINN method to alleviate this phenomenon.

\begin{figure}[htbp]
\centering
\includegraphics[width=12cm,height=8cm]{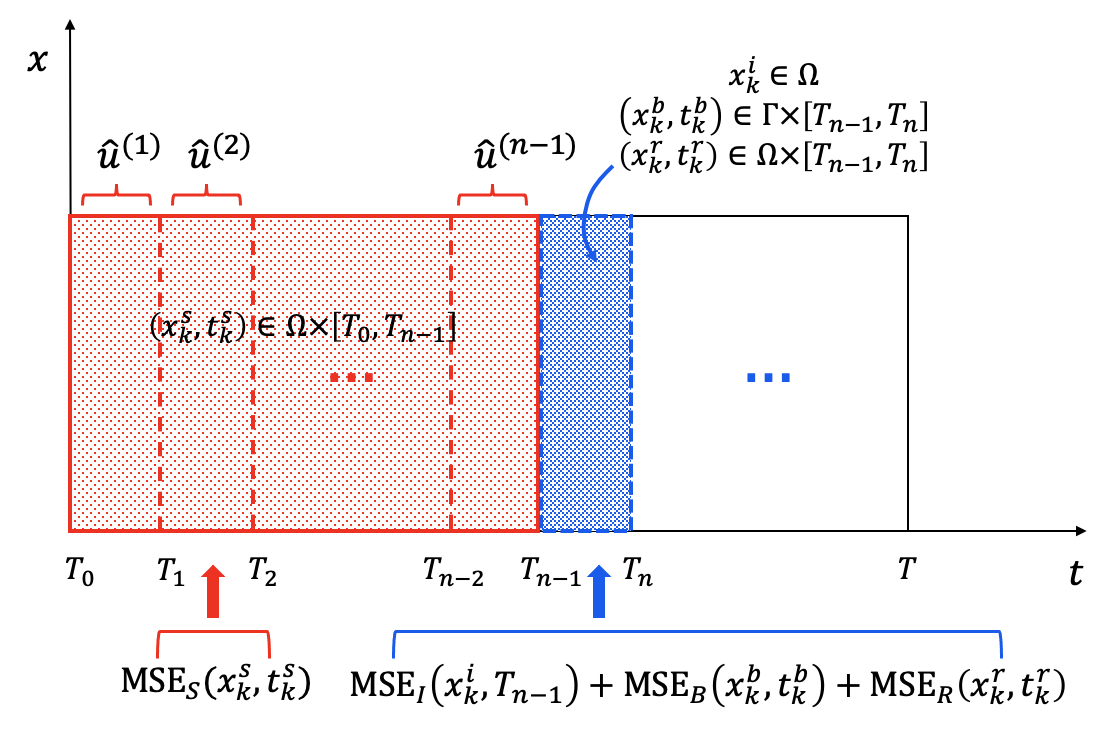}
\caption{(Color online) Illustration of the improved bc-PINN for the one-dimensional scenario.}
\label{fig2-2}
\end{figure}

In the sequential scheme, the neural network training of the posterior sub-domain is based on the training results of the anterior ones, which is reflected in the initial constraint $\operatorname{MSE}_I$ and the constraint $\operatorname{MSE}_S$ that ensures backward compatibility. Regarding $\operatorname{MSE}_I$, except for the initial values of the first sub-domain being accurate, those of all other regions are the predicted values of the previous trained neural network at the end of the time domain, known as the 'pseudo initial values'. However, this part of the error cannot be eliminated and can only be reduced by training the previous sub-domains with sufficient accuracy. But for $\operatorname{MSE}_S$, we can modify its form to reduce error accumulation. This term enforces the current predicted solution to approximate the solution obtained from the previous training at all previous time segments. Take $n=3$ as an example and the corresponding $\operatorname{MSE}_S$ is 
\begin{align}
\begin{gathered}
	\operatorname{MSE}_S\left(\boldsymbol{x}_k^s, t_k^s\right)=\frac{1}{N_s} \sum_{k=1}^{N_s} \left|\hat{u}^{(3)} (\boldsymbol{x}_k^s, t_k^s, \boldsymbol{\Theta}_{3})-\hat{u}^{(2)} (\boldsymbol{x}_k^s, t_k^s)\right|^2,\\
	\left(\boldsymbol{x}_k^s, t_k^s\right) \in \Omega \times\left[T_0, T_{2}\right]
\end{gathered}
\end{align}
where the acquisition of $\hat{u}^{(2)}$ is based on the predicted values of $\hat{u}^{(1)}$ in $[T_0, T_1]$, rather than the exact solution of PDE. Therefore, there is unavoidable error propagation in the process from $\hat{u}^{(1)}$ to $\hat{u}^{(2)}$, and it is reasonable to believe that the accuracy of $\hat{u}^{(2)}$ in $[T_0, T_1]$ is lower than that of $\hat{u}^{(1)}$ in the high probability. Thus, it is advisable to make some modifications to the form of $\operatorname{MSE}_S$ so that the currently trained solution $\hat{u}^{(3)}$ approaches the predicted values of $\hat{u}^{(1)}$ in $[T_0, T_1]$ and $\hat{u}^{(2)}$ in $(T_1, T_2]$. Similarly for the subsequent sub-domains, the prediction accuracy of $\hat{u}^{(n)}$ is probably higher than that of $\hat{u}^{(j)}, j=n+1, \cdots, n_{\max}$ in $\Omega \times\left(T_{n-1}, T_{n}\right]$. Accordingly, the improved bc-PINN makes the following modifications to $\operatorname{MSE}_S$, and its sequential scheme is shown in Fig. \ref{fig2-2}.

Define the following set of subscripts 
\begin{align}
\begin{gathered}
\tau_1 = \{k|(\boldsymbol{x}_k^s, t_k^s) \in \Omega \times [T_0,T_1] \},\\
\tau_j = \{k|(\boldsymbol{x}_k^s, t_k^s) \in \Omega \times (T_{j-1},T_j] \}, j=2,3,\cdots, n_{\max}-1	.
\end{gathered}
\end{align}
and then when $n=2, \ldots, n_{\text {max }}$, $\operatorname{MSE}_S$ is modified into
\begin{align}\label{IbcPINNMSES}
	\operatorname{MSE}_S\left(\boldsymbol{x}_k^s, t_k^s\right)=\frac{1}{N_s} \sum_{j=1}^{n-1} \sum_{k \in \tau_j} \left|\hat{u}^{(n)} (\boldsymbol{x}_k^s, t_k^s, \boldsymbol{\Theta}_{n})-\hat{u}^{(j)} (\boldsymbol{x}_k^s, t_k^s)\right|^2,
\end{align}
where $\hat{u}^{(j)} \triangleq \hat{u}^{(j)}(\boldsymbol{\Theta}_j^*), \boldsymbol{\Theta}_{j}^*= \arg \min \operatorname{MSE}_{\Delta T_{j}}(\boldsymbol{\Theta}_j)$.

To summarize, the loss function for the first sub-domain $\Omega \times [T_0, T_1]$ is given 
\begin{align}
\begin{gathered}
\operatorname{MSE}_{\Delta T_1}=w_i \operatorname{MSE}_I\left(\boldsymbol{x}_k^i, T_0\right)+w_b \operatorname{MSE}_B\left(\boldsymbol{x}_k^b, t_k^b\right)+w_r \operatorname{MSE}_R\left(\boldsymbol{x}_k^r, t_k^r\right) \\
\boldsymbol{x}_k^i \in \Omega, \quad\left(\boldsymbol{x}_k^b, t_k^b\right) \in \Gamma \times\left[T_0, T_1\right], \quad\left(\boldsymbol{x}_k^r, t_k^r\right) \in \Omega \times\left[T_0, T_1\right]
\end{gathered}	
\end{align}
where
\begin{align}
\begin{gathered}
 \operatorname{MSE}_I\left(\boldsymbol{x}_k^i, T_0\right)= \frac{1}{N_i} \sum_{k=1}^{N_i} \left|\hat{u}^{(1)}\left(\boldsymbol{x}_k^i, T_0, \boldsymbol{\Theta}_1\right)-u_k^i\right|^2,\\
 \operatorname{MSE}_B\left(\boldsymbol{x}_k^b, t_k^b\right)	= \frac{1}{N_b} \sum_{k=1}^{N_b} \left|\hat{u}^{(1)}\left(\boldsymbol{x}_k^b, t_k^b, \boldsymbol{\Theta}_1\right)-u_k^b\right|^2,\\
 \operatorname{MSE}_R\left(\boldsymbol{x}_k^r, t_k^r\right)=\frac{1}{N_r} \sum_{k=1}^{N_r} \left|R\left(\boldsymbol{x}_k^r, t_k^r, \boldsymbol{\Theta}_1\right)\right|^2,
\end{gathered}	
\end{align}
and that for subsequent sub-domain $\Omega \times \Delta T_n (n=2, \ldots, n_{\max })$ is changed into
\begin{align}
\begin{gathered}
\operatorname{MSE}_{\Delta T_n}=w_i \operatorname{MSE}_I\left(\boldsymbol{x}_k^i, T_{n-1}\right)+w_b \operatorname{MSE}_B\left(\boldsymbol{x}_k^b, t_k^b\right)+w_r \operatorname{MSE}_R\left(\boldsymbol{x}_k^r, t_k^r\right) 
+w_s \operatorname{MSE}_S\left(\boldsymbol{x}_k^s, t_k^s\right), \\
\boldsymbol{x}_k^i \in \Omega, \quad\left(\boldsymbol{x}_k^b, t_k^b\right) \in \Gamma \times\left[T_{n-1}, T_n\right] \\
\left(\boldsymbol{x}_k^r, t_k^r\right) \in \Omega \times\left[T_{n-1}, T_n\right], \quad\left(\boldsymbol{x}_k^s, t_k^s\right) \in \Omega \times\left[T_0, T_{n-1}\right]	
\end{gathered}
\end{align}
where
\begin{align}
\begin{gathered}
 \operatorname{MSE}_I\left(\boldsymbol{x}_k^i, T_{n-1}\right)= \frac{1}{N_i} \sum_{k=1}^{N_i} \left|\hat{u}^{(n)}\left(\boldsymbol{x}_k^i, T_{n-1}, \boldsymbol{\Theta}_{n}\right)-\hat{u}^{(n-1)} (\boldsymbol{x}_k^i, T_{n-1})\right|^2,\\
 \operatorname{MSE}_B\left(\boldsymbol{x}_k^b, t_k^b\right)	= \frac{1}{N_b} \sum_{k=1}^{N_b} \left|\hat{u}^{(n)}\left(\boldsymbol{x}_k^b, t_k^b, \boldsymbol{\Theta}_{n}\right)-u_k^b\right|^2,\\
 \operatorname{MSE}_R\left(\boldsymbol{x}_k^r, t_k^r\right)=\frac{1}{N_r} \sum_{k=1}^{N_r} \left|R\left(\boldsymbol{x}_k^r, t_k^r, \boldsymbol{\Theta}_{n}\right)\right|^2,\\
 \operatorname{MSE}_S\left(\boldsymbol{x}_k^s, t_k^s\right)=\frac{1}{N_s} \sum_{j=1}^{n-1} \sum_{k \in \tau_j} \left|\hat{u}^{(n)} (\boldsymbol{x}_k^s, t_k^s, \boldsymbol{\Theta}_{n})-\hat{u}^{(j)} (\boldsymbol{x}_k^s, t_k^s)\right|^2,\\
 \tau_1 = \{k|(\boldsymbol{x}_k^s, t_k^s) \in \Omega \times [T_0,T_1] \}, \quad
 \tau_j = \{k|(\boldsymbol{x}_k^s, t_k^s) \in \Omega \times (T_{j-1},T_j] \}, \quad j=2,3,\cdots, n_{\max}-1.
\end{gathered}
\end{align}
For the loss term $\operatorname{MSE}_I\left(\boldsymbol{x}_k^i, T_{n-1}\right)$, $\hat{u}^{(n-1)}$ serves the role of pseudo initial values at $t=T_{n-1}$, and for $\operatorname{MSE}_S\left(\boldsymbol{x}_k^s, t_k^s\right)$, $\hat{u}^{(1)},\hat{u}^{(2)},\cdots,\hat{u}^{(n-1)}$ act as pseudo reference solutions of the previous region $\Omega \times\left[T_0, T_{n-1}\right]$. The aforementioned weights $wi, w_b, w_r, w_s$ are used to scale the differences in the amplitude of each loss term. Note that the loss functions of bc-PINN and Ibc-PINN in the first two sub-domains are the same, leading to identical training outcomes.

What's more, the final form of the predicted solution has also been improved
\begin{align}\label{IbcPINNsolution}
\hat{u}(\boldsymbol{x}, t)=\sum_{n=1}^{n_{max}} \hat{u}^{(n)}(\boldsymbol{x}, t) \cdot \mathds{1}_{\mathcal{D}_n}(\boldsymbol{x}, t), \quad	(\boldsymbol{x}, t) \in \Omega \times [T_0,T]
\end{align}
where
\begin{align}
\mathcal{D}_n= \begin{cases}\Omega \times [T_0,T_1] & \text { if } n=1, \\ \Omega \times (T_{n-1},T_n] & \text { if } n=2,3,\cdots,n_{\max} \end{cases}	
\end{align}
That is, we adopt the concatenation of solutions obtained from individual subnetworks as the final form of the predicted solution, rather than relying solely on the solution learned by the last subnetwork like bc-PINN.

\begin{figure}[htbp]
\centering
\includegraphics[width=18cm,height=13cm]{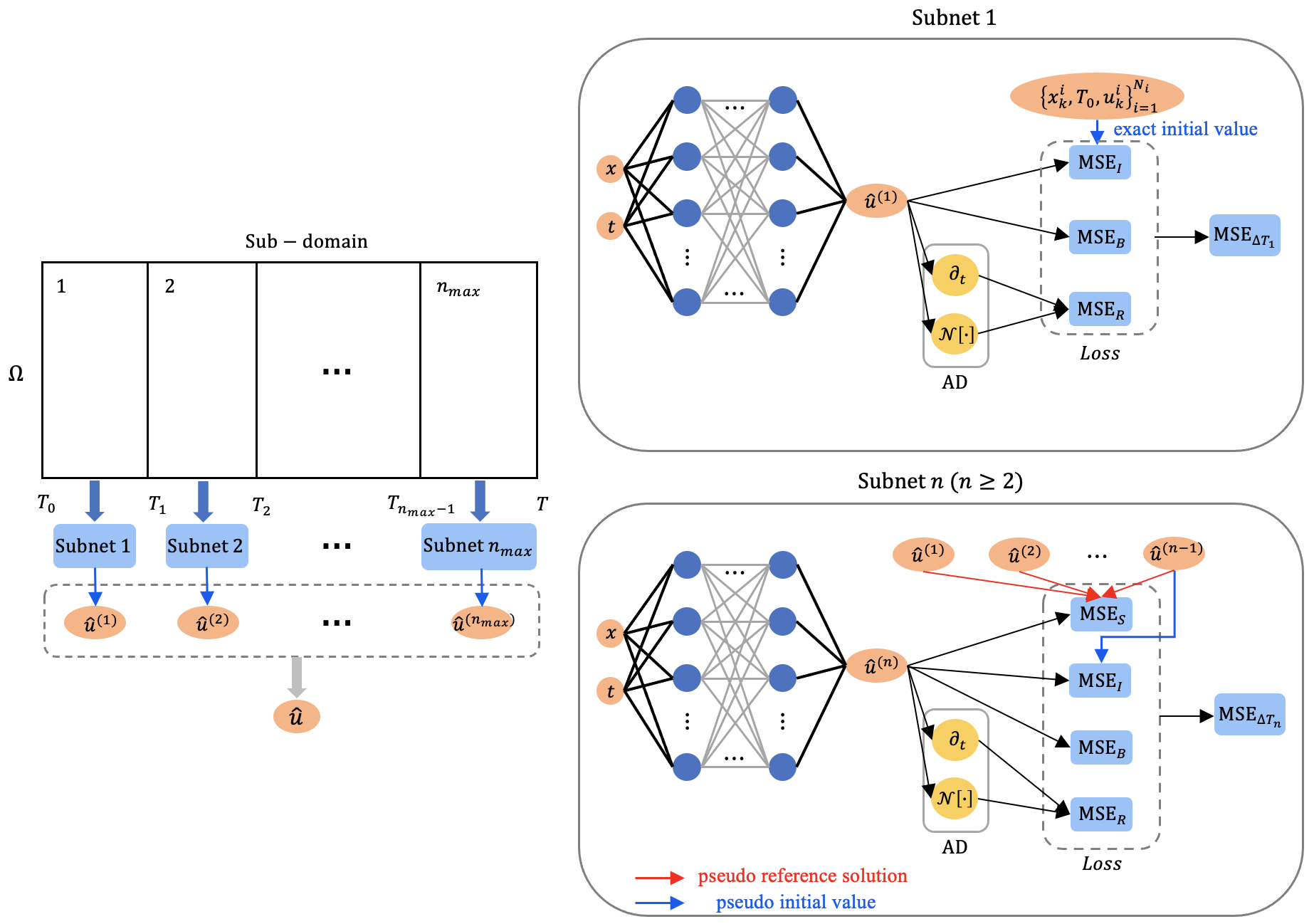}
\caption{(Color online) schematic diagram of Ibc-PINN.}
\label{fig2-3}
\end{figure}

Finally, the workflow of Ibc-PINN can be briefly summarized in Fig. \ref{fig2-3}, where the evolution equation is taken as an example, and $\mathcal{N}[\cdot]$ represents the nonlinear operator involving various order derivatives with respect to spatial variables.

\subsection{Comparison between bc-PINN and Ibc-PINN}
\quad

We compare the methods theoretically before and after improvement from the following perspectives and the specific effects of two methods in practical applications are shown based on the numerical experimental results in the following two sections.

\begin{itemize}
 \item \textbf{Accuracy}
Ibc-PINN is an improved methodology building upon the foundation of bc-PINN with the aim of enhancing accuracy. In bc-PINN, the loss term $\operatorname{MSE}_S$ is introduced to ensures backward compatibility of the solution. Its reference baseline is the predicted  solution obtained in the previous time segment, which can be considered as a pseudo reference solution across all prior regions with some margin of error. In addition, error propagation is unavoidable for algorithms performing time piecewise training to solve PDEs. Therefore, based on the characteristics of error propagation, we modify the loss term by selecting the earliest learned solution for each sub-domain as pseudo reference solution. It not only reduces the error of the pseudo reference solution used by Ibc-PINN in calculating the loss term $\operatorname{MSE}_S$ compared to bc-PINN, but also helps to decelerate the speed of error accumulation, which contributes to improving the accuracy of the solution in subsequent training.
\end{itemize}

\begin{itemize}
 \item \textbf{Efficiency}	
The main differences between Ibc-PINN and bc-PINN are reflected in two aspects: one is the data of pseudo reference solution used in computing $\operatorname{MSE}_S$, and the other is the final form of the predicted solution. All the data used originates from the training outcomes of intermediate subnetworks. Therefore, the improved method does not introduce almost any additional computational burden, and the impact on the efficiency of algorithm can be considered negligible. 
\end{itemize}

\begin{itemize}
 \item \textbf{Stability}
By adopting the concatenation of solutions obtained from individual subnetworks as the final form of the predicted solution, the Ibc-PINN is more stable since insufficient training in a specific subdomain only affects the accuracy of the predicted solutions from that subregion onwards. The successful prediction results in previous sub-domains remains unaffected and will not be rendered futile. However, the results of bc-PINN rely entirely on the success or failure of training in the last sub-domain. This will result in an immediate severe deterioration of accuracy across all trained regions if the network training performs poorly in a specific subdomain.

\end{itemize}

\begin{itemize}
 \item \textbf{Storage requirement of data}	
Unless there are additional analysis requirements, bc-PINN only needs to store the final weight matrices and bias vectors of the subnetwork from the last training. In contrast, Ibc-PINN needs to store the training results of all subnetworks, leading to larger data storage requirements.
\end{itemize}

\section{Numerical experiments}\label{experiments}

Rogue wave is a steep wave that is localized in both spatial and temporal directions, with an amplitude of three times the height of the background field. Due to the complexity of the shape and structure, many rogue wave solutions, such as high-order rogue waves, are always difficult to obtain by using standard PINN. Generally speaking, the geometric structure of high-order rogue wave solutions can be categorized into fundamental, triangular, pentagonal, heptagonal shapes, and even more complex patterns. Since high-order rogue waves may exhibit multiple peaks in a certain spatiotemporal region, it is appropriate to use domain decomposition methods to capture their intricate dynamic behaviors.

In this section, we applied the Ibc-PINN method to numerically simulate the first-order and higher-order rogue waves of the nonlinear Schr\"{o}dinger equation (NLS) equation and the AB system, and compared the results with bc-PINN to demonstrate the effectiveness of the improved method.

\subsection{The nonlinear Schr\"{o}dinger equation}
\quad

The nonlinear Schr\"{o}dinger equation is one of the most classical and well-known integrable equations, which contains many excellent properties. It can be employed to characterize the quantum behavior exhibited by microscopic particles in quantum mechanics \cite{NLS1926} as well as nonlinear phenomena in other physical domains \cite{NLS1973,NLS2003}. Numerous commonly employed techniques \cite{Peregrine,MatveevDT,YJKnonlinear}, including the Darboux transformation, the Riemann Hilbert method and so on, have been utilized to acquire localized wave solutions for the NLS equation.

\subsubsection{Data-driven first-order rogue waves}
\quad

In this subsection, we consider the nonlinear Schr\"{o}dinger equation
\begin{align}\label{NLS}
\mathrm{i} q_t+q_{x x}+2|q|^2 q=0,
\end{align}
where $q(x,t)$ is a complex-valued solution regarding spatial coordinate $x$ and temporal coordinate $t$. In optics, the nonlinear term arises from the material's intensity-dependent index. This equation can be reduced from the extended nonlinear Schr\"{o}dinger equation by setting $\beta$ to 0
\begin{align}\label{extendedNLS}
\mathrm{i} U_t+U_{x x}+2|U|^2 U+2 \mathrm{i} \beta\left(|U|^2 U_x-U^2 U_x^*\right) -2 \beta U \int\left(|U|^2\right)_t d x=0.
\end{align}
 Through the generalized Darboux transformation, the $N$th-order rogue wave $U[N]$ for the extended nonlinear Schr\"{o}dinger equation is derived in \cite{KE} and thus we can get the corresponding arbitrary-order rogue wave solution $q[N]$ of the standard NLS equation by taking $\beta=0$.

\begin{itemize}
 \item $\beta=0$	
\end{itemize}

By constructing the first-step generalized Darboux transformation, the first-order rogue wave solution for the extended NLS equation is presented
\begin{align}
U[1]=\exp [\mathrm{i} \theta] \frac{F_1+\mathrm{i} G_1}{D_1},	
\end{align}
where
\begin{align}\label{parameters}
\begin{gathered}
\theta=-2 \beta x+(4 \beta^2+2)t,\\
F_1=-4 x^2-32 \beta t x-16\left(4 \beta^2+1\right) t^2+3,\quad G_1=16 t,\\
D_1=4 x^2+32 \beta t x+16\left(4 \beta^2+1\right) t^2+1.
\end{gathered}
\end{align}
After choosing $\beta=0$, we arrive at the explicit expression of the corresponding first-order rogue wave $q(x,t)$ for the NLS equation.

The computational domain is taken as $\Omega =[x_0, x_1]=[-2,2]$ and $[T_0,T]=[-1,1]$.  Then the initial and boundary conditions are obtained based on the above exact solution. The time domain is divided into $n_{\max}=4$ segments:
\begin{align}
\left[T_0, T_1\right]=[-1,-0.5], \quad \left[T_1, T_2\right]=[-0.5,0], 	\quad \left[T_2, T_3\right]=[0,0.5], \quad \left[T_3, T\right]=[0,0.5]
\end{align}
We divide the whole spatial and time region into 512 and 201 discrete equidistance points, respectively. Then the solution is discretized into $512 \times 201$ data points in the grid points to generate the discretized dataset. In each sub-domain, $N_i=128$ and $N_b=50$ training points are randomly selected from the initial-boundary dataset. To reiterate, only the initial values of the first sub-domain are taken from the exact solution, while those of the other sub-domains are the predicted values of the solution obtained from the previous training. In addition, $N_r=20000$ collocation points are generated by means of Latin hypercube sampling method. It should be noted that as the trained sub-domain gradually move backwards, $N_s$ will increase since the range of previously trained regions expands. Here we set $N_s=128 \times 50 \times (n-1)$ for the $n$th sub-domain ($n=2,3,4$).

In view of the complexity of the structure of complex-valued solution $q(x,t)$, it is decomposed into real and imaginary parts, i.e., $q(x,t)=u(x,t)+\mathrm{i} v(x,t)$, which correspond to the two outputs of the neural network. After substituting it into \eqref{KE}, the PDE residual $R$ can be divided into two corresponding parts
\begin{align}
\begin{gathered}
R_u:=-v_t+u_{xx}+2(u^2+v^2)u,\\
R_v:=u_t+v_{xx}+2(u^2+v^2)v.
\end{gathered}
\end{align}
We choose the weights $w_i=w_b=w_r=w_s=1$ for all the examples presented in this paper to facilitate the analysis of the effect of the improvements. For the first sub-domain, the PINN method is applied to minimize the following loss function
\begin{align}\label{KE-loss1}
\operatorname{MSE}_{\Delta T_1}= \operatorname{MSE}_I\left(x_k^i, T_0\right)+ \operatorname{MSE}_B\left(x_k^b, t_k^b\right)+ \operatorname{MSE}_R\left(x_k^r, t_k^r\right) 
\end{align}
where
\begin{align}\label{KE-MSE}
\begin{gathered}
\operatorname{MSE}_I\left(x_k^i, T_0\right)= \frac{1}{N_i} \sum_{k=1}^{N_i} \left|\hat{u}^{(1)}\left(x_k^i, T_0, \boldsymbol{\Theta}_1\right)-u_k^i\right|^2+\frac{1}{N_i} \sum_{k=1}^{N_i} \left|\hat{v}^{(1)}\left(x_k^i, T_0, \boldsymbol{\Theta}_1\right)-v_k^i\right|^2,\\
 \operatorname{MSE}_B\left(x_k^b, t_k^b\right)	= \frac{1}{N_b} \sum_{k=1}^{N_b} \left|\hat{u}^{(1)}\left(x_k^b, t_k^b, \boldsymbol{\Theta}_1\right)-u_k^b\right|^2+\frac{1}{N_b} \sum_{k=1}^{N_b} \left|\hat{v}^{(1)}\left(x_k^b, t_k^b, \boldsymbol{\Theta}_1\right)-v_k^b\right|^2,\\
 \operatorname{MSE}_R\left(x_k^r, t_k^r\right)=\frac{1}{N_r} \sum_{k=1}^{N_r} \left|R_u\left(x_k^r, t_k^r, \boldsymbol{\Theta}_1\right)\right|^2+\frac{1}{N_r} \sum_{k=1}^{N_r} \left|R_v\left(x_k^r, t_k^r, \boldsymbol{\Theta}_1\right)\right|^2,
\end{gathered}
\end{align}
\begin{align}
T_0=-1,\quad x_k^i \in [-2,2], \quad\left(x_k^b, t_k^b\right) \in \{-2,2\} \times\left[-1, -0.5\right], \quad\left(x_k^r, t_k^r\right) \in [-2,2]\times\left[-1, -0.5\right]	
\end{align}
and $\hat{u}^{(1)}$ and $\hat{v}^{(1)}$ denote the predicted solutions by training subnet-$1$. $\{x_k^i, T_0, u_k^i, v_k^i \}_{k=1}^{128}$ and $\{x_b^i, t_b^i, u_b^i, v_b^i \}_{k=1}^{50}$ denote the initial and boundary data respectively. For subsequent three sub-domains ($n=2,3,4$), $\operatorname{MSE}_S$ is added into the loss function 
\begin{align}\label{KE-lossn}
\operatorname{MSE}_{\Delta T_n}= \operatorname{MSE}_I\left(x_k^i, T_{n-1}\right)+ \operatorname{MSE}_B\left(x_k^b, t_k^b\right)+ \operatorname{MSE}_R\left(x_k^r, t_k^r\right) 
+ \operatorname{MSE}_S\left(x_k^s, t_k^s\right),	
\end{align}
where
\begin{align}
\begin{gathered}
\operatorname{MSE}_I\left(x_k^i, T_{n-1}\right)= \frac{1}{N_i} \sum_{k=1}^{N_i} \left|\hat{u}^{(n)}\left(x_k^i, T_{n-1}, \boldsymbol{\Theta}_{n}\right)-\hat{u}^{(n-1)} (x_k^i, T_{n-1})\right|^2 +\frac{1}{N_i} \sum_{k=1}^{N_i} \left|\hat{v}^{(n)}\left(x_k^i, T_{n-1}, \boldsymbol{\Theta}_{n}\right)-\hat{v}^{(n-1)} (x_k^i, T_{n-1})\right|^2,\\
\operatorname{MSE}_B\left(x_k^b, t_k^b\right)	= \frac{1}{N_b} \sum_{k=1}^{N_b} \left|\hat{u}^{(n)}\left(x_k^b, t_k^b, \boldsymbol{\Theta}_{n}\right)-u_k^b\right|^2+\frac{1}{N_b} \sum_{k=1}^{N_b} \left|\hat{v}^{(n)}\left(x_k^b, t_k^b, \boldsymbol{\Theta}_{n}\right)-v_k^b\right|^2, \\
 \operatorname{MSE}_R\left(x_k^r, t_k^r\right)=\frac{1}{N_r} \sum_{k=1}^{N_r} \left|R_u\left(x_k^r, t_k^r, \boldsymbol{\Theta}_{n}\right)\right|^2+\frac{1}{N_r} \sum_{k=1}^{N_r} \left|R_v\left(x_k^r, t_k^r, \boldsymbol{\Theta}_{n}\right)\right|^2,\\
\operatorname{MSE}_S\left(x_k^s, t_k^s\right)=\frac{1}{N_s} \sum_{j=1}^{n-1} \sum_{k \in \tau_j} \left|\hat{u}^{(n)} (x_k^s, t_k^s, \boldsymbol{\Theta}_{n})-\hat{u}^{(j)} (x_k^s, t_k^s)\right|^2+\frac{1}{N_s} \sum_{j=1}^{n-1} \sum_{k \in \tau_j} \left|\hat{v}^{(n)} (x_k^s, t_k^s, \boldsymbol{\Theta}_{n})-\hat{v}^{(j)} (x_k^s, t_k^s)\right|^2,
\end{gathered}
\end{align}
\begin{align}
\begin{gathered}
x_k^i \in [-2,2], \quad\left(x_k^b, t_k^b\right) \in \{-2,2 \} \times\left[T_{n-1}, T_n\right], \quad \left(x_k^r, t_k^r\right) \in [-2,2] \times\left[T_{n-1}, T_n\right], \quad \left(x_k^s, t_k^s\right) \in [-2,2] \times\left[-1, T_{n-1}\right], \\
\tau_1 = \{k|(x_k^s, t_k^s) \in [-2,2] \times [-1,T_1] \}, \quad \tau_j = \{k|(x_k^s, t_k^s) \in [-2,2] \times (T_{j-1},T_j] \} \quad (j=2,\cdots, n-1).	
\end{gathered}
\end{align}
$\hat{u}^{(n)}$ and $\hat{v}^{(n)}$ are the predicted solutions by training subnet-$n$ in $n$th sub-domain.

Here we construct a 7-layer feedforward neural network with 128 neurons per hidden layer and use the $tanh$ activation function to learn the first-order rogue wave solution for the NLS equation. The loss functions of subnet-1 and subnet-$n$ ($n=2,3,4$) haven been given in \eqref{KE-loss1} and \eqref{KE-lossn} respectively. For subnet-1, the weights are initialized with Xavier initialization method and the biases is initialized to 0. For the subsequent subnetworks, we adopt transfer learning \cite{transferlearning} and weight freezing techniques. To be specific, the latter subnet inherits the saved weight matrixes and bias vectors of the previous subnet at the end of the iteration process as the initialization parameters, and thus the subsequent training is based on the previous results by leveraging the transfer learning technique instead of training from scratch. In addition, we use weight freezing to freeze the parameters of the first two layers to prevent them from updating in subsequent training, while only training the weights and biases of the subsequent layers. The advantage of this technique is that it can retain the features learned by the pre-trained model, reduce the training time and prevent the model from overfitting.

\begin{figure}[htbp]
\centering
\includegraphics[width=5.5cm,height=4cm]{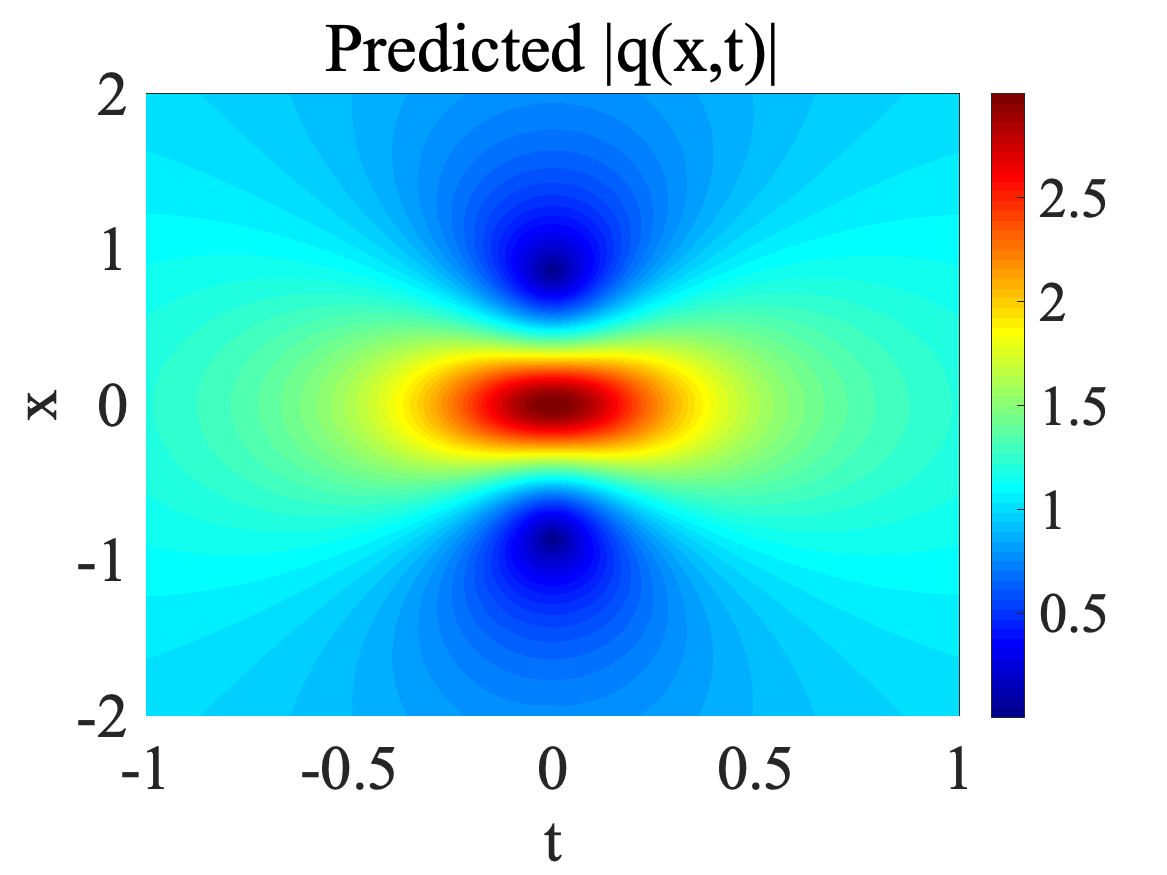}
$a$
\includegraphics[width=5.5cm,height=4cm]{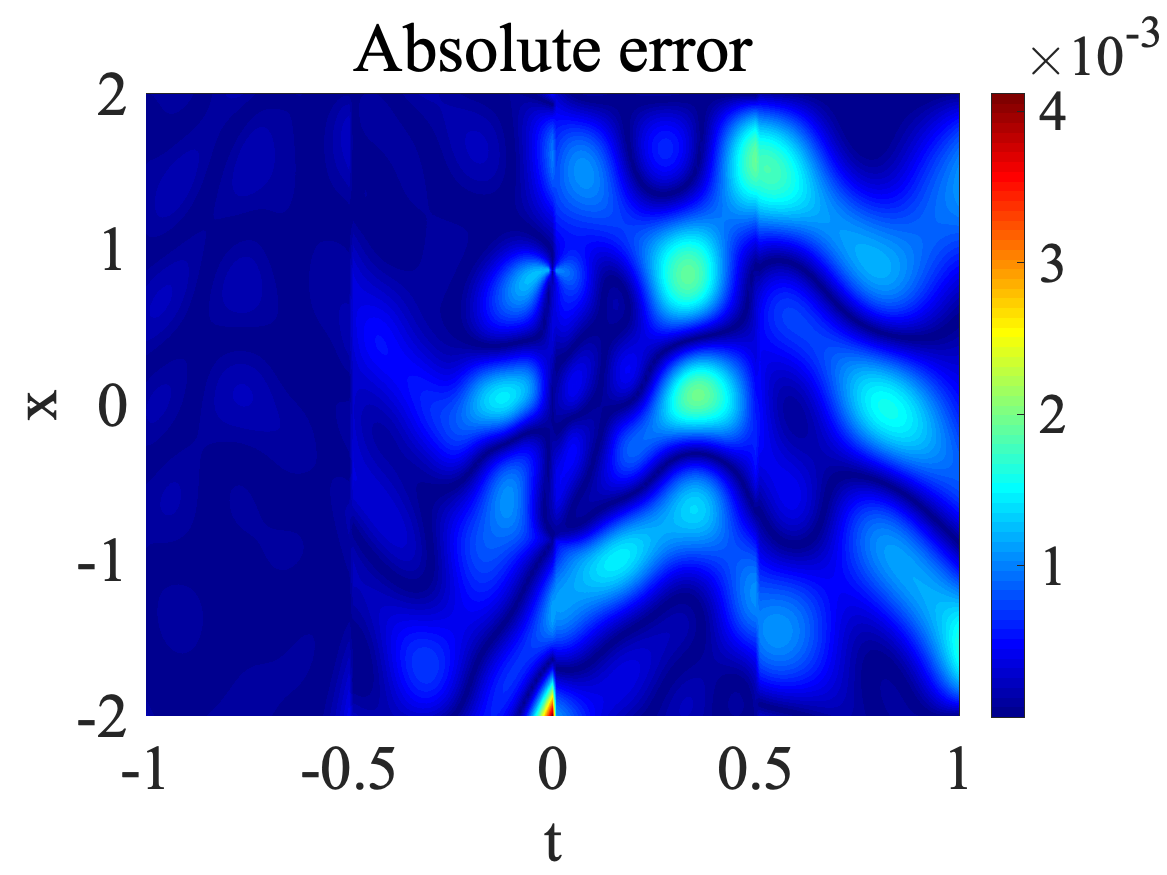}
$b$
\includegraphics[width=5.5cm,height=4cm]{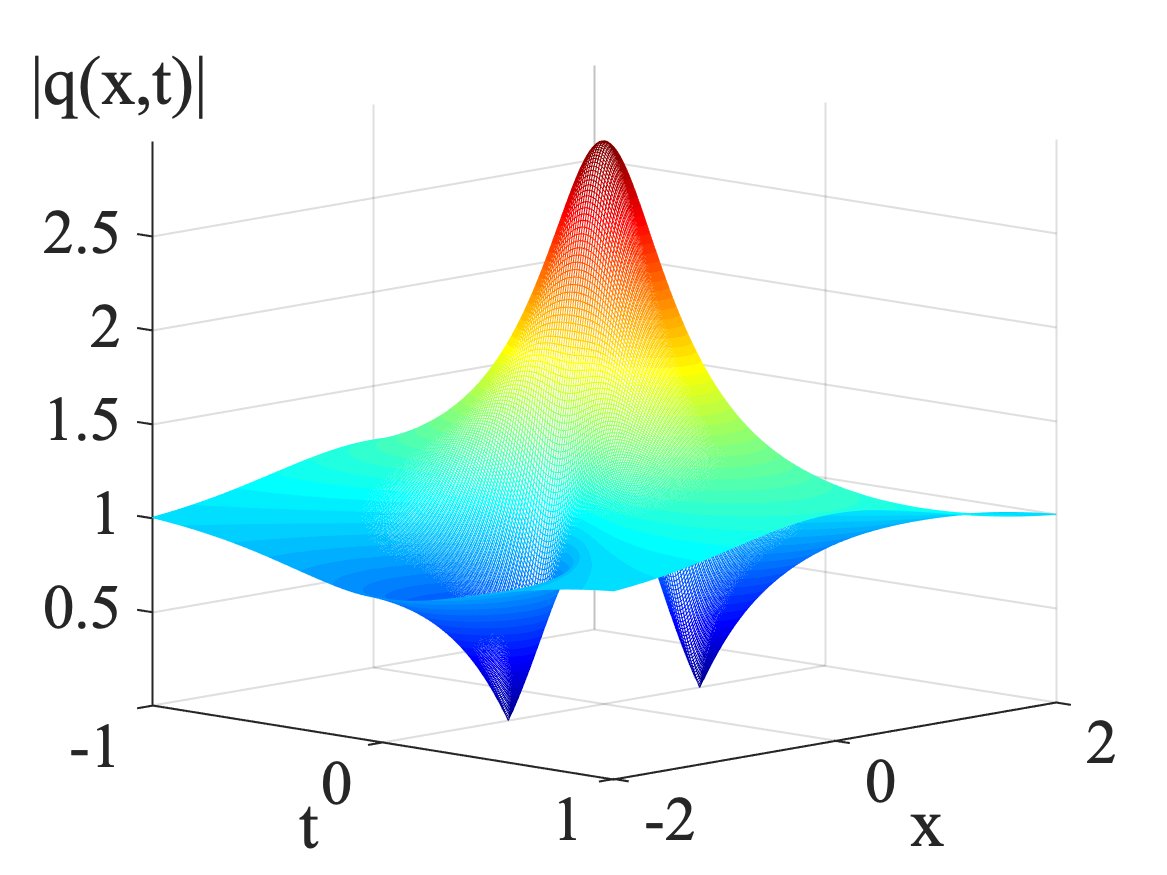}
$c$
\caption{(Color online) Data-driven first-order rogue wave ($\beta=0$) by Ibc-PINN: (a) The density diagram of the predicted solution $|q(x,t)|$; (b) The density diagram of absolute error; (c) The three-dimensional plot of the data-driven wave solution $|q(x,t)|$.}
\label{fig3-1}
\end{figure}

In this study, the first-order optimization algorithm (Adam) is first adopted to train the neural network with a certain number of iterations and then proceed by using the second-order algorithm (L-BFGS) to reach the ideal local optimal solution. This training strategy that combines first-order and second-order optimization algorithms can balance efficiency and accuracy. The number of iterations for the Adam optimizer is taken as 10000 here. Meanwhile, the initial condition guided learning (ICGL) technique \cite{bcPINN} is used here in order to better learn the initial conditions. The core idea is to employ a fraction of the overall iterations to train the neural network, focusing solely on matching the initial conditions of that specific time interval. Specifically, the loss function for ICGL in the first sub-domain is as follows
\begin{align}
	\operatorname{MSE}_{SI}=\frac{1}{N_{SI}} \sum_{k=1}^{N_{SI}} \left|\hat{u}^{(1)} (x_k^{si}, t_k^{si}, \boldsymbol{\Theta}_1)-u_k^{si}\right|^2+\frac{1}{N_{SI}} \sum_{k=1}^{N_{SI}} \left|\hat{v}^{(1)} (x_k^{si}, t_k^{si}, \boldsymbol{\Theta}_1)-v_k^{si}\right|^2,\quad (x_k^{si}, t_k^{si}) \in \Omega \times [T_{0},T_1]
\end{align}
while that in $n$th sub-domain ($n \geq 2$) is defined
\begin{align}
\begin{gathered}
	\operatorname{MSE}_{SI}=\frac{1}{N_{SI}} \sum_{k=1}^{N_{SI}} \left|\hat{u}^{(n)} (x_k^{si}, t_k^{si}, \boldsymbol{\Theta}_n)-\hat{u}^{(n-1)} (x_k^{si}, T_{n-1})\right|^2+\frac{1}{N_{SI}} \sum_{k=1}^{N_{SI}} \left|\hat{v}^{(n)} (x_k^{si}, t_k^{si}, \boldsymbol{\Theta}_n)-\hat{v}^{(n-1)} (x_k^{si}, T_{n-1})\right|^2,\\
	 (x_k^{si}, t_k^{si}) \in \Omega \times [T_{n-1},T_n]
\end{gathered}
\end{align}
where $\{u_k^{si}, v_k^{si} \}$ is the initial data at the point $(x_k^{si},T_0)$ and $\hat{u}^{(n-1)} (x_k^{si}, T_{n-1})$ ($\hat{v}^{(n-1)} (x_k^{si}, T_{n-1})$) denotes the value of the predicted solution $\hat{u}^{(n-1)}$ ($\hat{v}^{(n-1)}$) at $t=T_{n-1}$. The neural network is trained using 10\% of the total number of iterations of the Adam optimizer and we choose $N_{SI}=128*51$ to match only the initial condition of that time segment. The dynamic behavior of first-order rogue wave has been successfully learned by Ibc-PINN and the relative $\mathbb{L}_2$ error of $|q(x,t)|$ is 4.843e-04. The density plot, the absolute error and the 3d plot of the data-drive rogue wave solution are displayed in Fig. \ref{fig3-1}. It can be seen that there is one peak and two valleys, with the highest value of the peak roughly appearing near the point $(0,0)$.

\begin{itemize}
 \item $\beta=\frac{1}{3}$	
\end{itemize}

The NLS equation is a special case when $\beta=0$ of the Kundu-Eckhaus (KE) equation \cite{Kundu,Eckhaus}
\begin{align}\label{KE}
\mathrm{i} q_t+q_{x x}+2|q|^2 q+4 \beta^2|q|^4 q-4 \mathrm{i} \beta\left(|q|^2\right)_x q=0	,
\end{align}
which contains quintic nonlinearity and the Raman effect in nonlinear optics. The Kundu-Eckhaus equation can be converted to the extended NLS equation above by means of a gauge transformation. Therefore, the following formula 
\begin{align}
q[N]=U[N] \exp \left[2 \mathrm{i} \beta \int|U[N]|^2 d x\right].
\end{align}
gives rise to the corresponding $N$th-order rogue waves for the Kundu-Eckhaus equation after deriving the rogue wave solutions for the extended NLS equation through the generalized Darboux transformation \cite{KE}.

Here is the explicit first-order rogue wave solution of the Kundu-Eckhaus equation
\begin{align}
q[1]=U[1] \exp \left[2 \mathrm{i} \beta \int|U[1]|^2 d x\right],	
\end{align}
where $U[1]=\exp [\mathrm{i} \theta] \frac{F_1+\mathrm{i} G_1}{D_1}, \int|U[1]|^2 d x=\frac{H_1}{D_1}$ with $\theta, F_1, G_1, D_1$ shown in \eqref{parameters} and
\begin{align}
\begin{gathered}
H_1=4 x^3+16\left(4 \beta^2+1\right) t^2 x+9 x+32 \beta\left(x^2+1\right) t.
\end{gathered}
\end{align}
We select the spatiotemporal region as $\Omega \times\left[T_0, T\right]=[-1.5,1.5]\times [-0.5,0.5]$, $\beta=\frac{1}{3}$ and divide the whole spatiotemporal region into $n_{\max}=4$ sub-domains:
\begin{align}
\begin{gathered}
\Omega \times\left[T_0, T_1\right]=[-1.5,1.5]\times[-1,-0.5], \quad \Omega \times\left[T_1, T_2\right]=[-1.5,1.5]\times[-0.5,0], 	\\ \Omega \times\left[T_2, T_3\right]=[-1.5,1.5]\times[0,0.5], \quad \Omega \times \left[T_3, T\right]=[-1.5,1.5]\times[0,0.5].
\end{gathered}
\end{align}
Then the PDE residuals are changed into
\begin{align}
\begin{gathered}
R_u:=-v_t+u_{xx}+2(u^2+v^2)u+4 \beta^2(u^2+v^2)^2 u +4 \beta (2 u u_x+2vv_x)v,\\
R_v:=u_t+v_{xx}+2(u^2+v^2)v+4 \beta^2(u^2+v^2)^2 v-4 \beta (2 u u_x+2vv_x)u.
\end{gathered}
\end{align}
Other details are similar to that of $\beta=0$ except for changing the number of nodes in each hidden layer to 64, so we omit them here.

\begin{figure}[htbp]
\centering
\includegraphics[width=5.5cm,height=4cm]{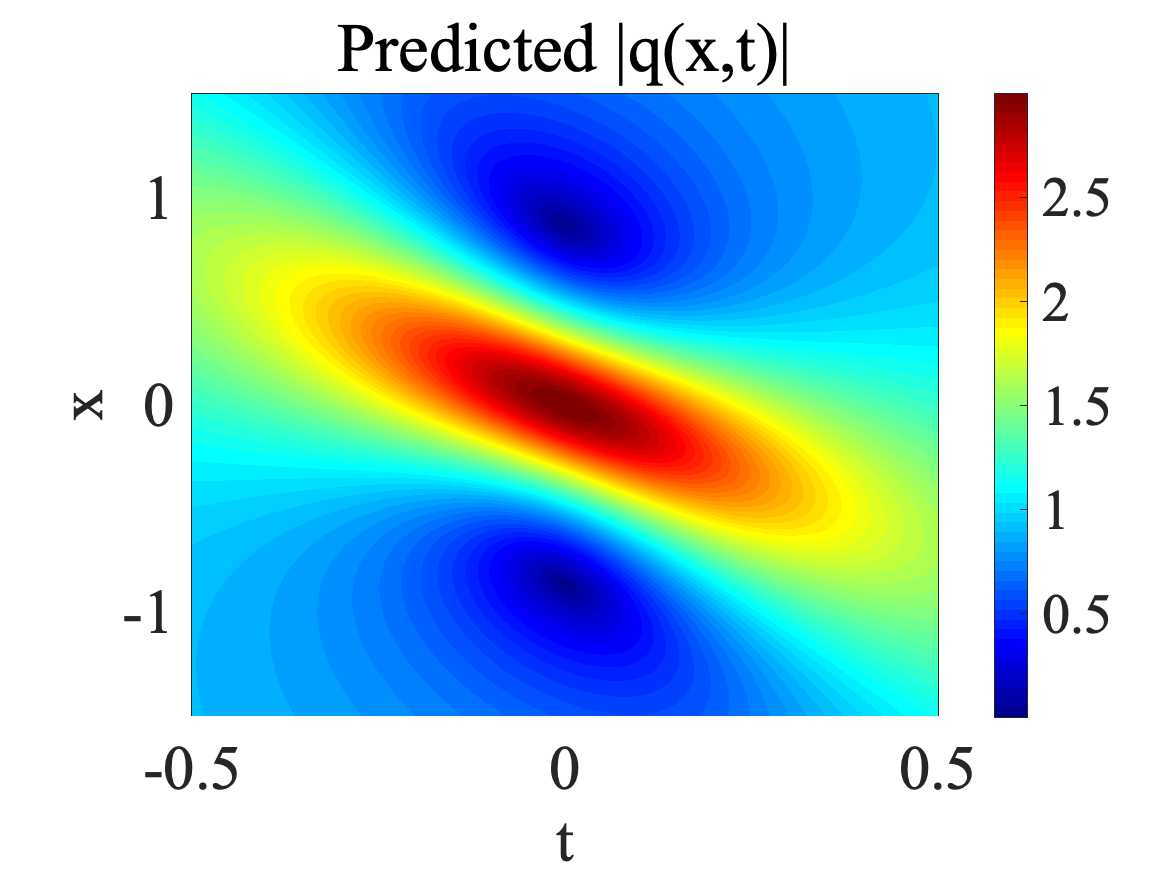}
$a$
\includegraphics[width=5.5cm,height=4cm]{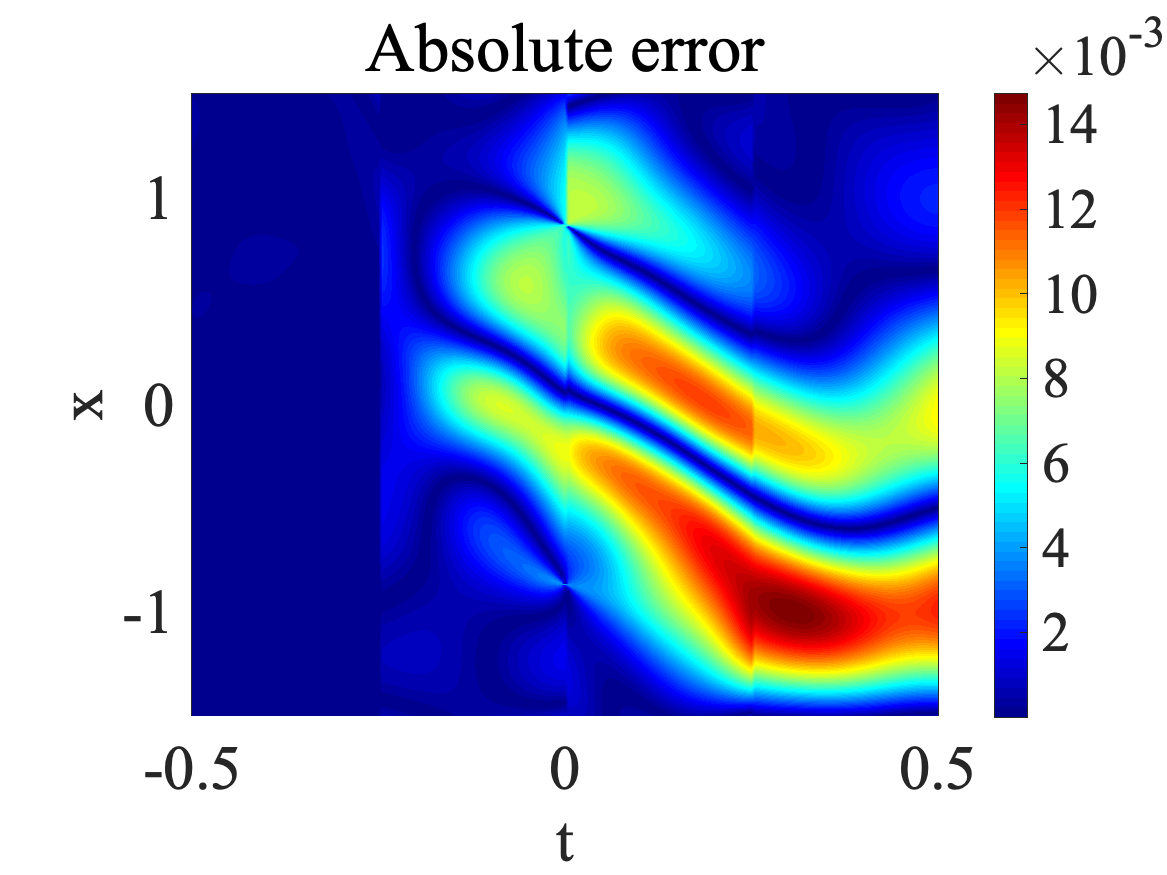}
$b$
\includegraphics[width=5.5cm,height=4cm]{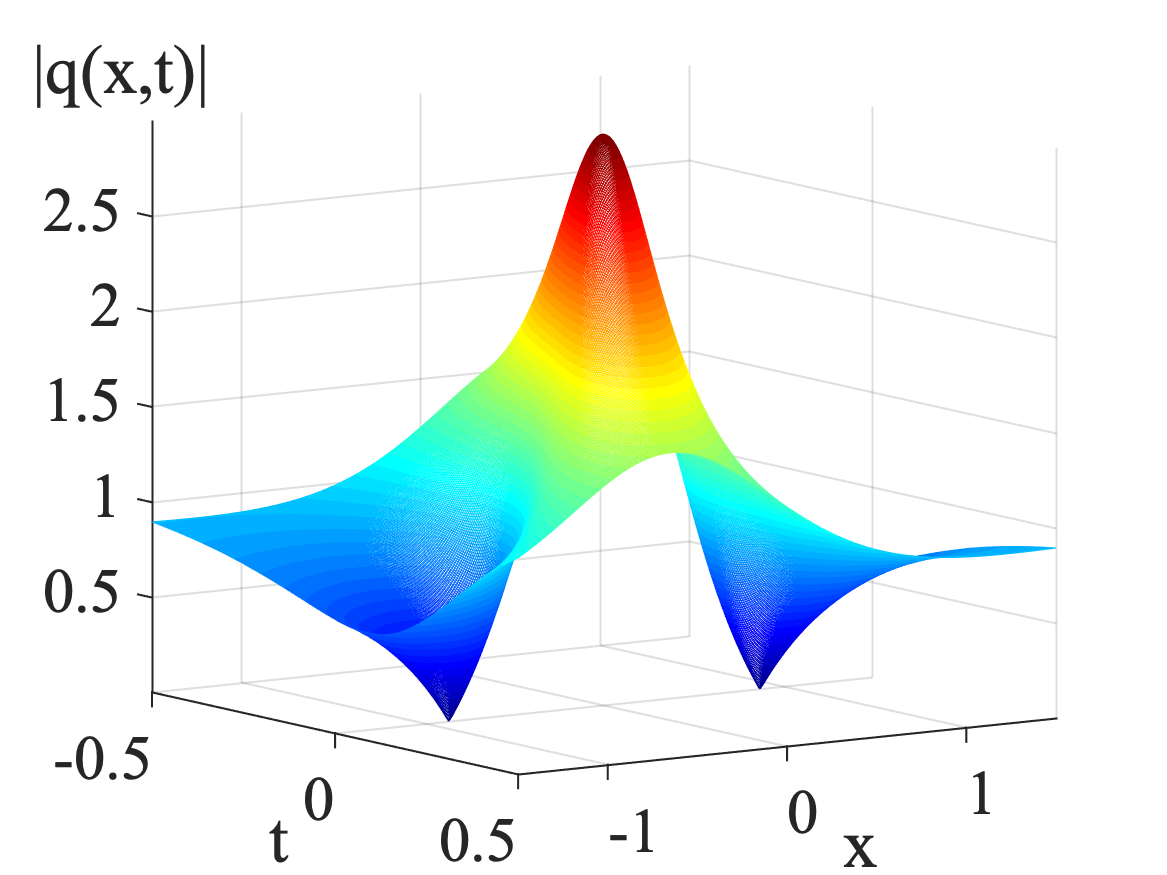}
$c$
\caption{(Color online) Data-driven first-order rogue wave ($\beta=\frac{1}{3}$) by Ibc-PINN: (a) The density diagram of the predicted solution $|q(x,t)|$; (b) The density diagram of absolute error; (c) The three-dimensional plot of the data-driven rogue wave solution $|q(x,t)|$.}
\label{fig3-2}
\end{figure}

The relative $\mathbb{L}_2$ error of the data-driven solution $|q(x,t)|$ for the KE equation obtained by Ibc-PINN is 3.789e-03. Although the network structure and other parameter selection of above two first-order rogue wave solutions are consistent on the whole, the data-driven solution with $\beta=\frac{1}{3}$ has lower precision even with smaller selected training domain. The reason behind this is that the form of the governing function is  more complex, thereby increasing the difficulty of training when $\beta \neq 0$, which involves quintic and Raman-effect nonlinear terms. According to Fig. \ref{fig3-2}, the shape of the rogue wave does not change drastically but it rotates a certain angle in a clockwise direction as $\beta$ changes from zero to non zero.

\begin{table}[htbp]
\caption{Relative $\mathbb{L}_2$ errors of the data-driven first-order rogue waves for the NLS and KE equations by bc-PINN and Ibc-PINN.}
\label{table3-1} 
\centering
\begin{tabular}{c|cccc}
\bottomrule
\multirow{2}{*}{} & \multicolumn{1}{c|}{\multirow{2}{*}{$\Omega$}} & \multicolumn{1}{c|}{\multirow{2}{*}{$[T_0,T]$}} & \multicolumn{2}{c}{relative $\mathbb{L}_2$ error of $|q(x,t)|$} \\ \cline{4-5} 
                  & \multicolumn{1}{c|}{}                   & \multicolumn{1}{c|}{}                   & bc-PINN   & Ibc-PINN   \\ \hline
$\beta=0$                & $[-2,2]$                                       & $[-1,1]$                                       & 1.062e-03         & 4.843e-04          \\ \hline
$\beta=\frac{1}{3}$ (KE)               & $[-1.5,1.5]$                                       & $[-0.5,0.5]$                                      & 4.471e-03         & 3.789e-03          \\ \toprule
\end{tabular}
\end{table}

Finally, relative $\mathbb{L}_2$ errors generated by bc-PINN and Ibc-PINN are summarized in Table \ref{table3-1} for intuitive comparison. The Ibc-PINN method has improved accuracy, especially for the case when $\beta=0$.

\subsubsection{Data-driven second-order rogue waves}
\quad

Based on the above derivation, the second-order rogue wave solution of the extended NLS equation naturally takes the form of
\begin{align}
U[2]=\exp [\mathrm{i} \theta] \frac{F_2+\mathrm{i} G_2}{D_2},	
\end{align}
where the concrete expressions of $F_2, G_2, D_2$ are not detailed here due to space limitations and they contain free parameters ($m_1$ and $n_1$) and were given in \cite{KE}. After taking $\beta=0$, we obtain the corresponding second-order rogue wave solution $q(x,t)$ for the NLS equation.

Once the spatiotemporal region is selected, the corresponding initial and boundary conditions can be derived from the exact solution. After acquiring the initial-boundary data obtained by taking two different sets of free parameters ($m_1$ and $n_1$), the Ibc-PINN method is utilized to learn second-order rogue wave solutions.

\begin{itemize}
 \item $m_1=0, n_1=0$	
\end{itemize}

Here the spatiotemporal region  $\Omega \times\left[T_0, T\right]=[-2,2]\times [-0.8,0.8]$ are also equally divided into four sub-domains. The loss function \eqref{KE-loss1} is employed for optimizing subnet-1 in the first time segment $[-0.8,-0.4]$, while the loss \eqref{KE-lossn} for subnet-$n$ ($n=2,3,4$) in subsequent ones ($[-0.4,0]$, $[0,0.4]$, $[0.4,0.8]$).

We establish a neural network with the depth of 7 and width of 128. Other parameters are selected as $N_i=128$, $N_b=50$, $N_r=20000$, and incremental $N_s=128 \times 50 \times (n-1)$ for the $n$th sub-domain. The effective techniques mentioned above have also been adopted, including the initial condition guided learning (ICGL), transfer learning and weight freezing, and specific details are consistent with the previous subsection.

\begin{figure}[htbp]
\centering
\includegraphics[width=5.5cm,height=4cm]{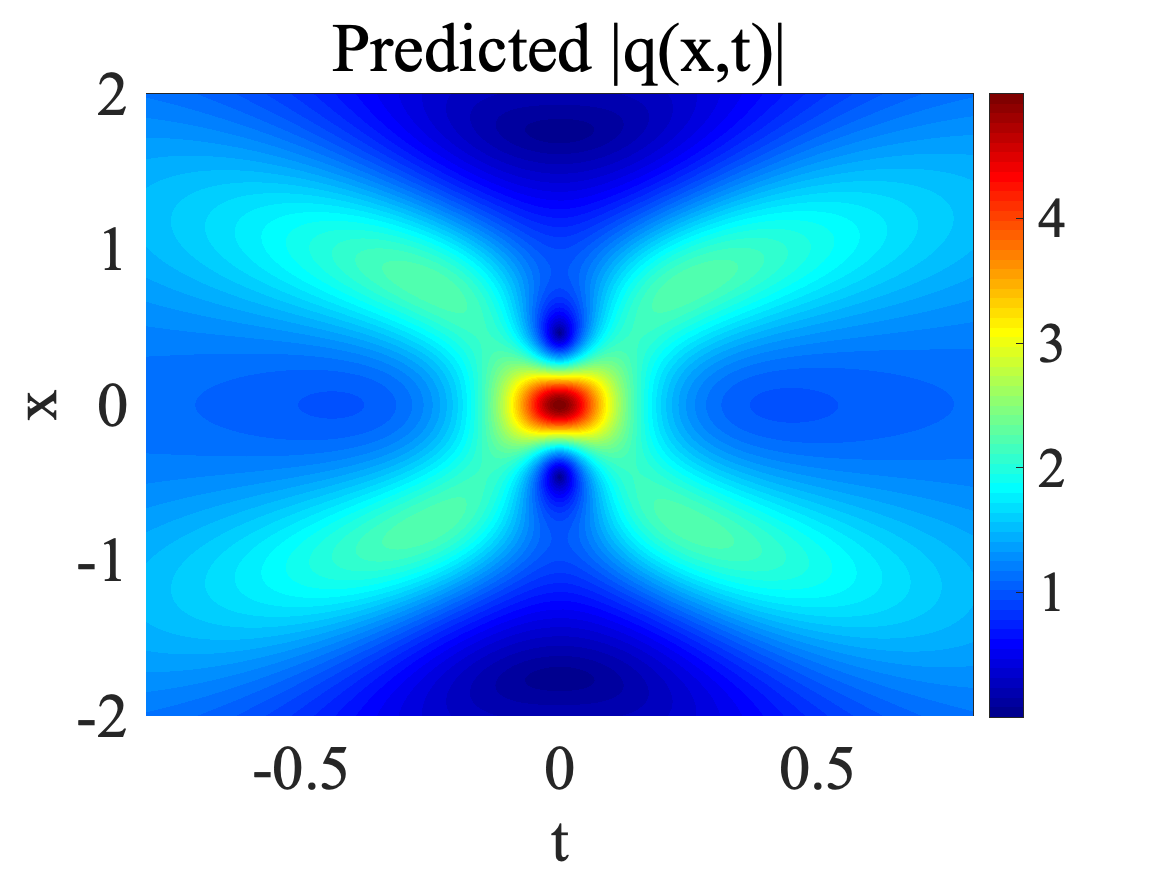}
$a$
\includegraphics[width=5.5cm,height=4cm]{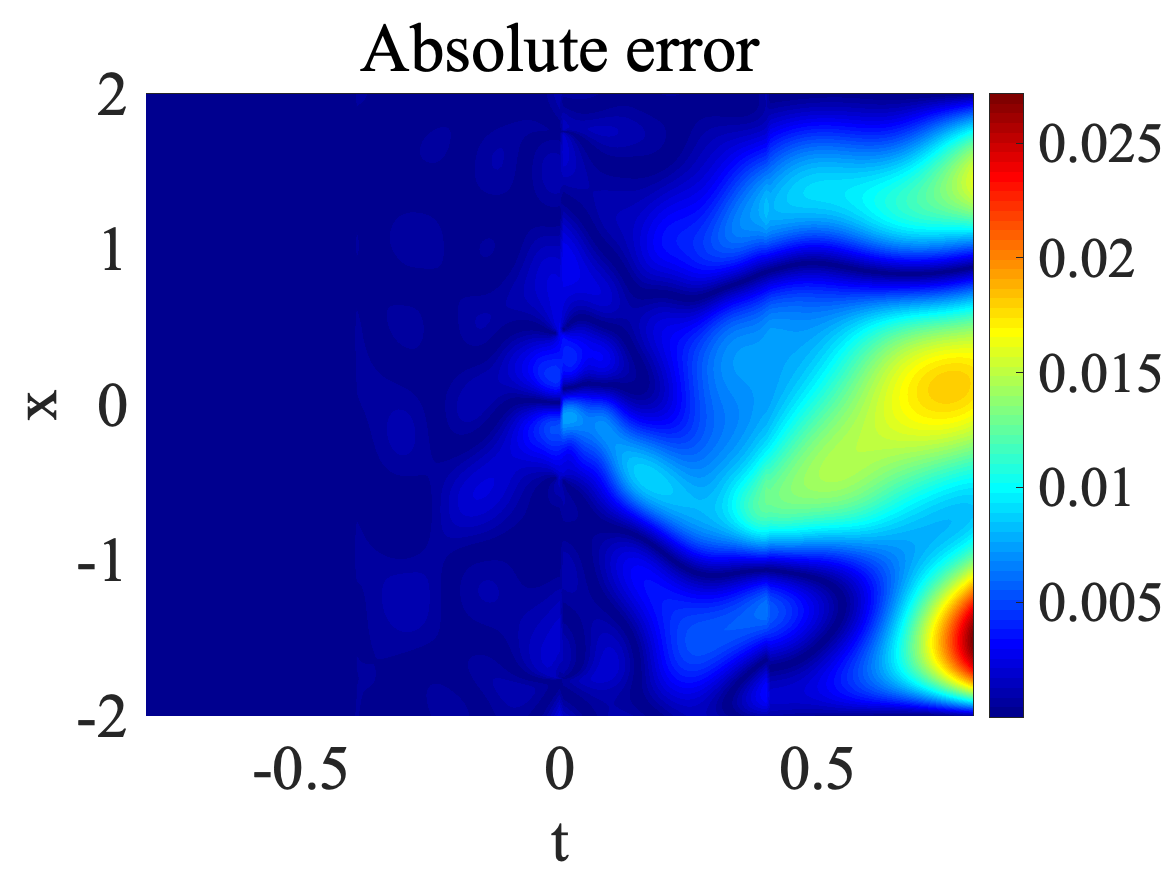}
$b$
\includegraphics[width=5.5cm,height=4cm]{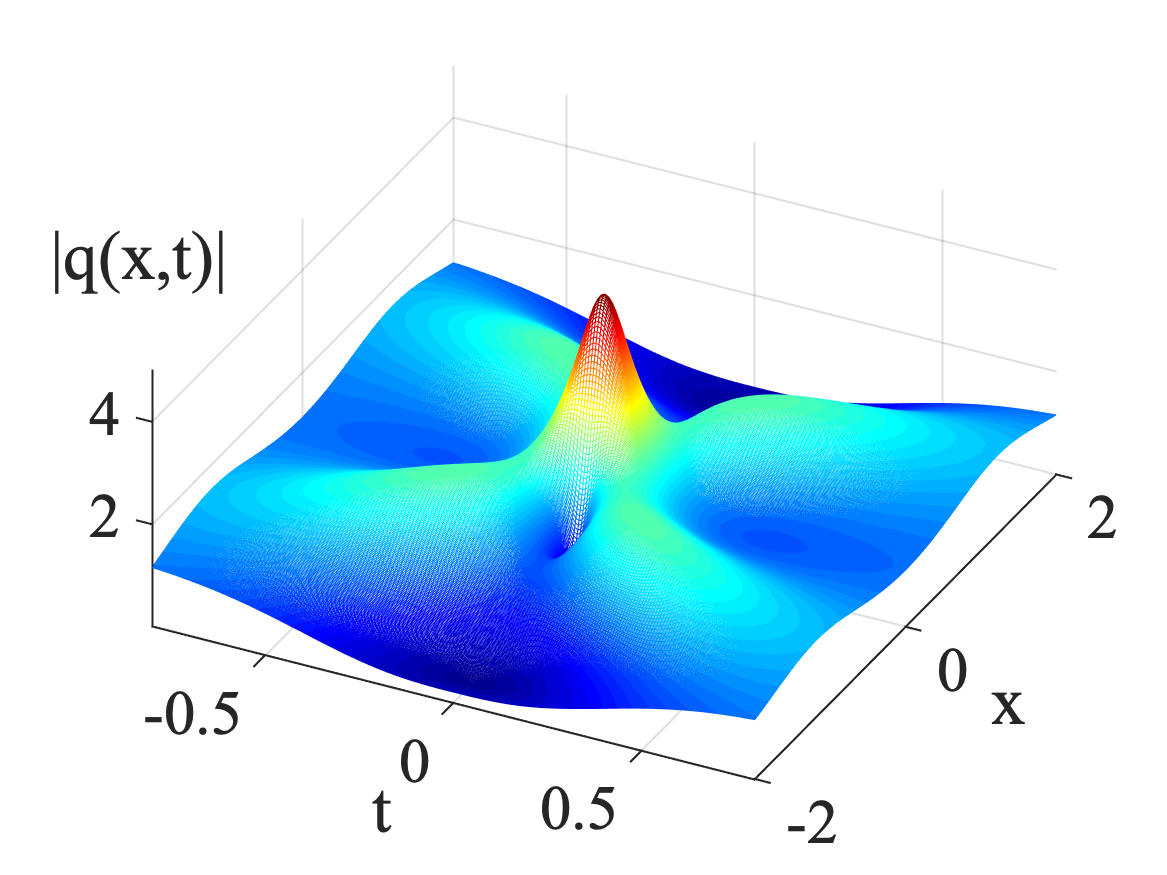}
$c$
\caption{(Color online) Data-driven second-order rogue wave ($m_1=0, n_1=0$) by Ibc-PINN: (a) The density diagram of the predicted solution $|q(x,t)|$; (b) The density diagram of absolute error; (c) The three-dimensional plot of the data-driven rogue wave solution $|q(x,t)|$.}
\label{fig3-3}
\end{figure}

When $m_1=0, n_1=0$, we successfully simulate the fundamental second-order rogue wave solution by Ibc-PINN shown in Fig. \ref{fig3-3}. The maximum value of $|q(x,t)|$ is approximately 5, reaching at point $(0,0)$.

\begin{itemize}
 \item $m_1=10, n_1=0$	
\end{itemize}

Under this set of parameter values, the morphology of the rogue wave solution becomes more complex. Thus, we will divide the selected time domain $[T_0, T]=[-1.5, 1.5]$ into $n_{\max}=6$ equal parts to apply the Ibc-PINNs. Take the space range as $\Omega =[-4,4]$ and choose $N_i=256$ due to the expansion of the spatial region. Accordingly, $N_s$ is changed to $256 \times 50 \times (n-1)$ and we select $N_{SI}=256*51$ for the $n$th sub-domain. Other parameter settings, including network structure, are consistent with $m_1=n_1=0$.

\begin{figure}[htbp]
\centering
\includegraphics[width=5.5cm,height=4cm]{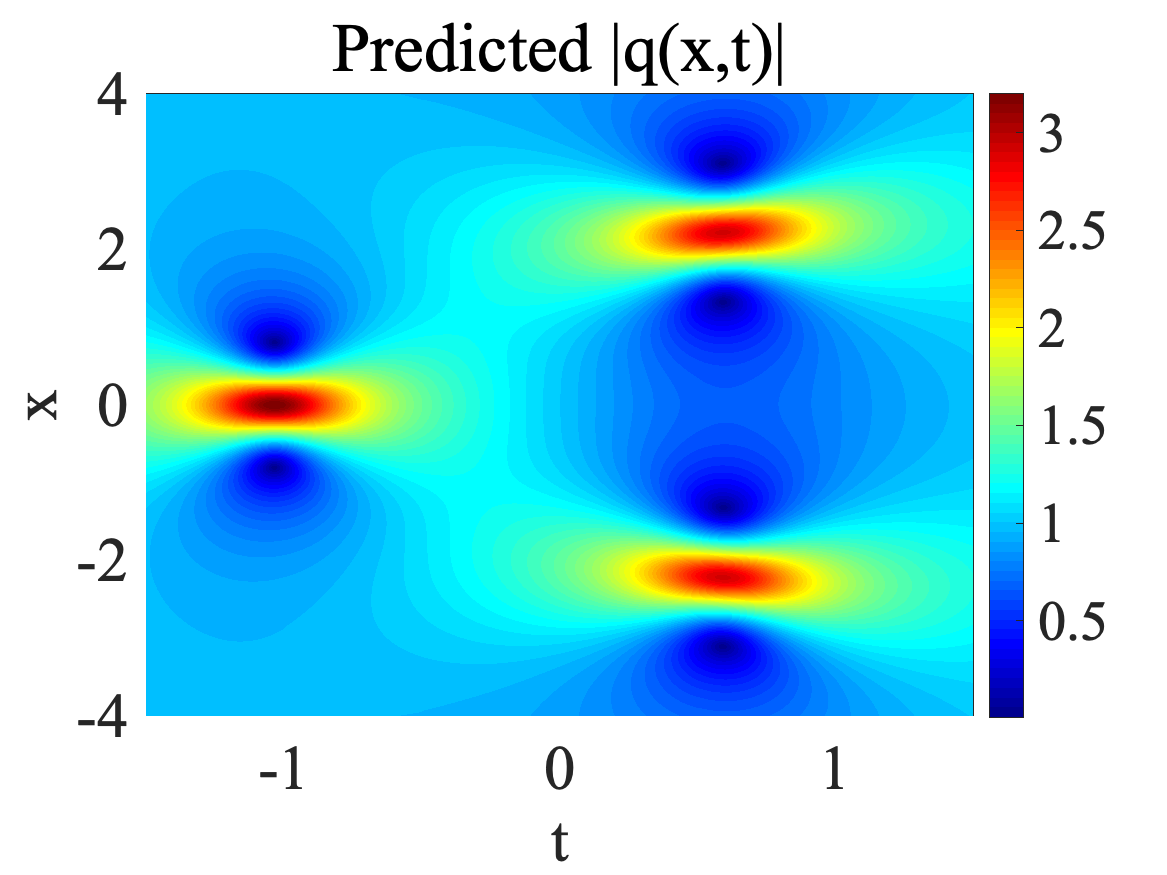}
$a$
\includegraphics[width=5.5cm,height=4cm]{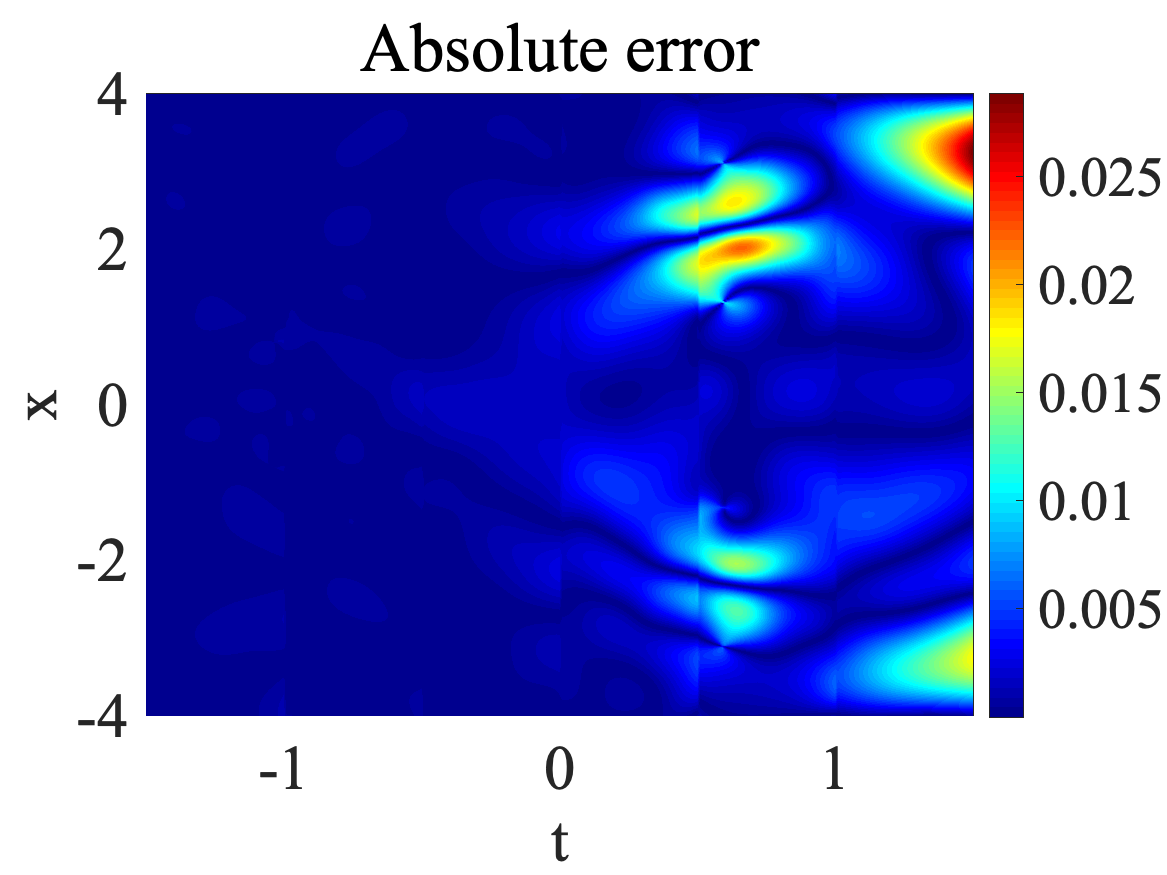}
$b$
\includegraphics[width=5.5cm,height=4cm]{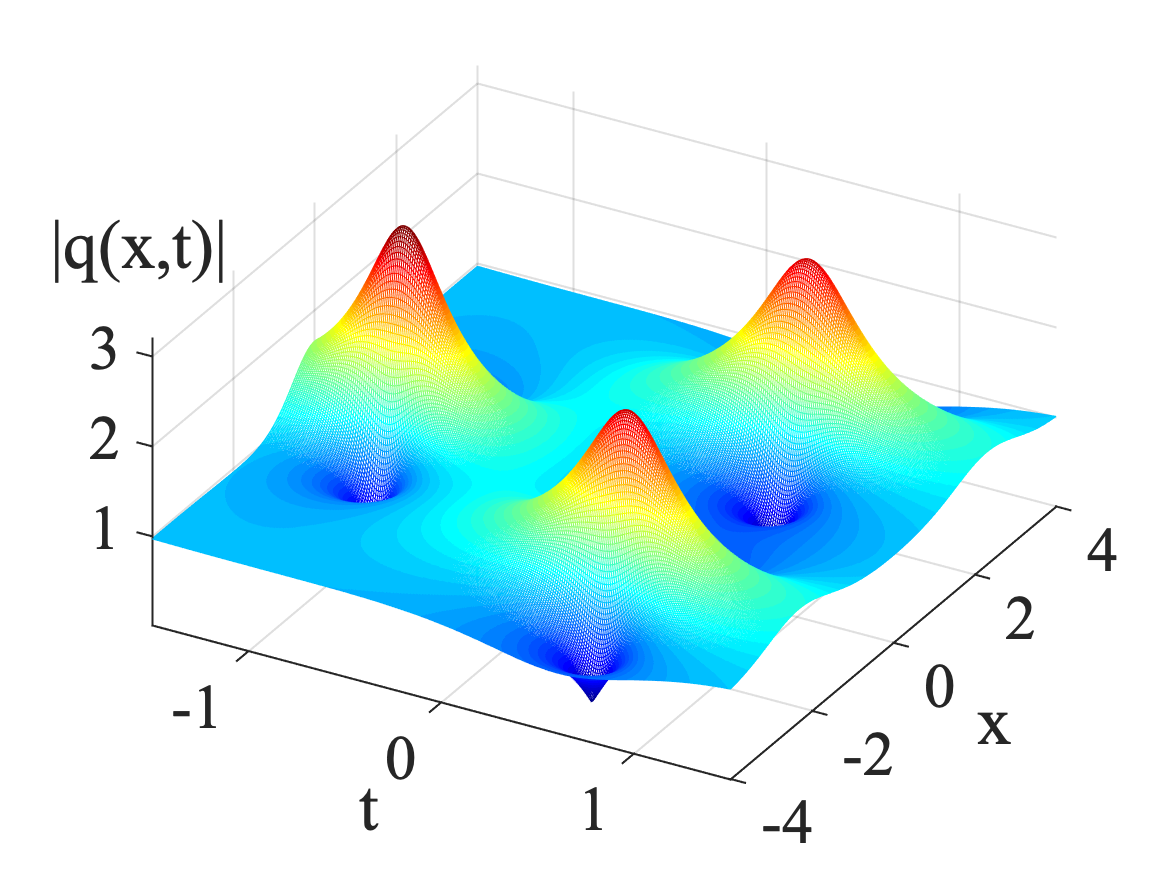}
$c$
\caption{(Color online) Data-driven second-order rogue wave ($m_1=10, n_1=0$) by Ibc-PINN: (a) The density diagram of the predicted solution $|q(x,t)|$; (b) The density diagram of absolute error; (c) The three-dimensional plot of the data-driven rogue wave solution $|q(x,t)|$.}
\label{fig3-4}
\end{figure}

By means of Ibc-PINN, data-driven second-order rogue wave is displayed in Fig. \ref{fig3-4}. It vividly illustrates that the fundamental second-order rogue wave is separated into three first-order rogue waves: a single and a double spatial hump. We also observe that the single hump appears at $t \approx -1$ and rapidly decays while  two spatial humps rise up simultaneously at $t \approx 0.6$. The spatial coordinates corresponding to the three peaks are $x=0$, $x\approx \pm 2$ separately.

Comparing the performance of the two methods shown in Table \ref{table3-2}, the error of bc-PINN is reduced to approximately half of its original value after improvement.

\begin{table}[htbp]
\caption{Relative $\mathbb{L}_2$ errors of the data-driven second-order rogue waves for the NLS equation by bc-PINN and Ibc-PINN.}
\label{table3-2} 
\centering
\begin{tabular}{c|cccc}
\bottomrule
\multirow{2}{*}{} & \multicolumn{1}{c|}{\multirow{2}{*}{$\Omega$}} & \multicolumn{1}{c|}{\multirow{2}{*}{$[T_0,T]$}} & \multicolumn{2}{c}{relative $\mathbb{L}_2$ error of $|q(x,t)|$} \\ \cline{4-5} 
                  & \multicolumn{1}{c|}{}                   & \multicolumn{1}{c|}{}                   & bc-PINN   & Ibc-PINN   \\ \hline
$m_1=0$                & $[-2,2]$                                       & $[-0.8,0.8]$                                       & 6.368e-03         & 3.690e-03          \\ \hline
$m_1=10$               & $[-4,4]$                                       & $[-1.5,1.5]$                                      & 6.580e-03         & 3.543e-03         \\ \toprule
\end{tabular}
\end{table}

\subsubsection{Data-driven third-order rogue waves}
\quad

Similarly, after denoting
\begin{align}
\Psi_1[l]= & \Psi_1^{[0]}+\sum_{j=1}^l T_1[j] \Psi_1^{[1]}+\sum_{j=1}^l \sum_{k=1}^{j-1} T_1[j] T_1[k] \Psi_1^{[2]} +\ldots+T_1[l] T_1[l-1] \ldots T_1[1] \Psi_1^{[l]}
\end{align}
the $N$th-step generalized Darboux transformation results in
\begin{align}
\Psi[N] & =T[N] T[N-1] \ldots T[1] \Psi, \\
T[l]& =\zeta I-H[l-1] \Lambda_l H[l-1]^{-1}, \\
U[N] & =U[0]-2 \mathrm{i}\left(\zeta_1-\zeta_1^*\right) \sum_{l=0}^{N-1} \frac{\psi_1[l] \phi_1[l]^*}{\left(\left|\psi_1[l]\right|^2+\left|\phi_1[l]\right|^2\right)},
\end{align}
where
\begin{align}
\left(\psi_1[l], \phi_1[l]\right)^T & =\Psi_1[l], \\
H[l-1] & =\left(\begin{array}{ll}
\psi_1[l-1] & -\phi_1[l-1]^* \\
\phi_1[l-1] & \psi_1[l-1]^*
\end{array}\right), \\
\Lambda_l & =\left(\begin{array}{cc}
\zeta_1 & 0 \\
0 & \zeta_1^*
\end{array}\right), 1 \leqslant l \leqslant N,
\end{align}
and can derive the $N$-order rogue wave solution $U[N]$ for the extended NLS equation. The specific details of the parameters involved have been presented in \cite{KE} and will not be elaborated here. Then we choose $\beta=0, N=3$ and simulate the third-order rogue wave solution $q(x,t)$ for the NLS equation by Ibc-PINN.

\begin{itemize}
 \item $m_1=10, n_1=0, m_2=0, n_2=0$	
\end{itemize}

We divide the spatiotemporal region $\Omega \times\left[T_0, T\right]=[-5,5]\times [-2,1.5]$ into $n_{\max}=7$ sub-domains with a time step of 0.5. The parameters of each dataset in the loss function are taken as $N_i=128, N_b=50, N_r=20000$ and $N_s=128 \times 50 \times (n-1)$ for the $n$th sub-domain. Afterwards, a neural network with a depth of 7 and a width of 64 is established to learn dynamic behaviors of the solution in each subregion. We opt for a blend of the Adam and L-BFGS optimizers, commencing with 10,000 iterations using the Adam optimization algorithm, followed by subsequent iterations employing the L-BFGS algorithm until convergence. Additionally, during the initial training phase in each subregion, we leverage ICGL technology to match the initial conditions. This step accounts for 10\% of the Adam iteration count and $N_{SI}$ is taken as $128*51$. Similarly, techniques such as transfer learning and weight freezing are employed to accelerate the training speed of the network, and the number of frozen weight layers is set to 2.

\begin{figure}[htbp]
\centering
\includegraphics[width=5.5cm,height=4cm]{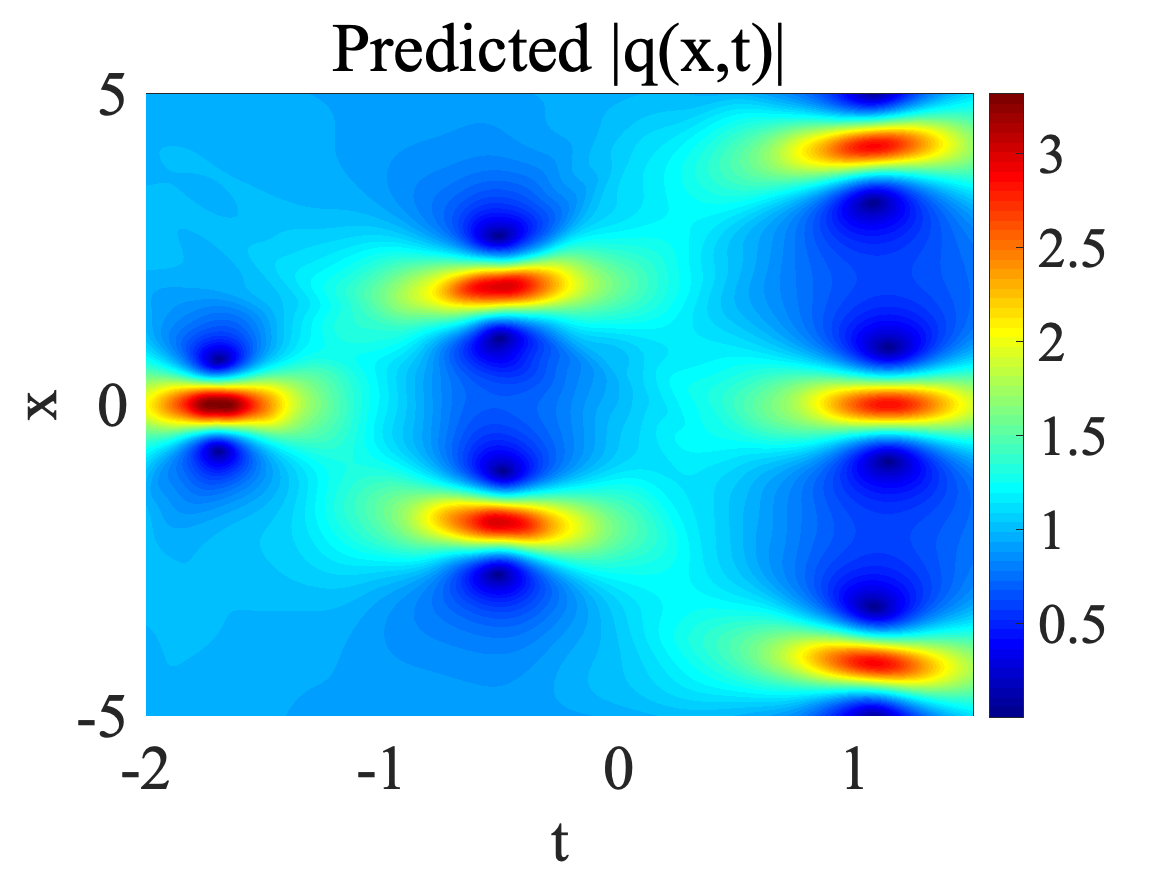}
$a$
\includegraphics[width=11cm,height=3cm]{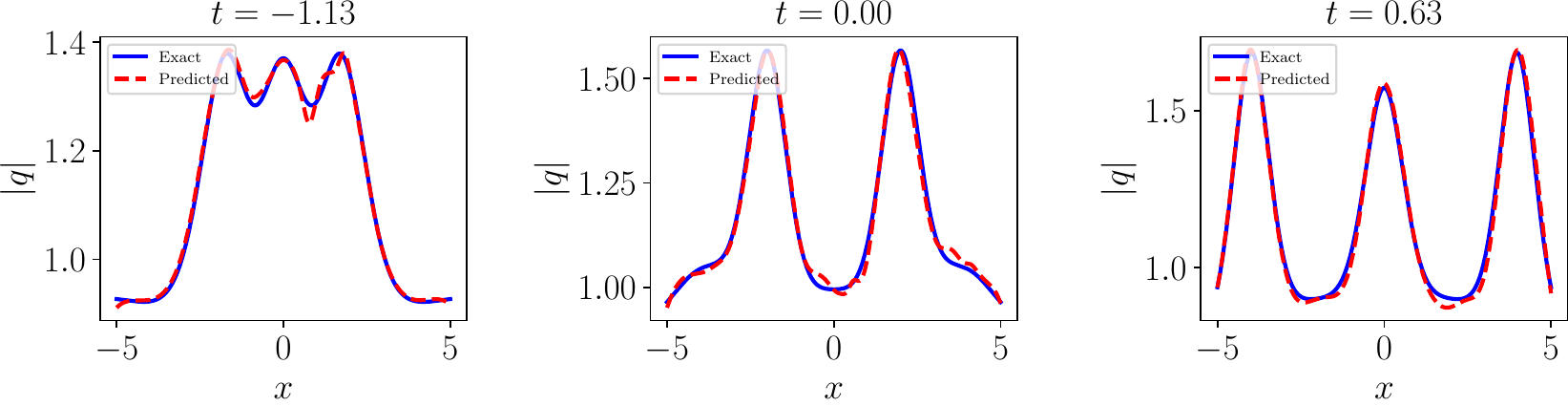}
$b$\\
\includegraphics[width=5.5cm,height=4cm]{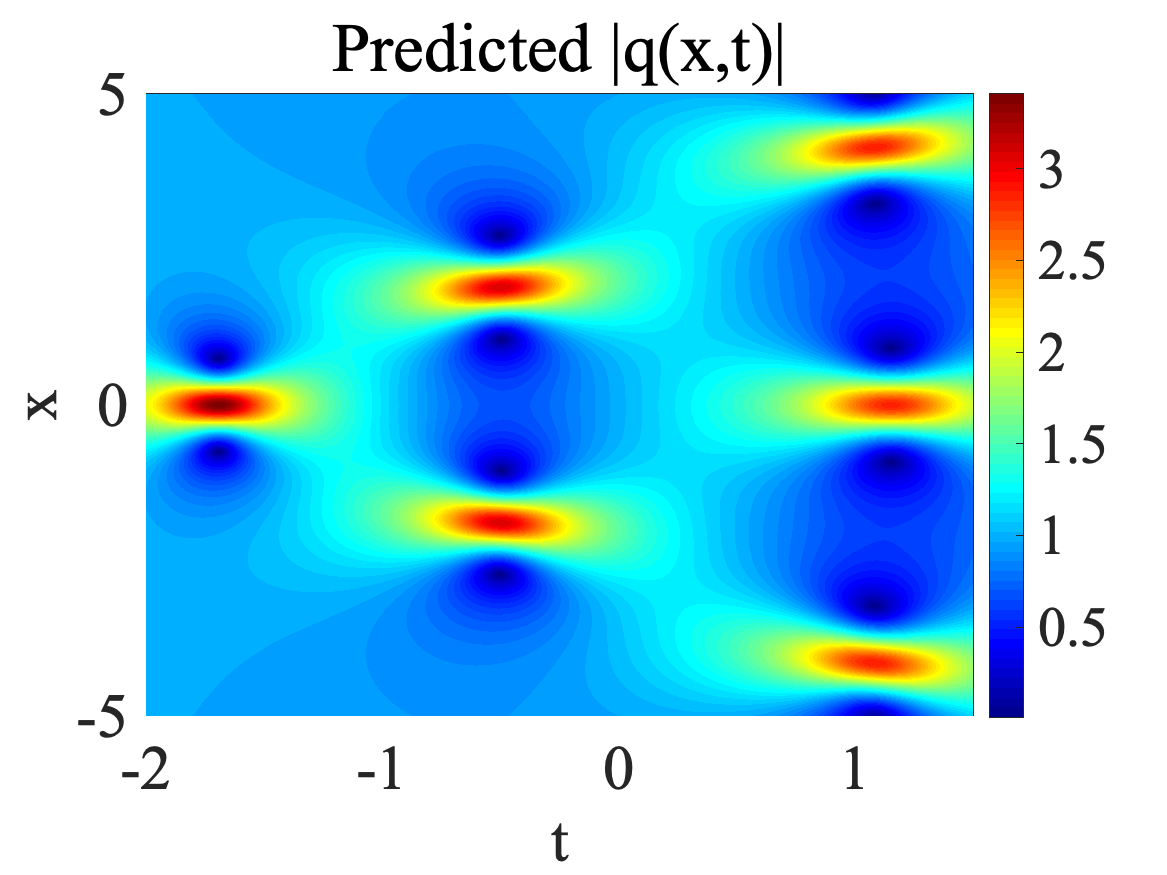}
$c$
\includegraphics[width=11cm,height=3cm]{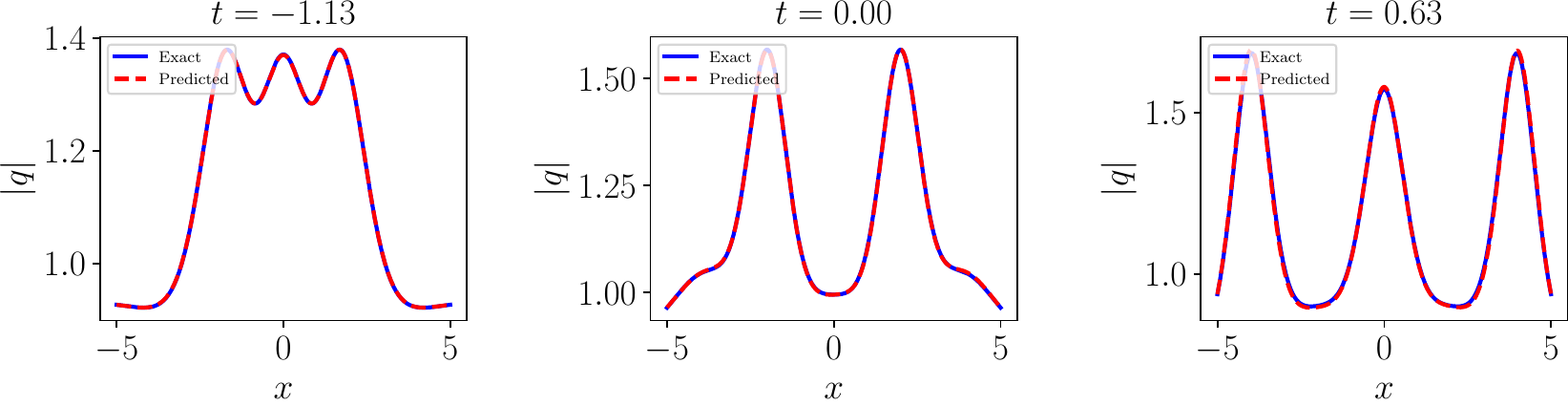}
$d$
\caption{(Color online) Data-driven third-order rogue wave ($m_1=10, n_1=m_2=n_2=0$): The density diagrams of the predicted solution $|q(x,t)|$: (a) by bc-PINN and (c) by Ibc-PINN; Comparison between the predicted and exact solutions at the three temporal snapshots of $|q(x,t)|$: (b) by bc-PINN and (d) by Ibc-PINN.}
\label{fig3-5}
\end{figure}

\begin{figure}[htbp]
\centering
\includegraphics[width=5.5cm,height=4cm]{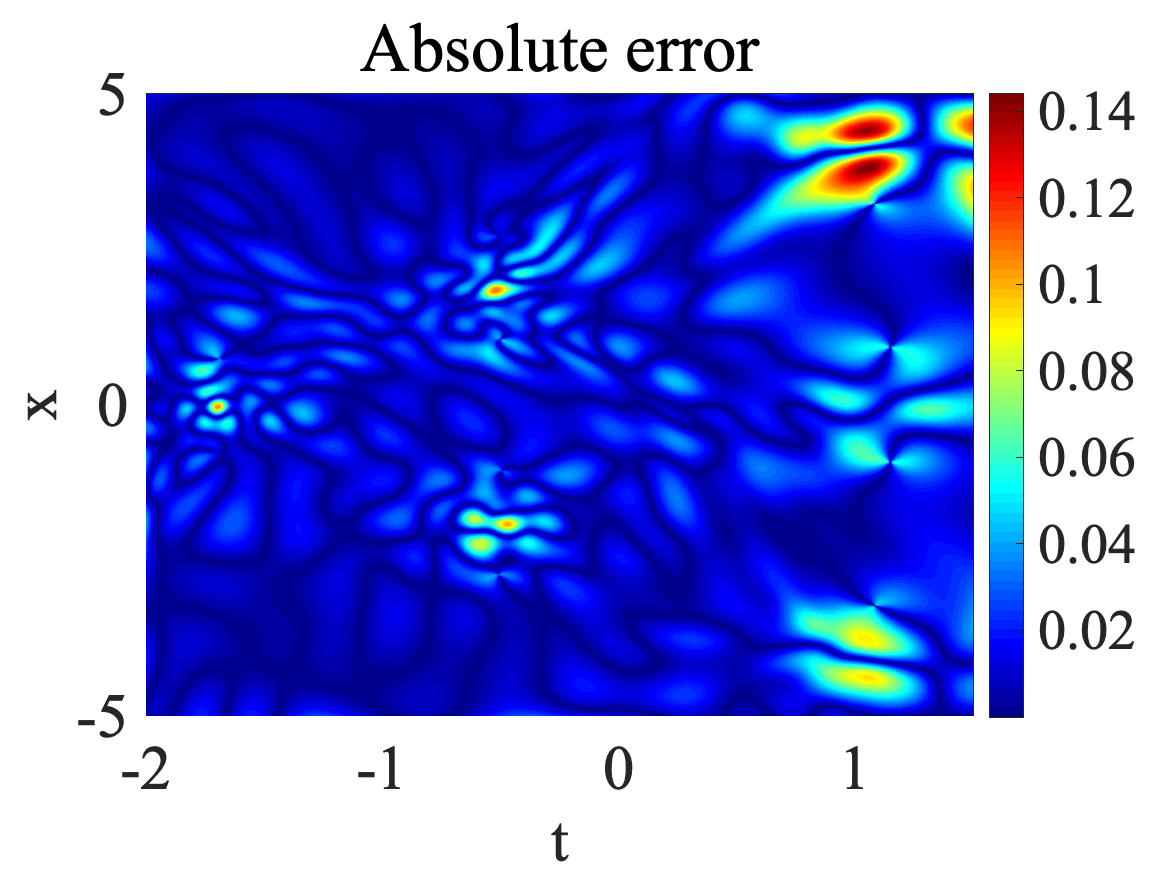}
$a$
\includegraphics[width=5.5cm,height=4cm]{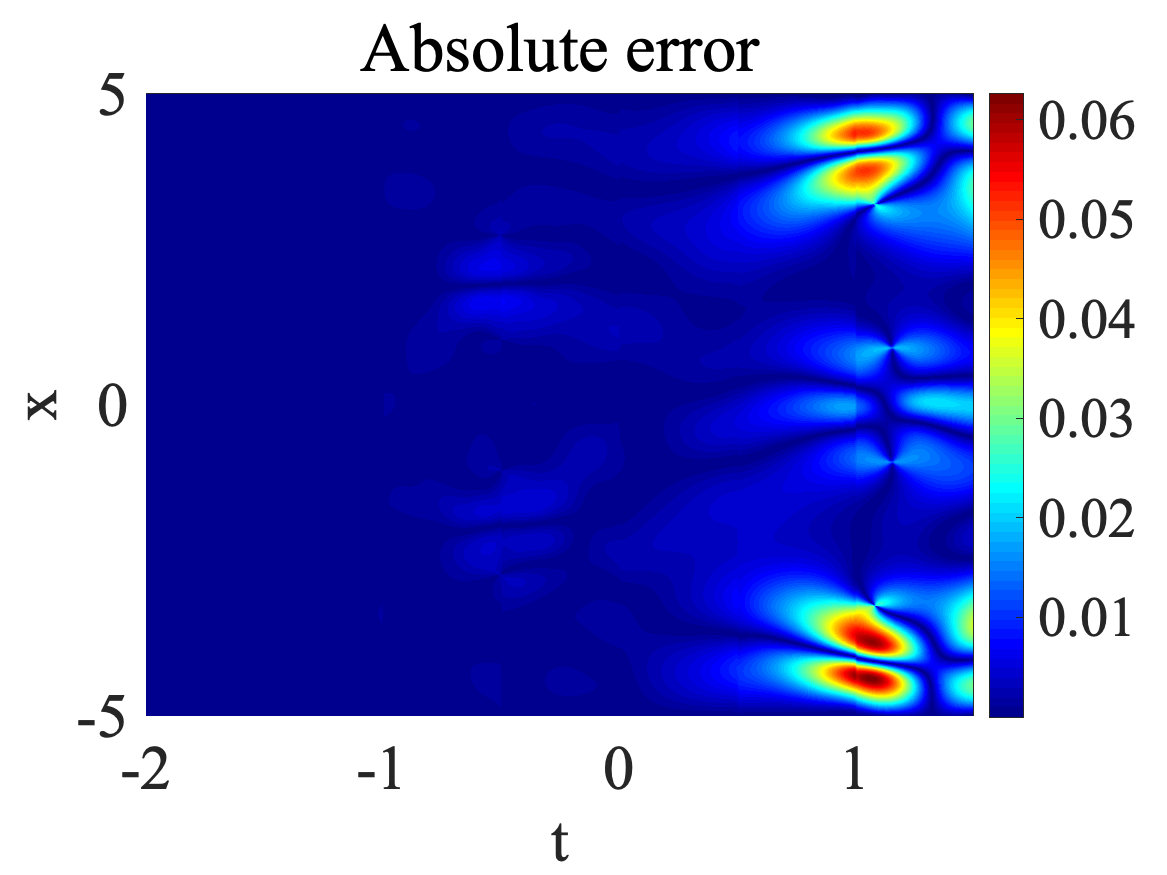}
$b$
\includegraphics[width=5.5cm,height=4cm]{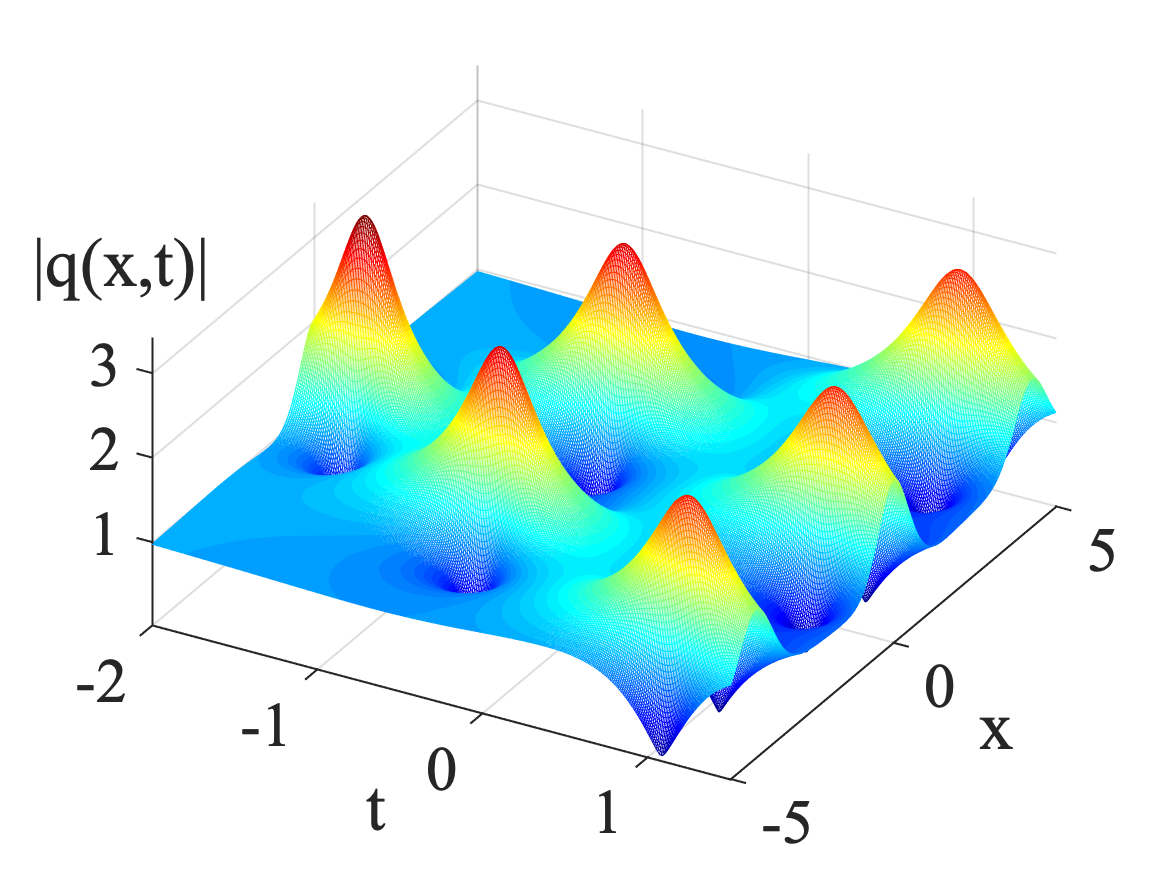}
$c$
\caption{(Color online) Data-driven third-order rogue wave ($m_1=10, n_1=m_2=n_2=0$): The density diagrams of absolute error: (a) by bc-PINN and (b) by Ibc-PINN; (c) The three-dimensional plot of the data-driven rogue wave solution $|q(x,t)|$ by Ibc-PINN.}
\label{fig3-6}
\end{figure}

The density diagrams of the predicted $|q(x,t)|$ obtained by bc-PINN and Ibc-PINN as well as the curve plots to show comparison between the predicted and exact solutions at the three temporal snapshots are depicted in Fig. \ref{fig3-5}. Under this parameter selection of $m_1=10, n_1=0, m_2=0, n_2=0$, the fundamental third-order rogue wave splits into a triangular arrangement of six first-order rogue waves. A single hump forms at $t \approx -1.75$, followed by the symmetric appearance of two rogue waves around $t \approx -0.5$, and a triple spatial hump rapidly rises at $t \approx 1.1$ and $t \approx 1.2$. We notice that the contour lines of the density plot for the third-order rogue wave obtained by bc-PINN are irregular and lack smoothness. Moreover, as shown in Fig. \ref{fig3-5}(b), there is a certain gap between the predicted solution and the exact solution curve. In contrast, the predictive outcomes of Ibc-PINN demonstrate a notable concordance with the reference solution. The absolute error plots in Fig. \ref{fig3-6} also reveals the difference in accuracy between these two methods.

\begin{itemize}
 \item $m_1=0, n_1=0, m_2=50, n_2=0$	
\end{itemize}

\begin{figure}[htbp]
\centering
\includegraphics[width=5.5cm,height=4cm]{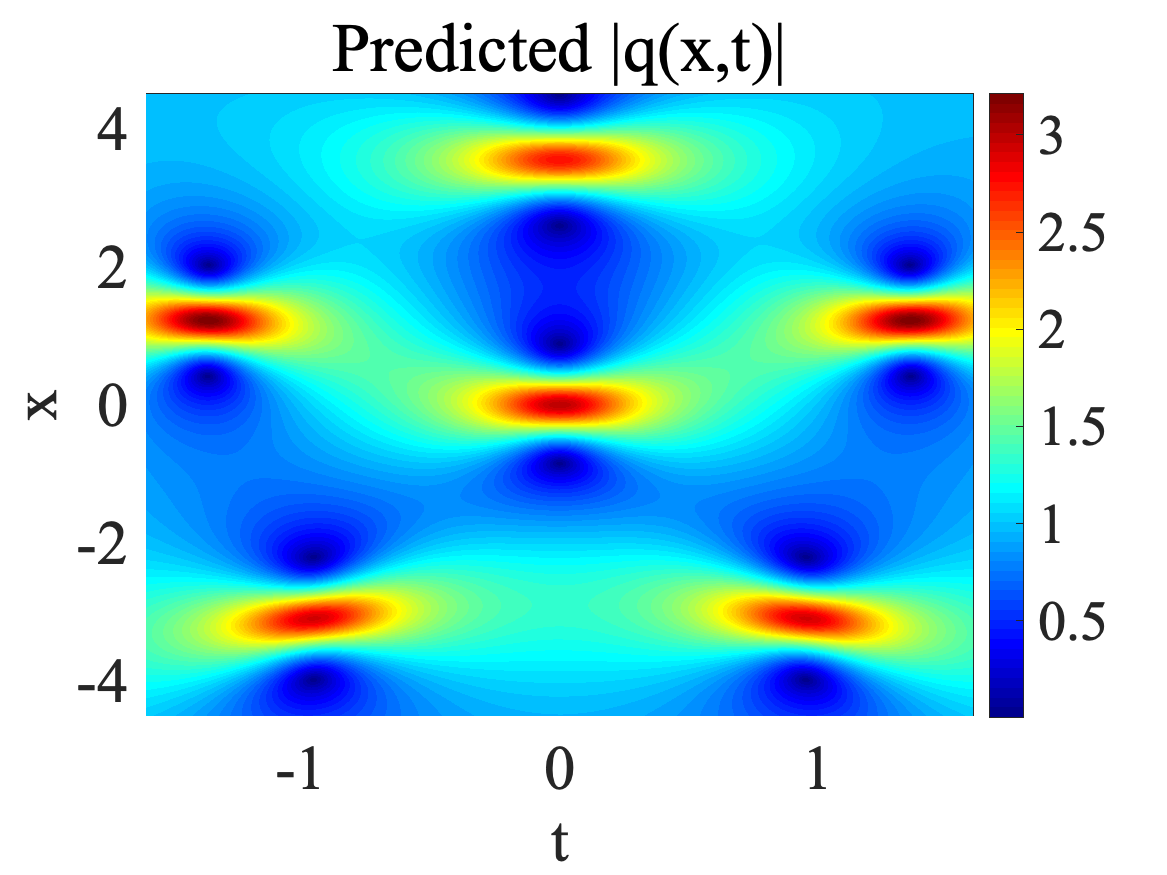}
$a$
\includegraphics[width=5.5cm,height=4cm]{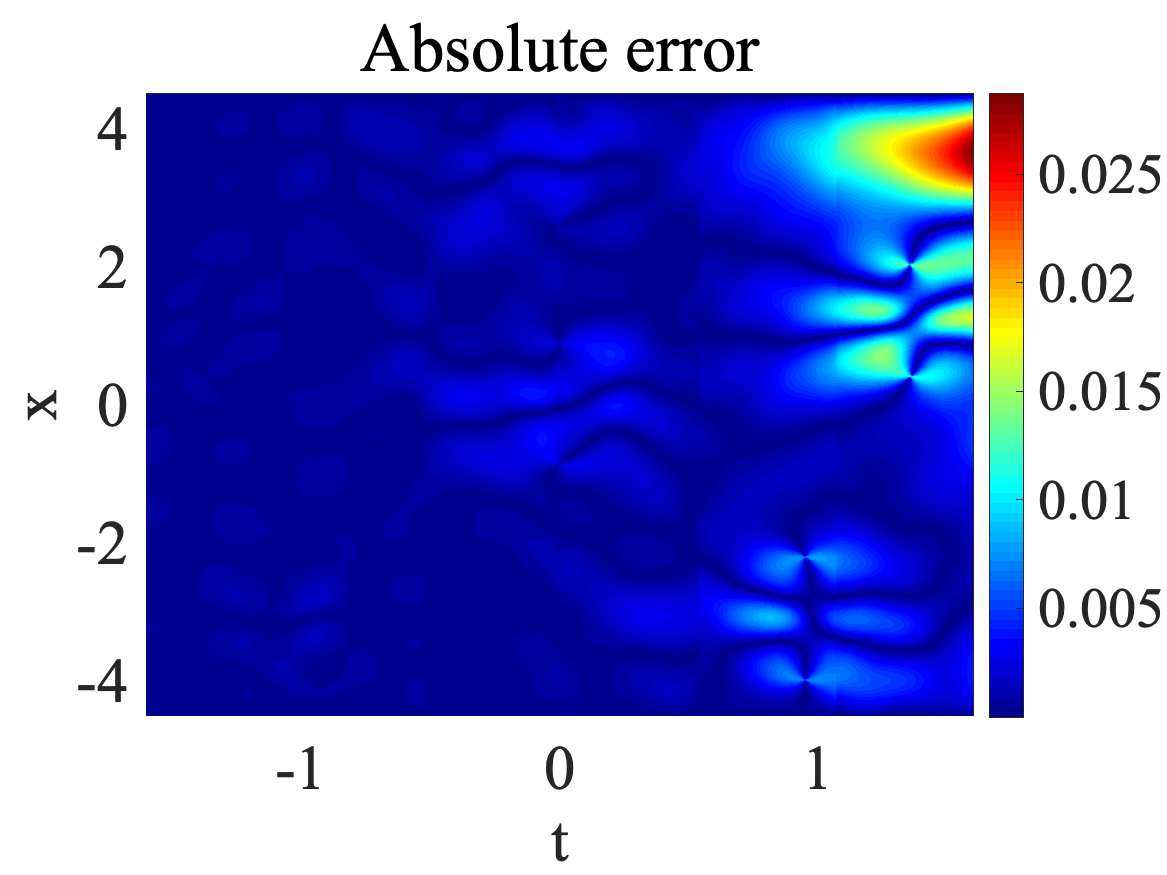}
$b$
\includegraphics[width=5.5cm,height=4cm]{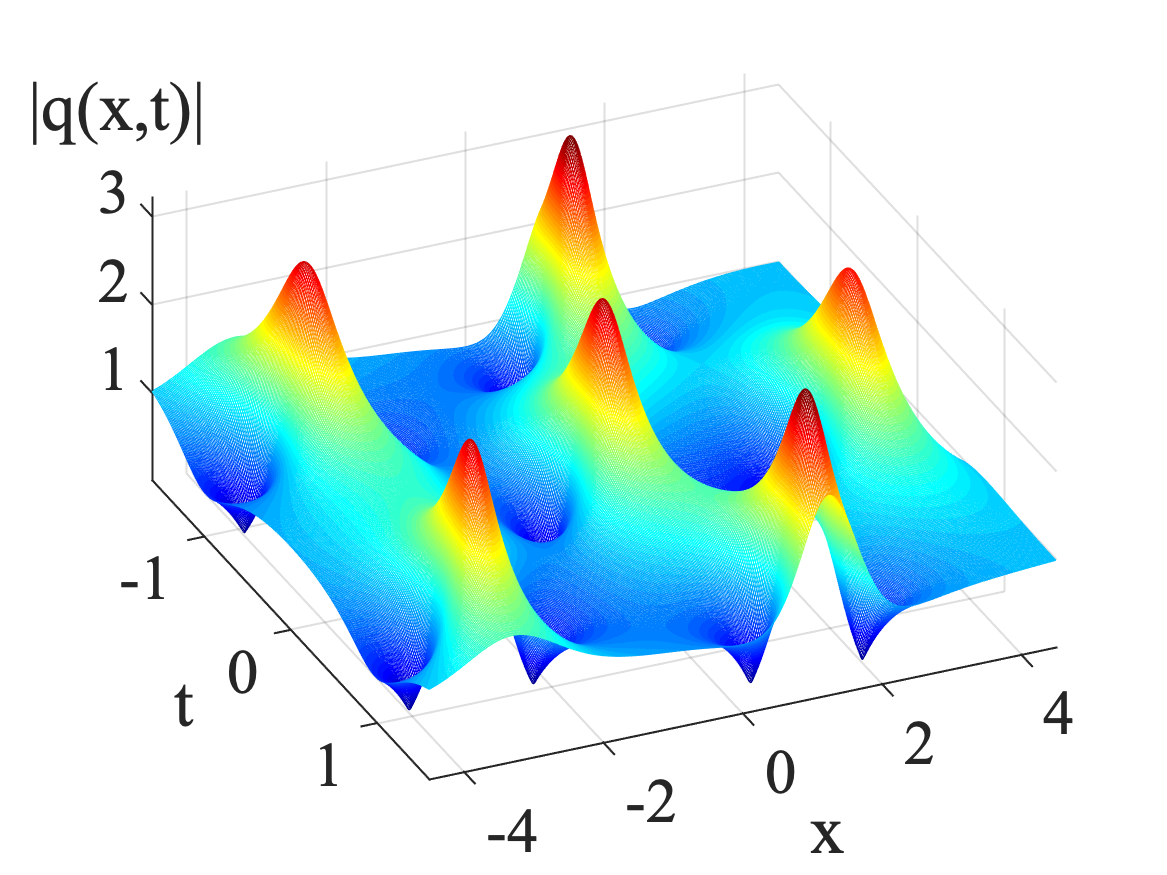}
$c$
\caption{(Color online) Data-driven third-order rogue wave ($m_1=n_1=n_2=0, m_2=50$) by Ibc-PINN: (a) The density diagram of the predicted solution $|q(x,t)|$; (b) The density diagram of absolute error; (c) The three-dimensional plot of the data-driven rogue wave solution $|q(x,t)|$.}
\label{fig3-7}
\end{figure}

Here we take the number of sub regions $n_{\max}$ as 6 and and each subnet contains 6 hidden layers, with 128 neurons per hidden layer. Other than that, all other configurations remain consistent with the previous example. Ultimately, Fig. \ref{fig3-7} illustrates third-order rogue wave of pentagonal pattern reconstructed by using Ibc-PINN. Table \ref{table3-3} summarizes the relative errors of data-driven third-order rogue waves for these two patterns, with Ibc-PINN maintaining higher accuracy compared to the unimproved method.

\begin{table}[htbp]
\caption{Relative $\mathbb{L}_2$ errors of the data-driven third-order rogue waves for the NLS equation by bc-PINN and Ibc-PINN.}
\label{table3-3} 
\centering
\begin{tabular}{c|cccc}
\bottomrule
\multirow{2}{*}{} & \multicolumn{1}{c|}{\multirow{2}{*}{$\Omega$}} & \multicolumn{1}{c|}{\multirow{2}{*}{$[T_0,T]$}} & \multicolumn{2}{c}{relative $\mathbb{L}_2$ error of $|q(x,t)|$} \\ \cline{4-5} 
                  & \multicolumn{1}{c|}{}                   & \multicolumn{1}{c|}{}                   & bc-PINN   & Ibc-PINN   \\ \hline
$m_1=10,m_2=0$                & [-5,5]                                       &   [-2,1.5]                                     & 1.932e-02        &  6.953e-03         \\ \hline
$m_1=0,m_2=50$               & [-4.5,4.5]      & [-1.6,1.6]                                &  6.419e-03                                     &   2.985e-03               \\ \toprule
\end{tabular}
\end{table}

\subsection{The AB system}
\quad

The AB system, initially proposed by Pedlosky \cite{Pedlosky} using singular perturbation theory, serves as model equations to depict marginally unstable baroclinic wave packets \cite{baroclinic1,baroclinic2} and ultra-short pulses in nonlinear optics \cite{opticalpulse}. The study of the dynamic properties of the AB system has yielded rich results, including the envelope solitary waves and periodic waves \cite{envelope}, the Lax pair \cite{ABLax}, the conservation laws, modulational instability and breather solutions via the Darboux transformation \cite{ABbreather}, the $N$-soliton solutions by using the dressing method \cite{ABdressing}, and rogue wave solutions via the generalized Darboux transformation \cite{ABrogue}.

\subsubsection{Data-driven first-order rogue waves}
\quad

We investigate the AB system in canonical form \cite{ABLax}
\begin{align}
	A_{xt}&=AB,\\
	B_x&=-\frac{1}{2}(|A|^2)_t.
\end{align}
Here, $A$ and $B$ are wave amplitudes satisfying the normalization condition
\begin{align}
|A_t|^2	+B^2=1,
\end{align}
and $x$ and $t$ represent semi-characteristic normalized coordinates. By using the generalized Darboux transformation, a unified formula of $N$th-order rogue wave solution for the AB system is given \cite{ABrogue}
\begin{align}
& A[N]=A[0]-4 \mathrm{i}\left(\lambda_1-\lambda_1^*\right) \sum_{j=0}^{N-1} \frac{\psi_1[j] \phi_1[j]^*}{\left(\left|\psi_1[j]\right|^2+\left|\phi_1[j]\right|^2\right)}\label{ABNrogue1} \\
& B[N]=B[0]-4 \mathrm{i}\left(\lambda_1-\lambda_1^*\right) \sum_{j=0}^{N-1} \frac{\left[\left|\psi_1[j]\right|^2\left(\left|\phi_1[j]\right|^2\right)_t-\left|\phi_1[j]\right|^2\left(\left|\psi_1[j]\right|^2\right)_t\right]}{\left(\left|\psi_1[j]\right|^2+\left|\phi_1[j]\right|^2\right)^2}\label{ABNrogue2} .
\end{align}

Specifically, the following plane wave solutions are taken as seed solutions
\begin{align}
A[0]=\mathrm{e}^{\mathrm{i} \theta},\quad B[0]=-\frac{1}{\sqrt{1+a^2}}	
\end{align}
with the real constant $a$ and $\theta=\frac{a\sqrt{1+a^2} x+t}{\sqrt{1+a^2}}$, and after substituting them into the Lax pair of the AB system, the first-order rogue wave is derived based on the the formulas \eqref{ABNrogue1}-\eqref{ABNrogue2} with $N=1$ and the fixed spectral parameter $\lambda_1=-\frac{a}{2}+\frac{\mathrm{i}}{2}$
\begin{align}
A[1]=\mathrm{e}^{\mathrm{i} \theta}\left(1+\frac{F_1+\mathrm{i} H_1}{D_1}\right), \quad B[1]=\frac{1}{\sqrt{1+a^2}} \frac{G_1}{D_1^2},
\end{align}
where
\begin{align}
F_1= & \left(2 a^4+4 a^2+2\right) x^2-4 \sqrt{1+a^2} a x t+2 t^2-2 a^4-4 a^2-2, \quad H_1=4 \sqrt{1+a^2} t, \nonumber\\
D_1= & -\left(a^4+2 a^2+1\right) x^2+2 a \sqrt{1+a^2} x t-t^2-a^4-2 a^2-1, \nonumber \\
G_1= & -a\left(a^2+1\right)^4 x^4+4 a^2\left(a^2+1\right)^{5 / 2} t x^3-2 a\left(a^2+1\right)\left(\left(3 a^2+1\right) t^2+a^6+5 a^4+7 a^2+3\right) x^2, \nonumber \\
& +4 \sqrt{1+a^2}\left(a^2 t^2+a^6+4 a^4+5 a^2+2\right) x t-a t^4-\left(2 a^5+8 a^3+6 a\right) t^2-a^9+6 a^5+8 a^3+3 a. \nonumber 
\end{align}

After selecting $a=\frac{1}{10}$ and the spatiotemporal region  $\Omega \times\left[T_0, T\right]=[-4,4]\times [-4,4]$, we can obtain the corresponding initial and boundary data. Then we divide the time domain into $n_{\max}=4$ segments and utilize the Ibc-PINN method to simulate the dynamic behavior of first-order rogue wave. The entire region is divided into $512 \times 201$ equidistant grid points, and the number of initial data points, boundary data points and residual configuration points is taken as $N_i=128$, $N_b=50$ and $N_r=20000$ separately. The increasing $N_s$ is selected as $N_s=128 \times 50 \times (n-1)$ for the $n$th sub-domain.

Decomposing the complex-valued solution $A(x,t)$ into real part $u(x,t)$ and imaginary part $v(x,t)$ and substituting them into the  AB system can derive three PDE residuals
\begin{align}
\begin{gathered}
R_1:=u_{xt}-uB,\\
R_2:=v_{xt}-vB,\\
R_3:=B_x+uu_t+vv_t.
\end{gathered}
\end{align}
Then the loss function of the first sub-domain is given
\begin{align}
\operatorname{MSE}_{\Delta T_1}= \operatorname{MSE}_I\left(x_k^i, T_0\right)+ \operatorname{MSE}_B\left(x_k^b, t_k^b\right)+ \operatorname{MSE}_R\left(x_k^r, t_k^r\right) 
\end{align}
\begin{align}
\begin{gathered}
\operatorname{MSE}_I\left(x_k^i, T_0\right)= \frac{1}{N_i} \sum_{k=1}^{N_i} \left|\hat{u}^{(1)}\left(x_k^i, T_0, \boldsymbol{\Theta}_1\right)-u_k^i\right|^2+\frac{1}{N_i} \sum_{k=1}^{N_i} \left|\hat{v}^{(1)}\left(x_k^i, T_0, \boldsymbol{\Theta}_1\right)-v_k^i\right|^2+\frac{1}{N_i} \sum_{k=1}^{N_i} \left|\hat{B}^{(1)}\left(x_k^i, T_0, \boldsymbol{\Theta}_1\right)-B_k^i\right|^2,\\
 \operatorname{MSE}_B\left(x_k^b, t_k^b\right)	= \frac{1}{N_b} \sum_{k=1}^{N_b} \left|\hat{u}^{(1)}\left(x_k^b, t_k^b, \boldsymbol{\Theta}_1\right)-u_k^b\right|^2+\frac{1}{N_b} \sum_{k=1}^{N_b} \left|\hat{v}^{(1)}\left(x_k^b, t_k^b, \boldsymbol{\Theta}_1\right)-v_k^b\right|^2+\frac{1}{N_b} \sum_{k=1}^{N_b} \left|\hat{B}^{(1)}\left(x_k^b, t_k^b, \boldsymbol{\Theta}_1\right)-B_k^b\right|^2,\\
 \operatorname{MSE}_R\left(x_k^r, t_k^r\right)=\frac{1}{N_r} \sum_{k=1}^{N_r} \left|R_1\left(x_k^r, t_k^r, \boldsymbol{\Theta}_1\right)\right|^2+\frac{1}{N_r} \sum_{k=1}^{N_r} \left|R_2\left(x_k^r, t_k^r, \boldsymbol{\Theta}_1\right)\right|^2+\frac{1}{N_r} \sum_{k=1}^{N_r} \left|R_3\left(x_k^r, t_k^r, \boldsymbol{\Theta}_1\right)\right|^2,
\end{gathered}
\end{align}
\begin{align}
T_0=-4,\quad x_k^i \in [-4,4], \quad\left(x_k^b, t_k^b\right) \in \{-4,4\} \times\left[-4, -2\right], \quad\left(x_k^r, t_k^r\right) \in [-4,-4]\times\left[-4, -2\right]	
\end{align}
and the PINN is applied to learn the real and imaginary parts ($\hat{u}^{(1)}$ and $\hat{v}^{(1)}$) of $\hat{A}^{(1)}$ and $\hat{B}^{(1)}$. In order to ensure the backward compatibility of the solution and simultaneously reduce the accumulation of errors, the modified loss term $\operatorname{MSE}_S$ is introduced into the loss function for subsequent three sub-domains ($n=2,3,4$)
\begin{align}
\operatorname{MSE}_{\Delta T_n}= \operatorname{MSE}_I\left(x_k^i, T_{n-1}\right)+ \operatorname{MSE}_B\left(x_k^b, t_k^b\right)+ \operatorname{MSE}_R\left(x_k^r, t_k^r\right) 
+ \operatorname{MSE}_S\left(x_k^s, t_k^s\right),	
\end{align}
where
\begin{align}
\operatorname{MSE}_I\left(x_k^i, T_{n-1}\right)= &\frac{1}{N_i} \sum_{k=1}^{N_i} \left|\hat{u}^{(n)}\left(x_k^i, T_{n-1}, \boldsymbol{\Theta}_{n}\right)-\hat{u}^{(n-1)} (x_k^i, T_{n-1})\right|^2 +\frac{1}{N_i} \sum_{k=1}^{N_i} \left|\hat{v}^{(n)}\left(x_k^i, T_{n-1}, \boldsymbol{\Theta}_{n}\right)-\hat{v}^{(n-1)} (x_k^i, T_{n-1})\right|^2\nonumber \\
&+\frac{1}{N_i} \sum_{k=1}^{N_i} \left|\hat{B}^{(n)}\left(x_k^i, T_{n-1}, \boldsymbol{\Theta}_{n}\right)-\hat{B}^{(n-1)} (x_k^i, T_{n-1})\right|^2,\nonumber\\
\operatorname{MSE}_B\left(x_k^b, t_k^b\right)	= &\frac{1}{N_b} \sum_{k=1}^{N_b} \left|\hat{u}^{(n)}\left(x_k^b, t_k^b, \boldsymbol{\Theta}_{n}\right)-u_k^b\right|^2+\frac{1}{N_b} \sum_{k=1}^{N_b} \left|\hat{v}^{(n)}\left(x_k^b, t_k^b, \boldsymbol{\Theta}_{n}\right)-v_k^b\right|^2+\frac{1}{N_b} \sum_{k=1}^{N_b} \left|\hat{B}^{(n)}\left(x_k^b, t_k^b, \boldsymbol{\Theta}_{n}\right)-B_k^b\right|^2,\nonumber\\
 \operatorname{MSE}_R\left(x_k^r, t_k^r\right)=&\frac{1}{N_r} \sum_{k=1}^{N_r} \left|R_1\left(x_k^r, t_k^r, \boldsymbol{\Theta}_{n}\right)\right|^2+\frac{1}{N_r} \sum_{k=1}^{N_r} \left|R_2\left(x_k^r, t_k^r, \boldsymbol{\Theta}_{n}\right)\right|^2+\frac{1}{N_r} \sum_{k=1}^{N_r} \left|R_3\left(x_k^r, t_k^r, \boldsymbol{\Theta}_{n}\right)\right|^2,\nonumber\\
\operatorname{MSE}_S\left(x_k^s, t_k^s\right)=&\frac{1}{N_s} \sum_{j=1}^{n-1} \sum_{k \in \tau_j} \left|\hat{u}^{(n)} (x_k^s, t_k^s, \boldsymbol{\Theta}_{n})-\hat{u}^{(j)} (x_k^s, t_k^s)\right|^2+\frac{1}{N_s} \sum_{j=1}^{n-1} \sum_{k \in \tau_j} \left|\hat{v}^{(n)} (x_k^s, t_k^s, \boldsymbol{\Theta}_{n})-\hat{v}^{(j)} (x_k^s, t_k^s)\right|^2 \nonumber \\
	&+\frac{1}{N_s} \sum_{j=1}^{n-1} \sum_{k \in \tau_j} \left|\hat{B}^{(n)} (x_k^s, t_k^s, \boldsymbol{\Theta}_{n})-\hat{B}^{(j)} (x_k^s, t_k^s)\right|^2,
\end{align}
\begin{align}
\begin{gathered}
x_k^i \in [-4,4], \quad\left(x_k^b, t_k^b\right) \in \{-4,4 \} \times\left[T_{n-1}, T_n\right], \quad \left(x_k^r, t_k^r\right) \in [-4,4] \times\left[T_{n-1}, T_n\right], \quad \left(x_k^s, t_k^s\right) \in [-4,4] \times\left[-4, T_{n-1}\right], \\
\tau_1 = \{k|(x_k^s, t_k^s) \in [-4,4] \times [-4,T_1] \}, \quad \tau_j = \{k|(x_k^s, t_k^s) \in [-4,4] \times (T_{j-1},T_j] \}	\quad (j=2,\cdots, n-1).
\end{gathered}
\end{align}
Here, $\hat{A}^{(n)}=\hat{u}^{(n)}+\mathrm{i} \hat{v}^{(n)}$ and $\hat{B}^{(n)}$ are the predicted solutions by training subnet-$n$ in $n$th sub-domain.

\begin{figure}[htbp]
\centering
\includegraphics[width=5.5cm,height=4cm]{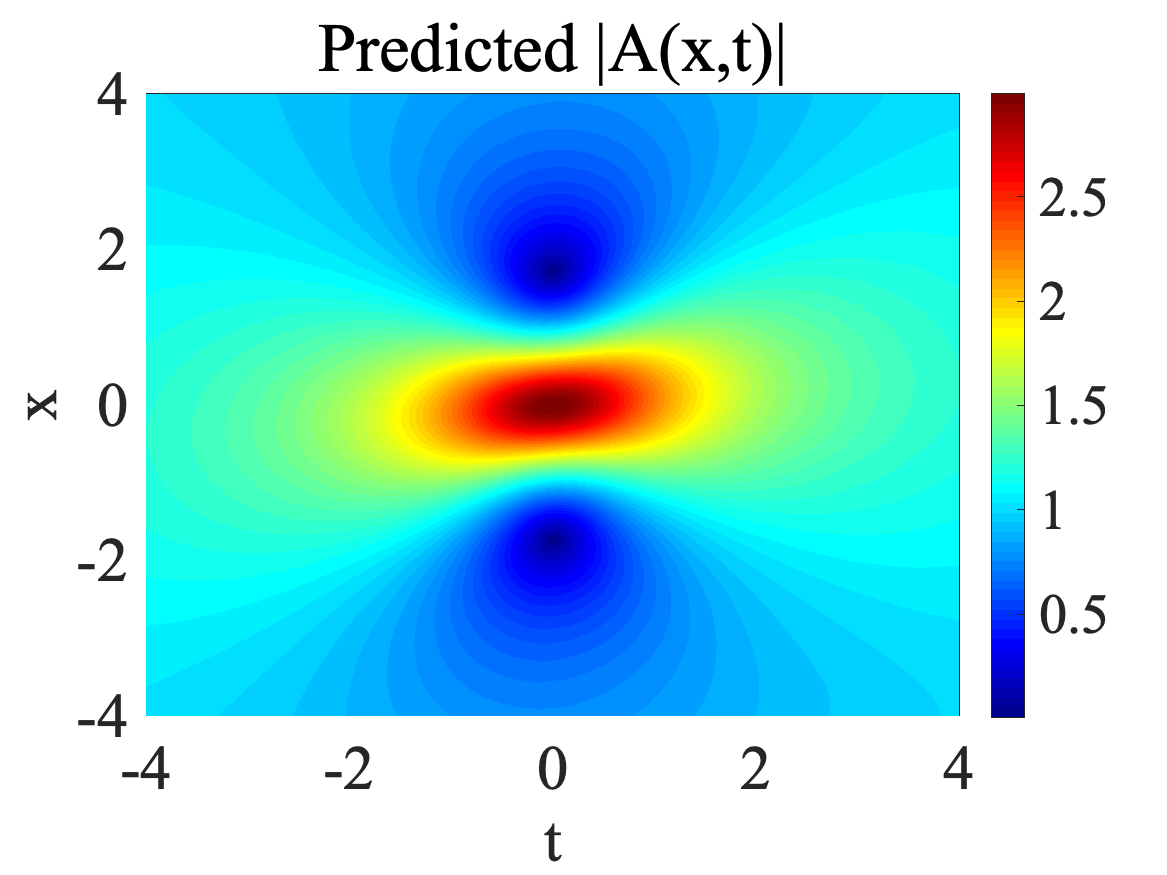}
$a$
\includegraphics[width=5.5cm,height=4cm]{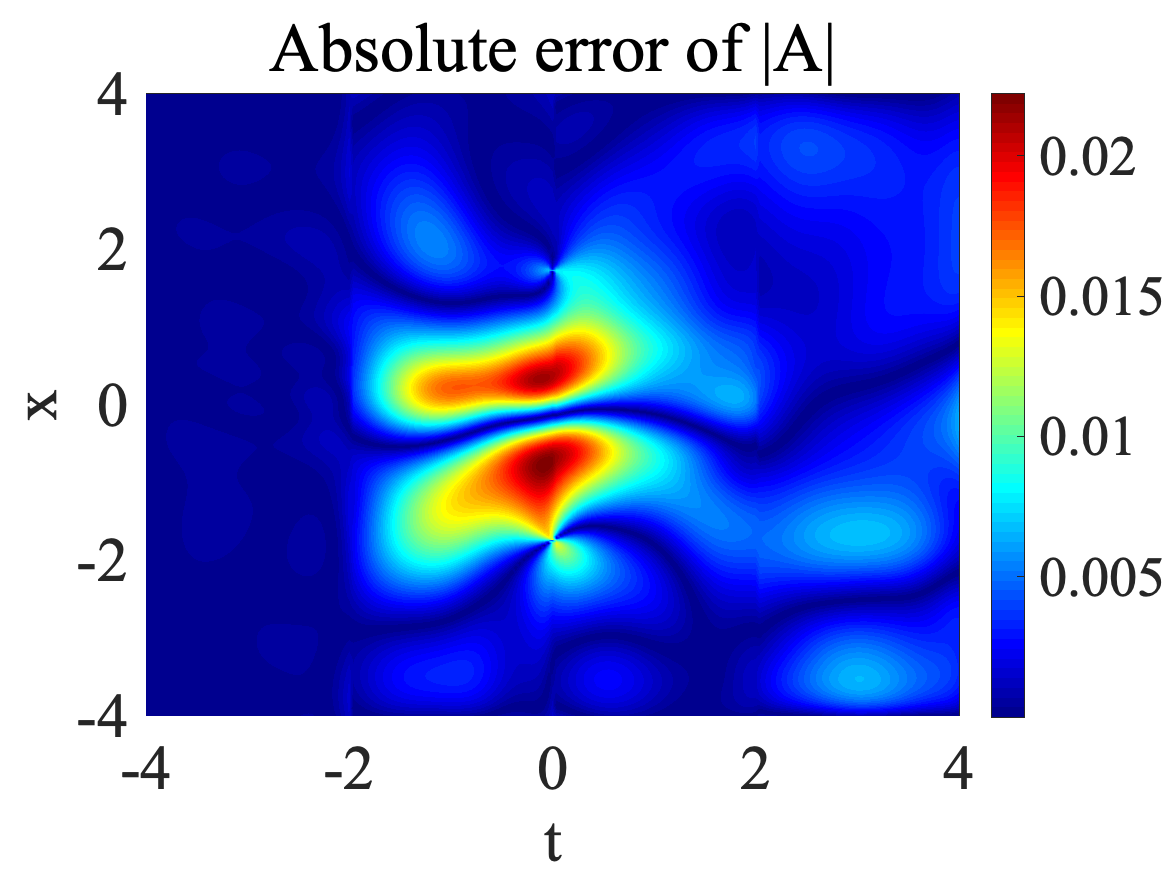}
$b$
\includegraphics[width=5.5cm,height=4cm]{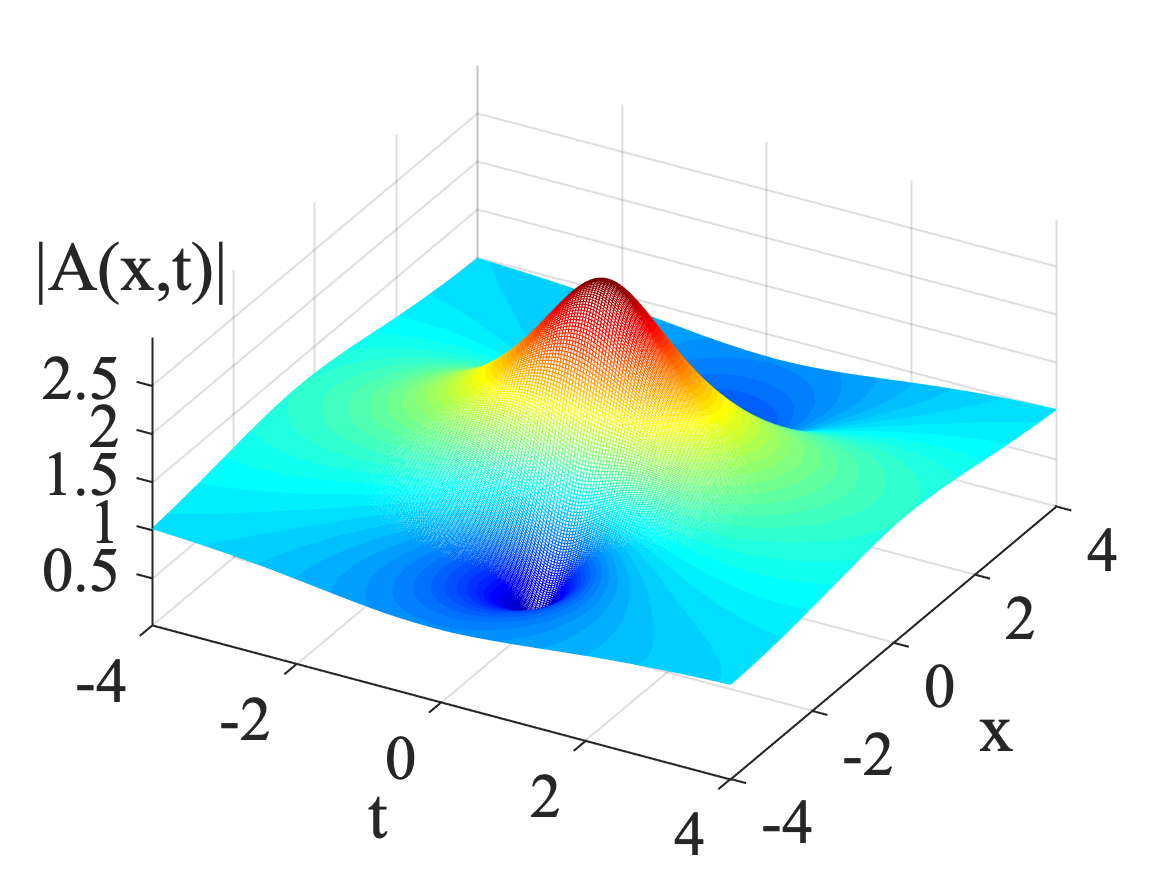}
$c$\\
\includegraphics[width=5.5cm,height=4cm]{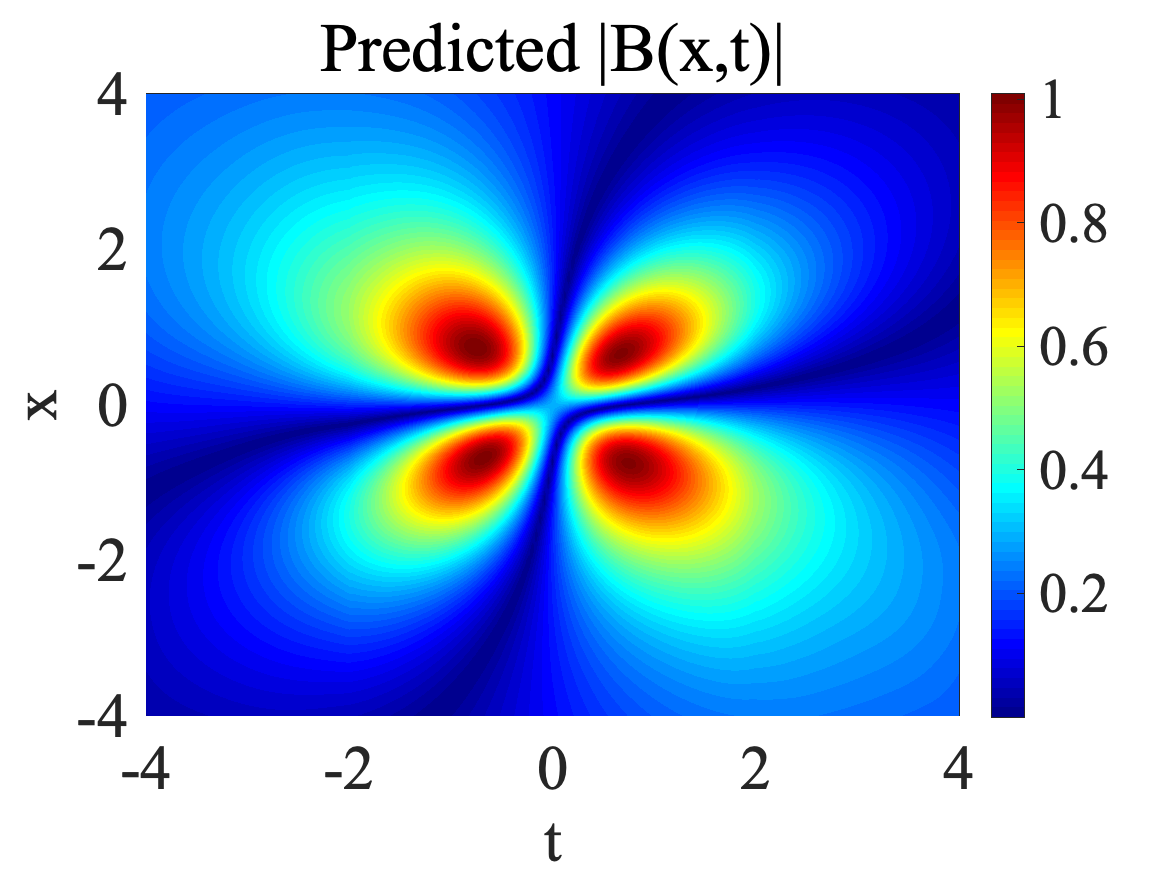}
$d$
\includegraphics[width=5.5cm,height=4cm]{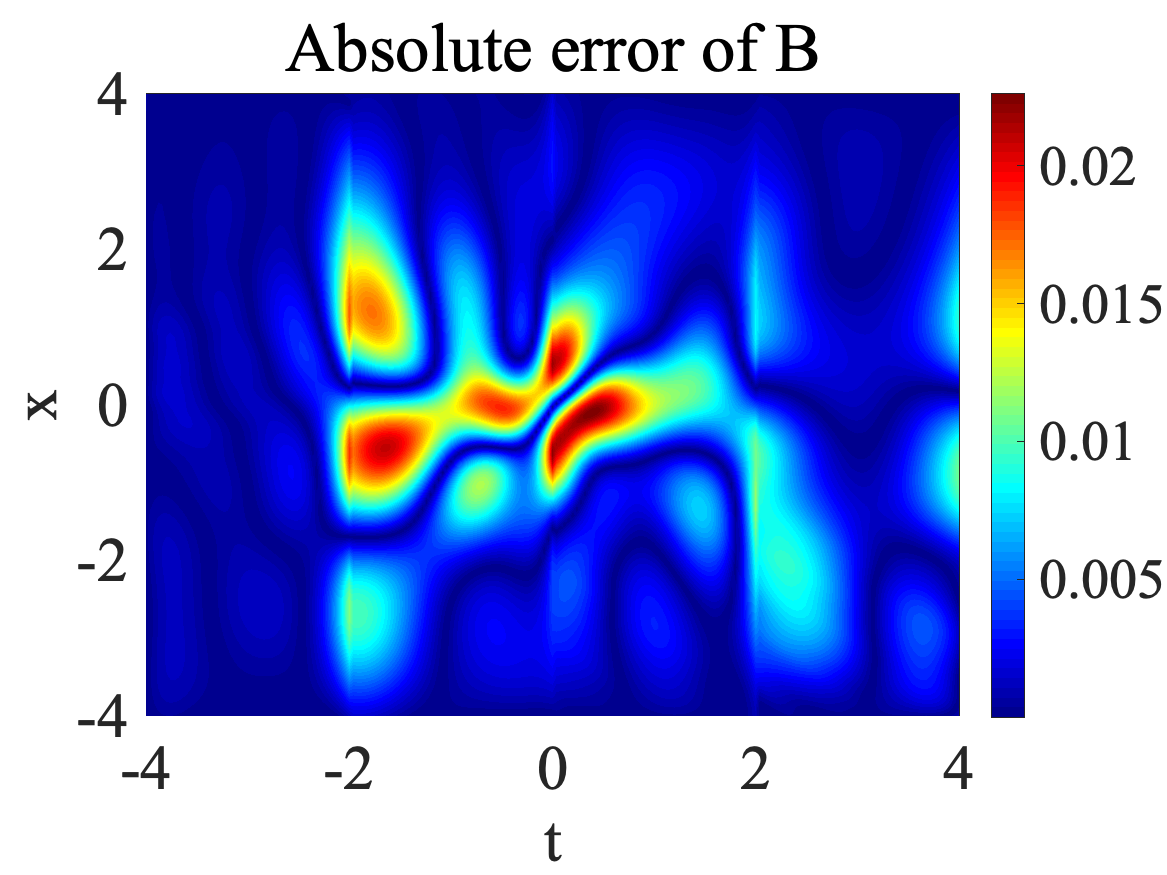}
$e$
\includegraphics[width=5.5cm,height=4cm]{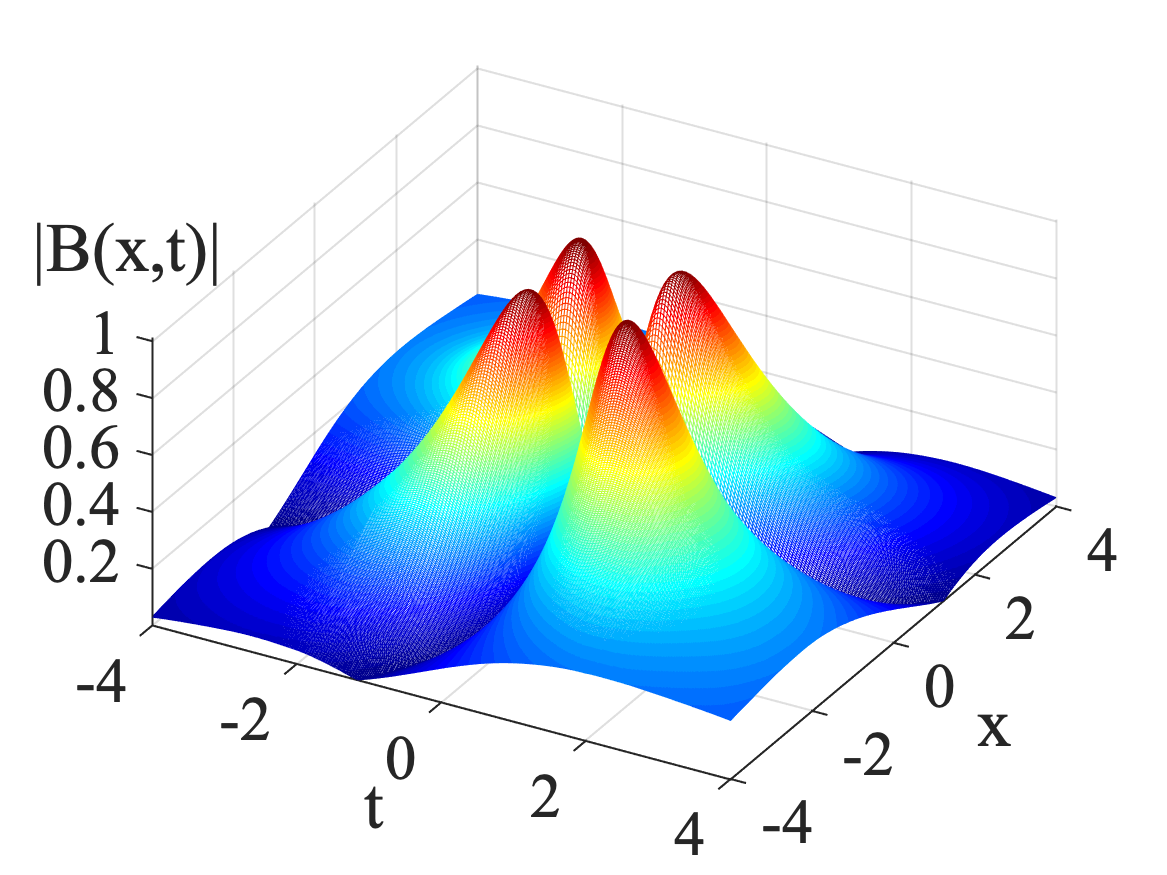}
$f$
\caption{(Color online) Data-driven first-order rogue wave by Ibc-PINN: The density diagrams of the predicted solutions: (a) $|A(x,t)|$ and (d) $|B(x,t)|$; The density diagrams of absolute error: (b) $|A(x,t)|$ and (e) $B(x,t)$; The three-dimensional plots of the data-driven wave solutions: (c) $|A(x,t)|$ and (f) $|B(x,t)|$.}
\label{fig4-1}
\end{figure}

For each subnetwork, the width and depth are taken as 7 and 128, respectively. The selection of optimizers and the application of ICGL and transfer learning techniques are similar to those used in the nonlinear Schr\"{o}dinger equation, and are not further elaborated upon here. The dynamic behaviors of the first-order rogue wave for the AB system are successfully reproduced by using Ibc-PINN, as displayed in Fig. \ref{fig4-1}. The waveform of $A$ component is the standard eye-shaped Peregrine soliton whereas the $B$ component exhibits the characteristic shape of four peaky-shaped rogue wave. Moreover, the results in Table \ref{table4-1} indicate that under this network structure configuration, the network training of bc-PINN fails and there is still a considerable error compared to the exact solution.

\subsubsection{Data-driven second-order rogue waves}
\quad

Naturally, the second-order rogue wave solution for the AB system with free parameters $a_1, m_1$ and $n_1$ can be derived \cite{ABrogue} by performing the second-step generalized DT. Here we refrain from providing the explicit expressions in this context due to the intricate nature of their representation. The value of $a_1$ is fixed at $\frac{1}{10}$, and once the spatiotemporal domain is chosen and the remaining two free parameters are determined, we can acquire the corresponding initial and boundary conditions.

The time domain $[-4, 4]$ is divided into $n_{\max}=6$ equidistant segments. For the $n$th sub-domain, we choose $N_i=128, N_b=50, N_r=2000$ and $N_s=128 \times 50 \times (n-1)$ and establish a sub network with 6 hidden layers, each consisting of 64 neurons.

\begin{figure}[htbp]
\centering
\includegraphics[width=5.5cm,height=4cm]{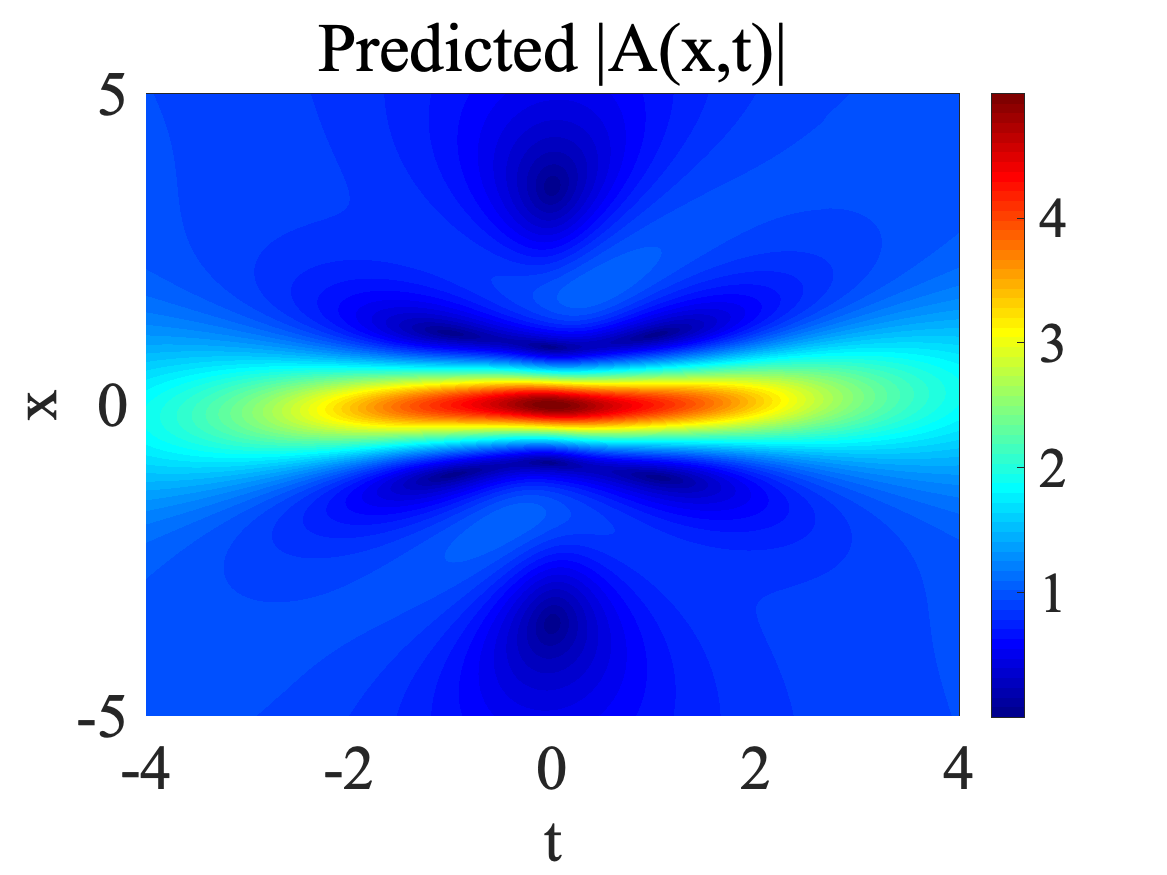}
$a$
\includegraphics[width=5.5cm,height=4cm]{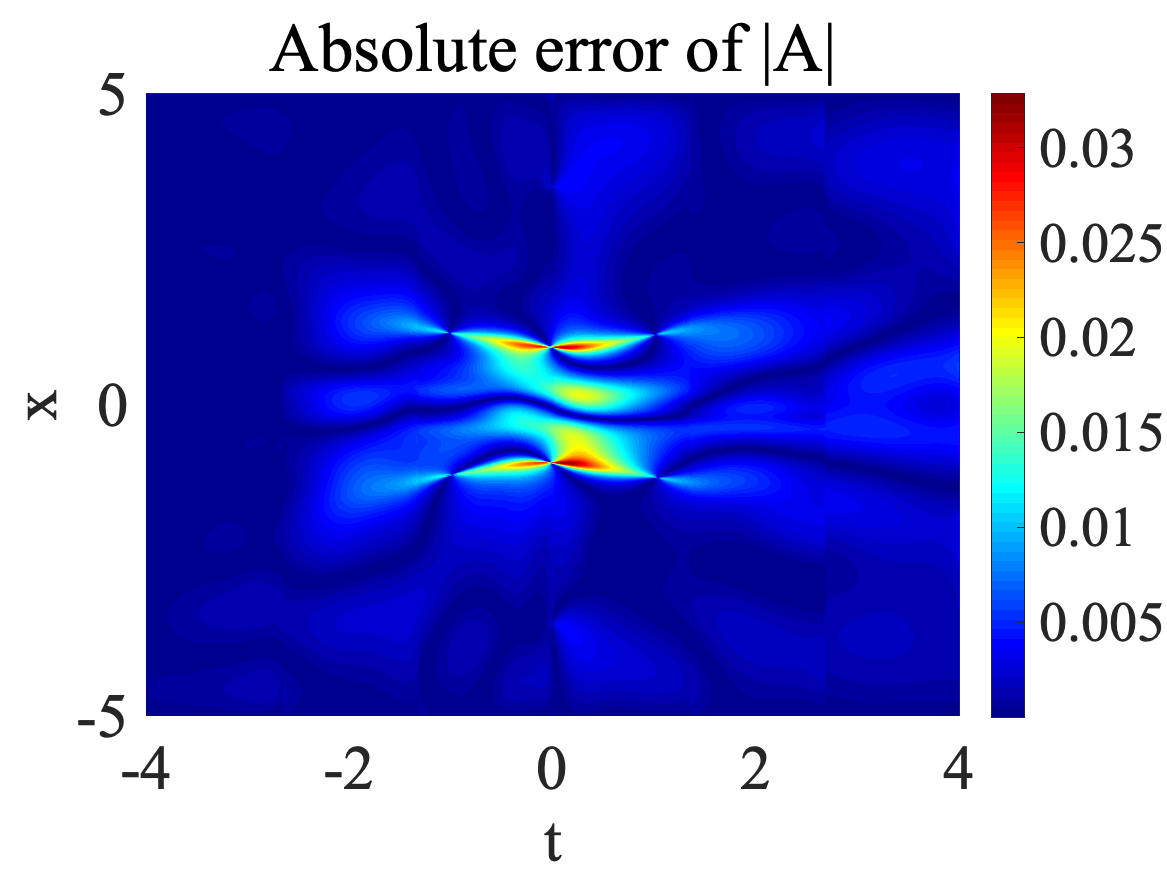}
$b$
\includegraphics[width=5.5cm,height=4cm]{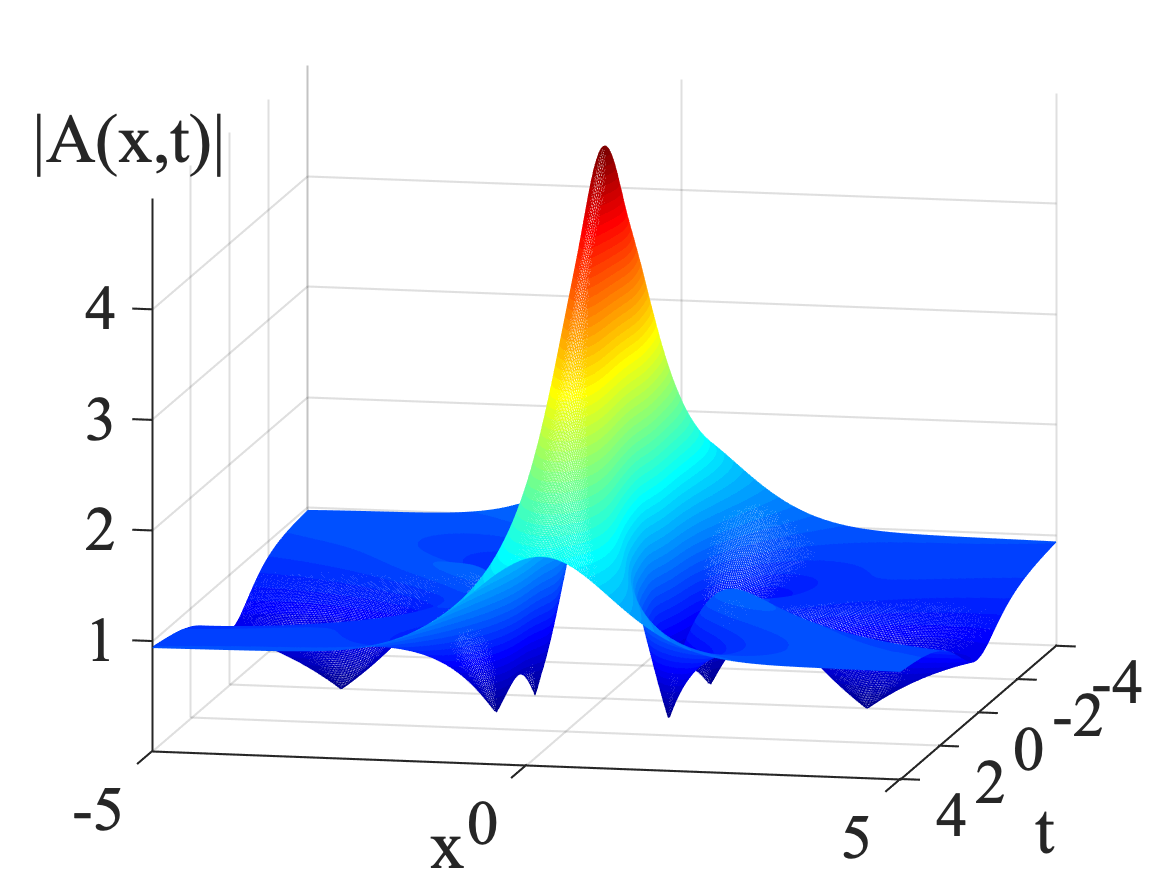}
$c$\\
\includegraphics[width=5.5cm,height=4cm]{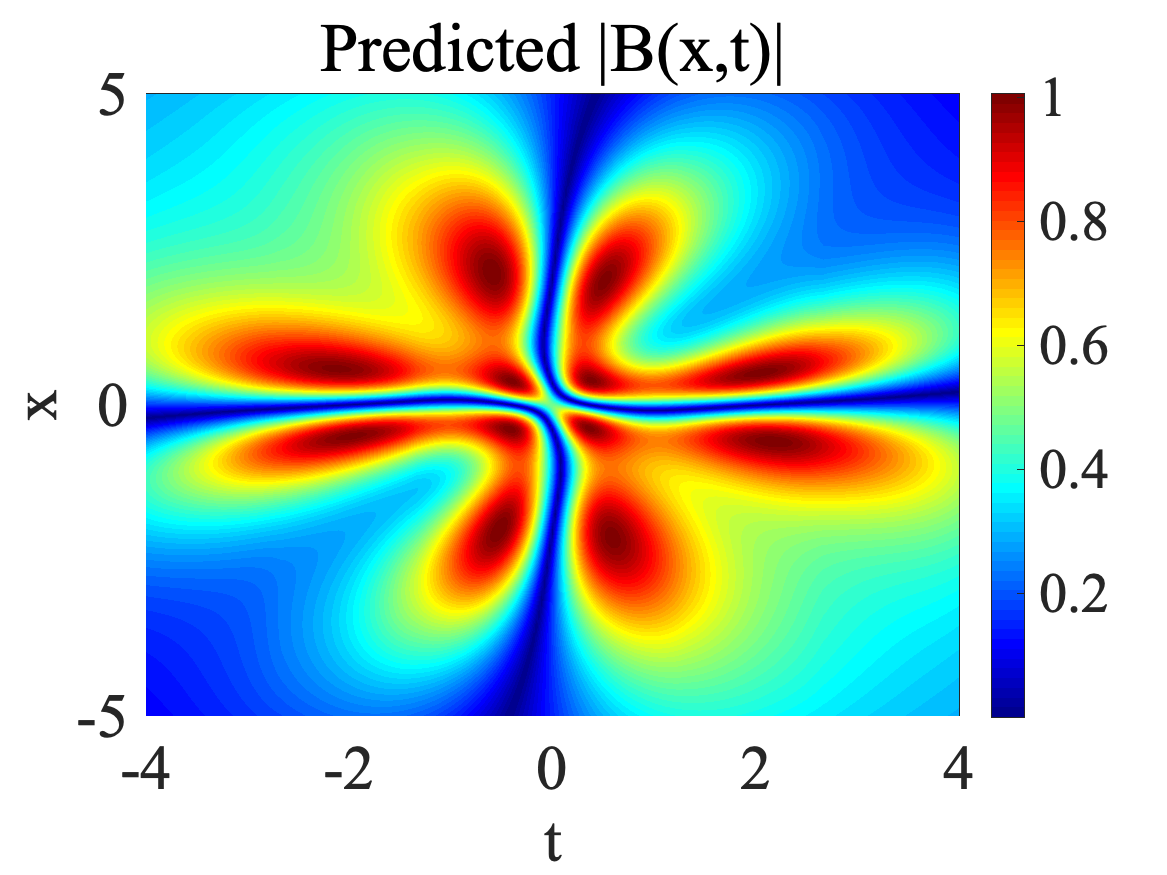}
$d$
\includegraphics[width=5.5cm,height=4cm]{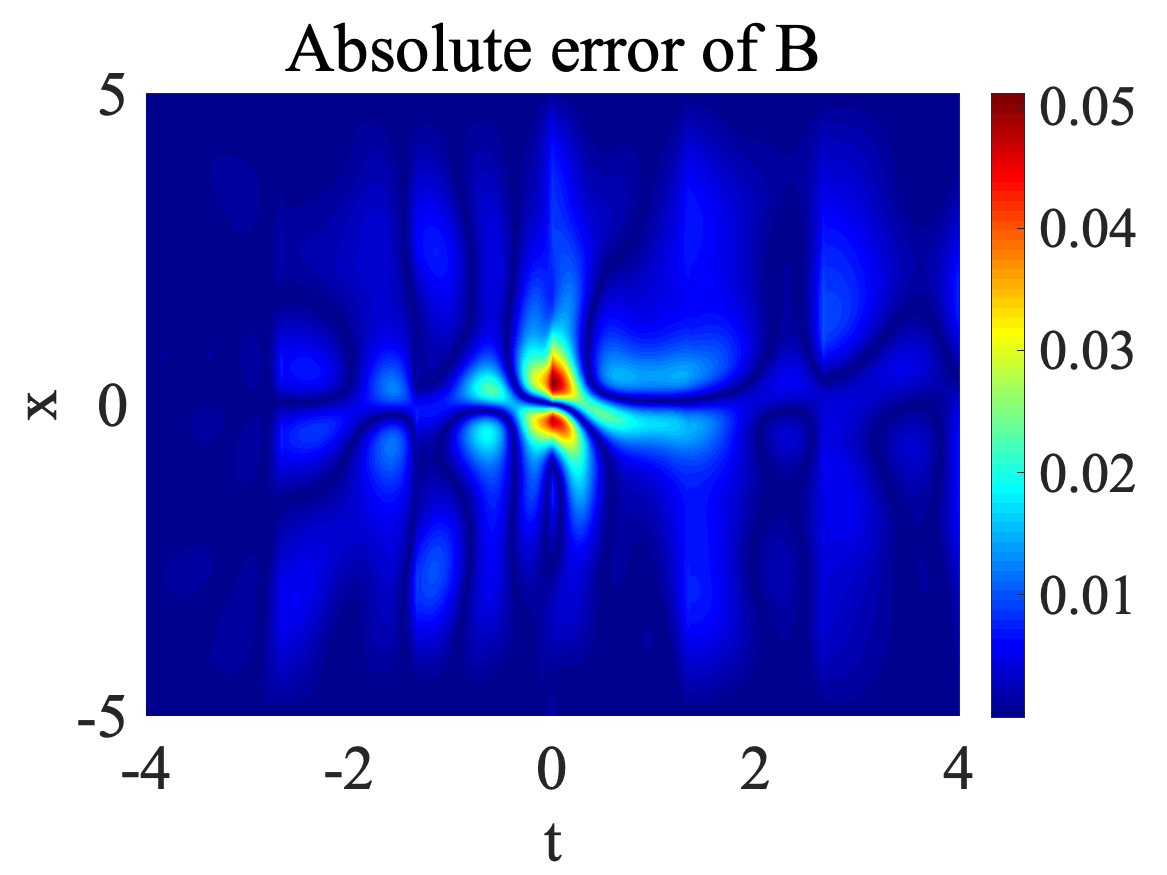}
$e$
\includegraphics[width=5.5cm,height=4cm]{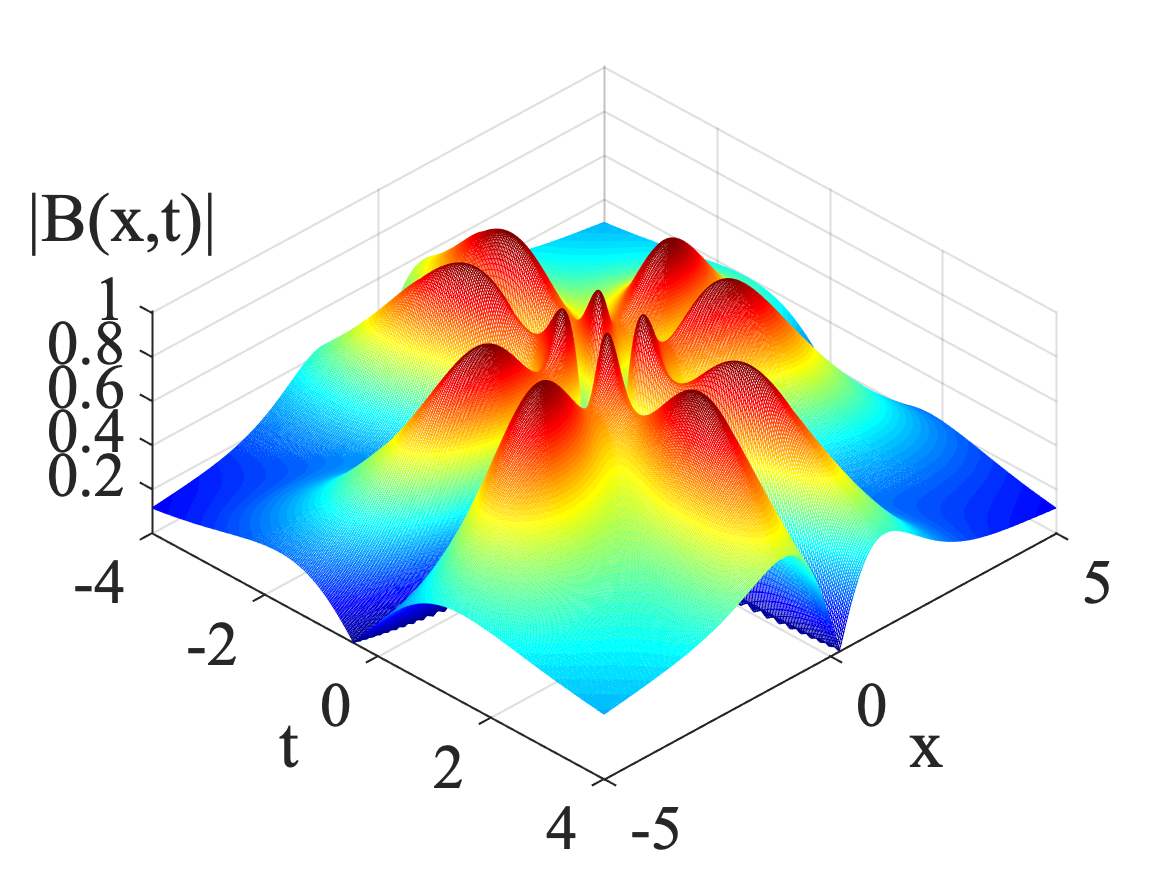}
$f$
\caption{(Color online) Data-driven second-order rogue wave by Ibc-PINN: The density diagrams of the predicted solutions: (a) $|A(x,t)|$ and (d) $|B(x,t)|$; The density diagrams of absolute error: (b) $|A(x,t)|$ and (e) $B(x,t)$; The three-dimensional plots of the data-driven wave solutions: (c) $|A(x,t)|$ and (f) $|B(x,t)|$.}
\label{fig4-2}
\end{figure}

The selection of free parameters $m_1=0, n_1=0$ corresponds to the fundamental second-order rogue wave, which is simulated by Ibc-PINN shown in Fig. \ref{fig4-2}. Obviously, the maximum amplitude of predicted $|A(x,t)|$ occurs at the center (0,0) and it is roughly five times the amplitude of the background wave. For $B$ component, there are twelve peaks with a peak value of 1 around the center. Finally, Table \ref{table4-1} provides the relative error of solutions obtained by using two methods, and the accuracy of Ibc-PINN still remains higher.

\begin{table}[htbp]
\caption{Relative $\mathbb{L}_2$ errors of the data-driven rogue waves for the AB system by bc-PINN and Ibc-PINN.}
\label{table4-1} 
\centering
\begin{tabular}{cc|cc}
\bottomrule
\multicolumn{2}{c|}{}                                  & bc-PINN & Ibc-PINN \\ \hline
\multicolumn{1}{c|}{\multirow{2}{*}{First-order}}  & $|A(x,t)|$ & 2.083e-01       & 4.238e-03        \\ \cline{2-2}
\multicolumn{1}{c|}{}                              & $B(x,t)$ & 5.070e-01       & 1.442e-02        \\ \cline{1-2}
\multicolumn{1}{c|}{\multirow{2}{*}{Second-order}} & $|A(x,t)|$ & 3.309e-03       & 2.671e-03        \\ \cline{2-2}
\multicolumn{1}{c|}{}                              & $B(x,t)$ & 1.212e-02       & 9.891e-03        \\ \toprule
\end{tabular}
\end{table}

\section{Analysis and discussion}\label{analysis}

\subsection{Analysis of error accumulation}
\quad

In this part, we take the second and third-order rogue waves for the nonlinear Schr\"{o}dinger equation as examples to analyze the cumulative effect of errors by calculating the relative $\mathbb{L}_2$ errors of bc-PINN and Ibc-PINN in each sub-domain.

For the second-order ($m_1=0, n_1=0$) and third-order rogue waves ($m_1=n_1=n_2=0, m_2=50$), the entire regions were divided into four and six sub-domains, respectively, and a total of four and six subnets were trained. We calculate the relative $\mathbb{L}_2$ errors on $N_x \times N_t=512 \times 51$ grid points for each sub-domain to evaluate the accuracy of models and the specific results are listed in Table \ref{table5-1} - Table \ref{table5-4}.

\begin{table}[htbp]
\caption{Relative $\mathbb{L}_2$ errors of the data-driven second-order rogue wave ($m_1=0, n_1=0$) for the NLS equation in each sub-domain by bc-PINN.}
\label{table5-1} 
\centering
\begin{tabular}{cc|cccc}
\bottomrule
\multicolumn{2}{c|}{}                                                                                                          & subnet-1 & subnet-2 & subnet-3 & subnet-4 \\ \hline
\multicolumn{1}{c|}{\multirow{4}{*}{\begin{tabular}[c]{@{}c@{}}relative $\mathbb{L}_2$  \\ error of $u(x,t)$\end{tabular}}} & sub-domain 1 & \pmb{1.015e-04}         & 2.543e-03         & 1.026e-02         & \pmb{8.885e-03}        \\ \cline{2-2}
\multicolumn{1}{c|}{}                                                                                           & sub-domain 2 & \rotatebox[origin=c]{40}{\textbackslash}        & \pmb{8.519e-04}        & 6.831e-03         & \pmb{5.008e-03}         \\ \cline{2-2}
\multicolumn{1}{c|}{}                                                                                           & sub-domain 3 & \rotatebox[origin=c]{40}{\textbackslash}         & \rotatebox[origin=c]{40}{\textbackslash}         & \pmb{7.742e-03}         & \pmb{7.764e-03}         \\ \cline{2-2}
\multicolumn{1}{c|}{}                                                                                           & sub-domain 4 & \rotatebox[origin=c]{40}{\textbackslash}         & \rotatebox[origin=c]{40}{\textbackslash}         &  \rotatebox[origin=c]{40}{\textbackslash}        &\pmb{4.013e-02}          \\ \hline
\multicolumn{1}{c|}{\multirow{4}{*}{\begin{tabular}[c]{@{}c@{}}relative $\mathbb{L}_2$  \\ error of $v(x,t)$\end{tabular}}}                                                                         & sub-domain 1 & \pmb{6.315e-05}         & 1.571e-03         & 5.985e-03         &  \pmb{5.162e-03}        \\ \cline{2-2}
\multicolumn{1}{c|}{}                                                                                           & sub-domain 2 & \rotatebox[origin=c]{40}{\textbackslash}         &  \pmb{1.260e-03}        &  1.018e-02        & \pmb{7.282e-03}         \\ \cline{2-2}
\multicolumn{1}{c|}{}                                                                                           & sub-domain 3 & \rotatebox[origin=c]{40}{\textbackslash}         & \rotatebox[origin=c]{40}{\textbackslash}         & \pmb{1.366e-02}         & \pmb{1.371e-02}         \\ \cline{2-2}
\multicolumn{1}{c|}{}                                                                                           & sub-domain 4 &  \rotatebox[origin=c]{40}{\textbackslash}        & \rotatebox[origin=c]{40}{\textbackslash}         &  \rotatebox[origin=c]{40}{\textbackslash}        &\pmb{1.382e-02}          \\ \hline
\multicolumn{1}{c|}{\multirow{4}{*}{\begin{tabular}[c]{@{}c@{}}relative $\mathbb{L}_2$  \\ error of $|q(x,t)|$\end{tabular}}}                                                                        & sub-domain 1 & \pmb{5.543e-05}         & 1.307e-03         & 5.293e-03         &\pmb{4.656e-03}          \\\cline{2-2}
\multicolumn{1}{c|}{}                                                                                           & sub-domain 2 &  \rotatebox[origin=c]{40}{\textbackslash}        & \pmb{6.178e-04}         &  5.443e-03        & \pmb{3.975e-03}         \\\cline{2-2}
\multicolumn{1}{c|}{}                                                                                           & sub-domain 3 &  \rotatebox[origin=c]{40}{\textbackslash}        & \rotatebox[origin=c]{40}{\textbackslash}         &  \pmb{3.868e-03}        &\pmb{3.990e-03}          \\\cline{2-2}
\multicolumn{1}{c|}{}                                                                                           & sub-domain 4 &  \rotatebox[origin=c]{40}{\textbackslash}        & \rotatebox[origin=c]{40}{\textbackslash}        &  \rotatebox[origin=c]{40}{\textbackslash}        &\pmb{1.071e-02}          \\ \toprule
\end{tabular}
\end{table}

\begin{table}[htbp]
\caption{Relative $\mathbb{L}_2$ errors of the data-driven second-order rogue wave ($m_1=0, n_1=0$) for the NLS equation in each sub-domain by Ibc-PINN.}
\label{table5-2} 
\centering
\begin{tabular}{cc|cccc}
\bottomrule
\multicolumn{2}{c|}{}                                                                                                          & subnet-1 & subnet-2 & subnet-3 & subnet-4 \\ \hline
\multicolumn{1}{c|}{\multirow{4}{*}{\begin{tabular}[c]{@{}c@{}}relative $\mathbb{L}_2$  \\ error of $u(x,t)$\end{tabular}}} & sub-domain 1 & \pmb{1.015e-04}         & 2.543e-03         & 7.961e-03        & \pmb{2.167e-03}        \\ \cline{2-2}
\multicolumn{1}{c|}{}                                                                                           & sub-domain 2 & \rotatebox[origin=c]{40}{\textbackslash}        & \pmb{8.519e-04}        & 5.842e-03         & \pmb{1.991e-03}         \\ \cline{2-2}
\multicolumn{1}{c|}{}                                                                                           & sub-domain 3 & \rotatebox[origin=c]{40}{\textbackslash}         & \rotatebox[origin=c]{40}{\textbackslash}         & \pmb{6.108e-03}         & \pmb{6.465e-03}         \\ \cline{2-2}
\multicolumn{1}{c|}{}                                                                                           & sub-domain 4 & \rotatebox[origin=c]{40}{\textbackslash}         & \rotatebox[origin=c]{40}{\textbackslash}         &  \rotatebox[origin=c]{40}{\textbackslash}        &\pmb{3.268e-02}          \\ \hline
\multicolumn{1}{c|}{\multirow{4}{*}{\begin{tabular}[c]{@{}c@{}}relative $\mathbb{L}_2$  \\ error of $v(x,t)$\end{tabular}}}                                                                         & sub-domain 1 & \pmb{6.315e-05}         & 1.571e-03         & 5.609e-03         &  \pmb{1.058e-03}        \\ \cline{2-2}
\multicolumn{1}{c|}{}                                                                                           & sub-domain 2 & \rotatebox[origin=c]{40}{\textbackslash}         &  \pmb{1.260e-03}        & 9.123e-03         & \pmb{2.894e-03}         \\ \cline{2-2}
\multicolumn{1}{c|}{}                                                                                           & sub-domain 3 & \rotatebox[origin=c]{40}{\textbackslash}         & \rotatebox[origin=c]{40}{\textbackslash}         & \pmb{1.103e-02}         & \pmb{1.132e-02}         \\ \cline{2-2}
\multicolumn{1}{c|}{}                                                                                           & sub-domain 4 &  \rotatebox[origin=c]{40}{\textbackslash}        & \rotatebox[origin=c]{40}{\textbackslash}         &  \rotatebox[origin=c]{40}{\textbackslash}        &\pmb{1.020e-02}          \\ \hline
\multicolumn{1}{c|}{\multirow{4}{*}{\begin{tabular}[c]{@{}c@{}}relative $\mathbb{L}_2$  \\ error of $|q(x,t)|$\end{tabular}}}                                                                        & sub-domain 1 & \pmb{5.543e-05}         & 1.307e-03         & 4.602e-03        &\pmb{9.862e-04}          \\\cline{2-2}
\multicolumn{1}{c|}{}                                                                                           & sub-domain 2 &  \rotatebox[origin=c]{40}{\textbackslash}        & \pmb{6.178e-04}         & 5.347e-03         & \pmb{1.610e-03}         \\\cline{2-2}
\multicolumn{1}{c|}{}                                                                                           & sub-domain 3 &  \rotatebox[origin=c]{40}{\textbackslash}        & \rotatebox[origin=c]{40}{\textbackslash}         &  \pmb{2.684e-03}        &\pmb{3.189e-03}          \\\cline{2-2}
\multicolumn{1}{c|}{}                                                                                           & sub-domain 4 &  \rotatebox[origin=c]{40}{\textbackslash}        & \rotatebox[origin=c]{40}{\textbackslash}        &  \rotatebox[origin=c]{40}{\textbackslash}        &\pmb{7.022e-03}          \\ \toprule
\end{tabular}
\end{table}

\begin{table}[htbp]
\caption{Relative $\mathbb{L}_2$ errors of the data-driven third-order rogue wave ($m_1=n_1=n_2=0, m_2=50$) for the NLS equation in each sub-domain by bc-PINN.}
\label{table5-3} 
\centering
\begin{tabular}{cc|cccccc}
\bottomrule
\multicolumn{2}{c|}{}                                                                                                          & subnet-1 & subnet-2 & subnet-3 & subnet-4 & subnet-5 & subnet-6 \\ \hline
\multicolumn{1}{c|}{\multirow{6}{*}{\begin{tabular}[c]{@{}c@{}}relative $\mathbb{L}_2$  \\ error of $u(x,t)$\end{tabular}}} & sub-domain 1 &  \pmb{3.980e-04}        & 1.327e-03         &  1.933e-03        &  3.123e-03        & 4.536e-03         & \pmb{6.979e-03}         \\ \cline{2-2}
\multicolumn{1}{c|}{}                                                                                           & sub-domain 2 &  \rotatebox[origin=c]{40}{\textbackslash}        & \pmb{5.857e-04}         & 1.670e-03         & 3.097e-03         & 3.270e-03         & \pmb{5.652e-03}         \\ \cline{2-2}
\multicolumn{1}{c|}{}                                                                                           & sub-domain 3 & \rotatebox[origin=c]{40}{\textbackslash}         & \rotatebox[origin=c]{40}{\textbackslash}         & \pmb{2.065e-03}         &          3.365e-03 & 4.531e-03         & \pmb{6.939e-03}         \\ \cline{2-2}
\multicolumn{1}{c|}{}                                                                                           & sub-domain 4 &  \rotatebox[origin=c]{40}{\textbackslash}        & \rotatebox[origin=c]{40}{\textbackslash}         & \rotatebox[origin=c]{40}{\textbackslash}         &  \pmb{4.830e-03}        & 5.716e-03          & \pmb{8.763e-03}         \\ \cline{2-2}
\multicolumn{1}{c|}{}                                                                                           & sub-domain 5 &  \rotatebox[origin=c]{40}{\textbackslash}        & \rotatebox[origin=c]{40}{\textbackslash}         & \rotatebox[origin=c]{40}{\textbackslash}         & \rotatebox[origin=c]{40}{\textbackslash}         &  \pmb{7.516e-03}        & \pmb{1.019e-02}         \\ \cline{2-2}
\multicolumn{1}{c|}{}                                                                                           & sub-domain 6 &  \rotatebox[origin=c]{40}{\textbackslash}        & \rotatebox[origin=c]{40}{\textbackslash}         & \rotatebox[origin=c]{40}{\textbackslash}         & \rotatebox[origin=c]{40}{\textbackslash}         &          \rotatebox[origin=c]{40}{\textbackslash} & \pmb{1.990e-02}         \\ \hline
\multicolumn{1}{c|}{\multirow{6}{*}{\begin{tabular}[c]{@{}c@{}}relative $\mathbb{L}_2$  \\ error of $v(x,t)$\end{tabular}}}                                                                         & sub-domain 1 & \pmb{3.917e-04}         & 1.085e-03         & 2.179e-03         & 3.199e-03         &  4.393e-03        & \pmb{6.568e-03}         \\ \cline{2-2}
\multicolumn{1}{c|}{}                                                                                           & sub-domain 2 & \rotatebox[origin=c]{40}{\textbackslash}         & \pmb{5.577e-04}         &1.713e-03          & 3.522e-03         & 3.851e-03         &  \pmb{6.530e-03}        \\ \cline{2-2}
\multicolumn{1}{c|}{}                                                                                           & sub-domain 3 &  \rotatebox[origin=c]{40}{\textbackslash}        &  \rotatebox[origin=c]{40}{\textbackslash}        & \pmb{2.333e-03}         &3.738e-03          &4.244e-03          &  \pmb{7.770e-03}        \\ \cline{2-2}
\multicolumn{1}{c|}{}                                                                                           & sub-domain 4 &  \rotatebox[origin=c]{40}{\textbackslash}        & \rotatebox[origin=c]{40}{\textbackslash}         & \rotatebox[origin=c]{40}{\textbackslash}         &  \pmb{3.909e-03}        & 4.776e-03          & \pmb{8.020e-03}         \\ \cline{2-2}
\multicolumn{1}{c|}{}                                                                                           & sub-domain 5 &  \rotatebox[origin=c]{40}{\textbackslash}        & \rotatebox[origin=c]{40}{\textbackslash}         & \rotatebox[origin=c]{40}{\textbackslash}         & \rotatebox[origin=c]{40}{\textbackslash}         & \pmb{1.284e-02}         & \pmb{1.521e-02}         \\ \cline{2-2}
\multicolumn{1}{c|}{}                                                                                           & sub-domain 6 &  \rotatebox[origin=c]{40}{\textbackslash}        & \rotatebox[origin=c]{40}{\textbackslash}         & \rotatebox[origin=c]{40}{\textbackslash}         & \rotatebox[origin=c]{40}{\textbackslash}         &          \rotatebox[origin=c]{40}{\textbackslash} & \pmb{2.120e-02}         \\ \hline
\multicolumn{1}{c|}{\multirow{6}{*}{\begin{tabular}[c]{@{}c@{}}relative $\mathbb{L}_2$  \\ error of $|q(x,t)|$\end{tabular}}}                                                                         & sub-domain 1 &  \pmb{3.033e-04}        & 8.866e-04         & 1.476e-03         & 2.226e-03         & 3.306e-03         & \pmb{4.996e-03}         \\ \cline{2-2}
\multicolumn{1}{c|}{}                                                                                           & sub-domain 2 & \rotatebox[origin=c]{40}{\textbackslash}         & \pmb{4.405e-04}         & 1.183e-03         & 2.204e-03         & 2.521e-03         &  \pmb{3.904e-03}        \\ \cline{2-2}
\multicolumn{1}{c|}{}                                                                                           & sub-domain 3 &  \rotatebox[origin=c]{40}{\textbackslash}        & \rotatebox[origin=c]{40}{\textbackslash}         & \pmb{1.014e-03}         & 2.181e-03         & 2.884e-03         & \pmb{5.141e-03}         \\ \cline{2-2}
\multicolumn{1}{c|}{}                                                                                           & sub-domain 4 &  \rotatebox[origin=c]{40}{\textbackslash}        & \rotatebox[origin=c]{40}{\textbackslash}         & \rotatebox[origin=c]{40}{\textbackslash}         & \pmb{1.243e-03}         &           2.510e-03 &  \pmb{5.460e-03}        \\ \cline{2-2}
\multicolumn{1}{c|}{}                                                                                           & sub-domain 5 &  \rotatebox[origin=c]{40}{\textbackslash}        & \rotatebox[origin=c]{40}{\textbackslash}         & \rotatebox[origin=c]{40}{\textbackslash}         & \rotatebox[origin=c]{40}{\textbackslash}         &  \pmb{4.065e-03}        & \pmb{6.620e-03}         \\ \cline{2-2}
\multicolumn{1}{c|}{}                                                                                           & sub-domain 6 &  \rotatebox[origin=c]{40}{\textbackslash}        & \rotatebox[origin=c]{40}{\textbackslash}         & \rotatebox[origin=c]{40}{\textbackslash}         & \rotatebox[origin=c]{40}{\textbackslash}         &          \rotatebox[origin=c]{40}{\textbackslash} & \pmb{1.045e-02}         \\ \toprule
\end{tabular}
\end{table}

\begin{table}[htbp]
\caption{Relative $\mathbb{L}_2$ errors of the data-driven third-order rogue wave ($m_1=n_1=n_2=0, m_2=50$) for the NLS equation in each sub-domain by Ibc-PINN.}
\label{table5-4} 
\centering
\begin{tabular}{cc|cccccc}
\bottomrule
\multicolumn{2}{c|}{}                                                                                                          & subnet-1 & subnet-2 & subnet-3 & subnet-4 & subnet-5 & subnet-6 \\ \hline
\multicolumn{1}{c|}{\multirow{6}{*}{\begin{tabular}[c]{@{}c@{}}relative $\mathbb{L}_2$  \\ error of $u(x,t)$\end{tabular}}} & sub-domain 1 &  \pmb{3.980e-04}        & 1.327e-03         & 2.021e-03         &2.663e-03          & 4.578e-03         & \pmb{4.301e-03}         \\ \cline{2-2}
\multicolumn{1}{c|}{}                                                                                           & sub-domain 2 &  \rotatebox[origin=c]{40}{\textbackslash}        & \pmb{5.857e-04}         & 1.720e-03        & 2.473e-03        & 3.848e-03        & \pmb{3.552e-03}         \\ \cline{2-2}
\multicolumn{1}{c|}{}                                                                                           & sub-domain 3 & \rotatebox[origin=c]{40}{\textbackslash}         & \rotatebox[origin=c]{40}{\textbackslash}         & \pmb{1.906e-03}         & 3.883e-03          & 3.836e-03         & \pmb{5.183e-03}         \\ \cline{2-2}
\multicolumn{1}{c|}{}                                                                                           & sub-domain 4 &  \rotatebox[origin=c]{40}{\textbackslash}        & \rotatebox[origin=c]{40}{\textbackslash}         & \rotatebox[origin=c]{40}{\textbackslash}         &  \pmb{3.546e-03}        & 4.918e-03          & \pmb{5.269e-03}         \\ \cline{2-2}
\multicolumn{1}{c|}{}                                                                                           & sub-domain 5 &  \rotatebox[origin=c]{40}{\textbackslash}        & \rotatebox[origin=c]{40}{\textbackslash}         & \rotatebox[origin=c]{40}{\textbackslash}         & \rotatebox[origin=c]{40}{\textbackslash}         &  \pmb{6.795e-03}        & \pmb{7.909e-03}         \\ \cline{2-2}
\multicolumn{1}{c|}{}                                                                                           & sub-domain 6 &  \rotatebox[origin=c]{40}{\textbackslash}        & \rotatebox[origin=c]{40}{\textbackslash}         & \rotatebox[origin=c]{40}{\textbackslash}         & \rotatebox[origin=c]{40}{\textbackslash}         &          \rotatebox[origin=c]{40}{\textbackslash} & \pmb{1.292e-02}         \\ \hline
\multicolumn{1}{c|}{\multirow{6}{*}{\begin{tabular}[c]{@{}c@{}}relative $\mathbb{L}_2$  \\ error of $v(x,t)$\end{tabular}}}                                                                         & sub-domain 1 & \pmb{3.917e-04}         & 1.085e-03         & 2.328e-03         & 2.604e-03         &  4.039e-03        & \pmb{4.657e-03}         \\ \cline{2-2}
\multicolumn{1}{c|}{}                                                                                           & sub-domain 2 & \rotatebox[origin=c]{40}{\textbackslash}         & \pmb{5.577e-04}         & 2.059e-03         &2.821e-03         & 4.325e-03         &  \pmb{4.296e-03}        \\ \cline{2-2}
\multicolumn{1}{c|}{}                                                                                           & sub-domain 3 &  \rotatebox[origin=c]{40}{\textbackslash}        &  \rotatebox[origin=c]{40}{\textbackslash}        & \pmb{2.151e-03}         & 4.180e-03        & 4.304e-03         &  \pmb{4.726e-03}        \\ \cline{2-2}
\multicolumn{1}{c|}{}                                                                                           & sub-domain 4 &  \rotatebox[origin=c]{40}{\textbackslash}        & \rotatebox[origin=c]{40}{\textbackslash}         & \rotatebox[origin=c]{40}{\textbackslash}         &  \pmb{3.113e-03}        & 4.728e-03          & \pmb{4.669e-03}         \\ \cline{2-2}
\multicolumn{1}{c|}{}                                                                                           & sub-domain 5 &  \rotatebox[origin=c]{40}{\textbackslash}        & \rotatebox[origin=c]{40}{\textbackslash}         & \rotatebox[origin=c]{40}{\textbackslash}         & \rotatebox[origin=c]{40}{\textbackslash}         & \pmb{9.264e-03}         & \pmb{1.024e-02}         \\ \cline{2-2}
\multicolumn{1}{c|}{}                                                                                           & sub-domain 6 &  \rotatebox[origin=c]{40}{\textbackslash}        & \rotatebox[origin=c]{40}{\textbackslash}         & \rotatebox[origin=c]{40}{\textbackslash}         & \rotatebox[origin=c]{40}{\textbackslash}         &          \rotatebox[origin=c]{40}{\textbackslash} & \pmb{1.340e-02}         \\ \hline
\multicolumn{1}{c|}{\multirow{6}{*}{\begin{tabular}[c]{@{}c@{}}relative $\mathbb{L}_2$  \\ error of $|q(x,t)|$\end{tabular}}}                                                                         & sub-domain 1 &  \pmb{3.033e-04}        & 8.866e-04         & 1.578e-03         &  1.829e-03       & 3.040e-03         & \pmb{3.062e-03}         \\ \cline{2-2}
\multicolumn{1}{c|}{}                                                                                           & sub-domain 2 & \rotatebox[origin=c]{40}{\textbackslash}         & \pmb{4.405e-04}         & 1.408e-03         & 1.768e-03         & 2.736e-03         &  \pmb{2.566e-03}        \\ \cline{2-2}
\multicolumn{1}{c|}{}                                                                                           & sub-domain 3 &  \rotatebox[origin=c]{40}{\textbackslash}        & \rotatebox[origin=c]{40}{\textbackslash}         & \pmb{9.628e-04}         & 2.656e-03         & 2.590e-03         & \pmb{3.453e-03}         \\ \cline{2-2}
\multicolumn{1}{c|}{}                                                                                           & sub-domain 4 &  \rotatebox[origin=c]{40}{\textbackslash}        & \rotatebox[origin=c]{40}{\textbackslash}         & \rotatebox[origin=c]{40}{\textbackslash}         & \pmb{1.059e-03}         &2.877e-03            &  \pmb{2.909e-03}        \\ \cline{2-2}
\multicolumn{1}{c|}{}                                                                                           & sub-domain 5 &  \rotatebox[origin=c]{40}{\textbackslash}        & \rotatebox[origin=c]{40}{\textbackslash}         & \rotatebox[origin=c]{40}{\textbackslash}         & \rotatebox[origin=c]{40}{\textbackslash}         &  \pmb{2.793e-03}        & \pmb{4.114e-03}         \\ \cline{2-2}
\multicolumn{1}{c|}{}                                                                                           & sub-domain 6 &  \rotatebox[origin=c]{40}{\textbackslash}        & \rotatebox[origin=c]{40}{\textbackslash}         & \rotatebox[origin=c]{40}{\textbackslash}         & \rotatebox[origin=c]{40}{\textbackslash}         &          \rotatebox[origin=c]{40}{\textbackslash} & \pmb{6.690e-03}         \\ \toprule
\end{tabular}
\end{table}

According to Table \ref{table5-1} and Table \ref{table5-3}, we first analyze the error accumulation phenomenon of bc-PINN. Obviously, the relative $\mathbb{L}_2$ error of subnet-2 on sub-domain 1 has been almost an order of magnitude larger than that of subnet-1, and as the sequential number of subnets increases, the error on sub-domain 1 roughly shows an upward trend. There is also a similar phenomenon of error accumulation in other sub-domains. The values in the last column are the relative $\mathbb{L}_2$ errors of the solution by training the last sub network (i.e. the final form of the predicted solution obtained by bc-PINN shown in \eqref{bcPINNsolution}) in each sub-domain. Those on the diagonal are the errors generated by the predicted solution of the $n$th subnet on the $n$th sub-domain, while noting that the accuracy of this solution is the highest for this sub-domain. In addition, by comparing the values in the last column of the table with those on the diagonal, it can be intuitively seen that subsequent training of the subnets will inevitably be affected by the accumulation of errors, leading to a decrease in accuracy.

The performance of Ibc-PINN in various sub regions is shown in Table \ref{table5-2} and Table \ref{table5-4}. When calculating $\operatorname{MSE}_S$, we modified the original form in bc-PINN to reduce error accumulation. To reiterate, with regard to subnet-1 and subnet-2, the training results of bc-PINN and Ibc-PINN are the same, and there are differences in the results between the two methods starting from subnet-3. By comparing the values on the diagonal and the last column, we observe that the error accumulation speed of Ibc-PINN is significantly slower than that of bc-PINN. After comparing the errors on the diagonal in Table \ref{table5-1} - Table \ref{table5-4}, the highest accuracy that Ibc-PINN can achieve in each sub-domain is also higher than that of bc-PINN attributed to the reduction of error accumulation. Moreover, not to mention that bc-PINN is associated with errors in the last column, which are even greater than those on the diagonal.

\begin{figure}[htbp]
\centering
\includegraphics[width=8cm,height=6cm]{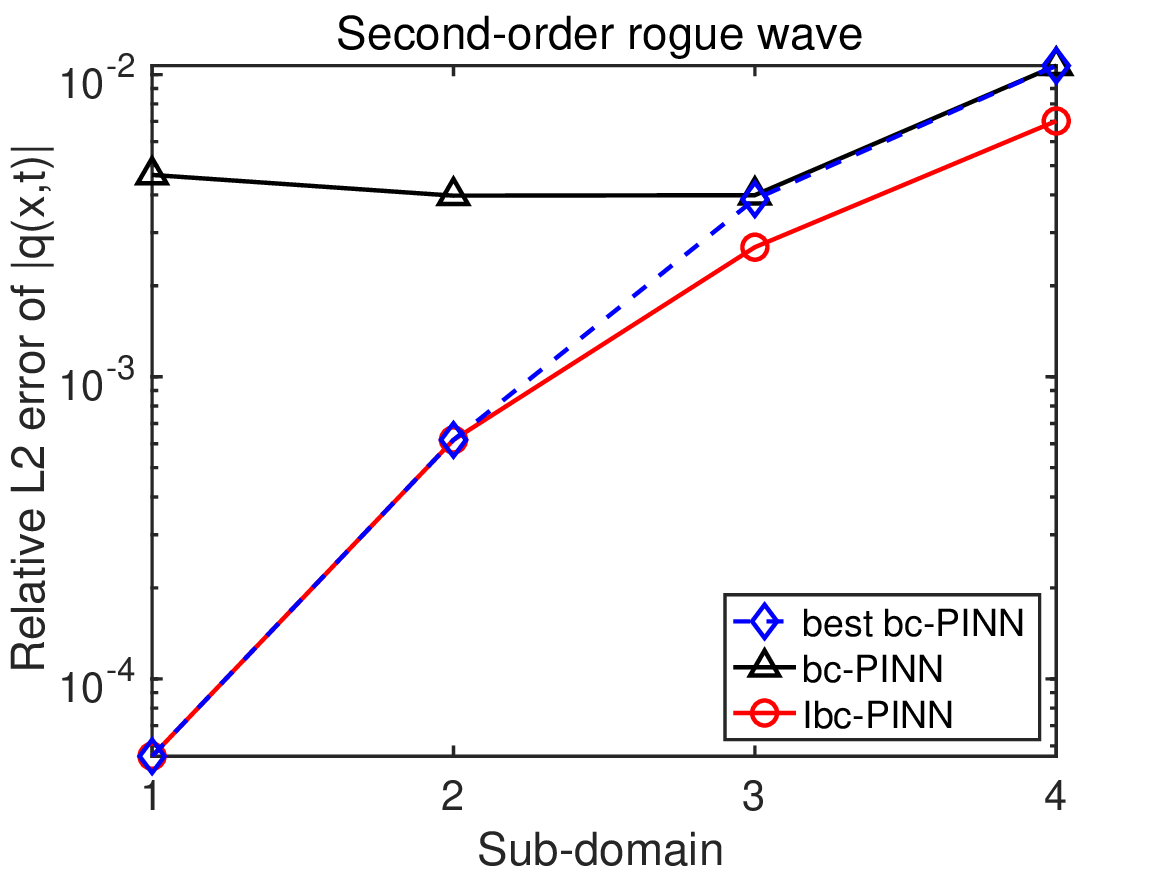}
$a$
\includegraphics[width=8cm,height=6cm]{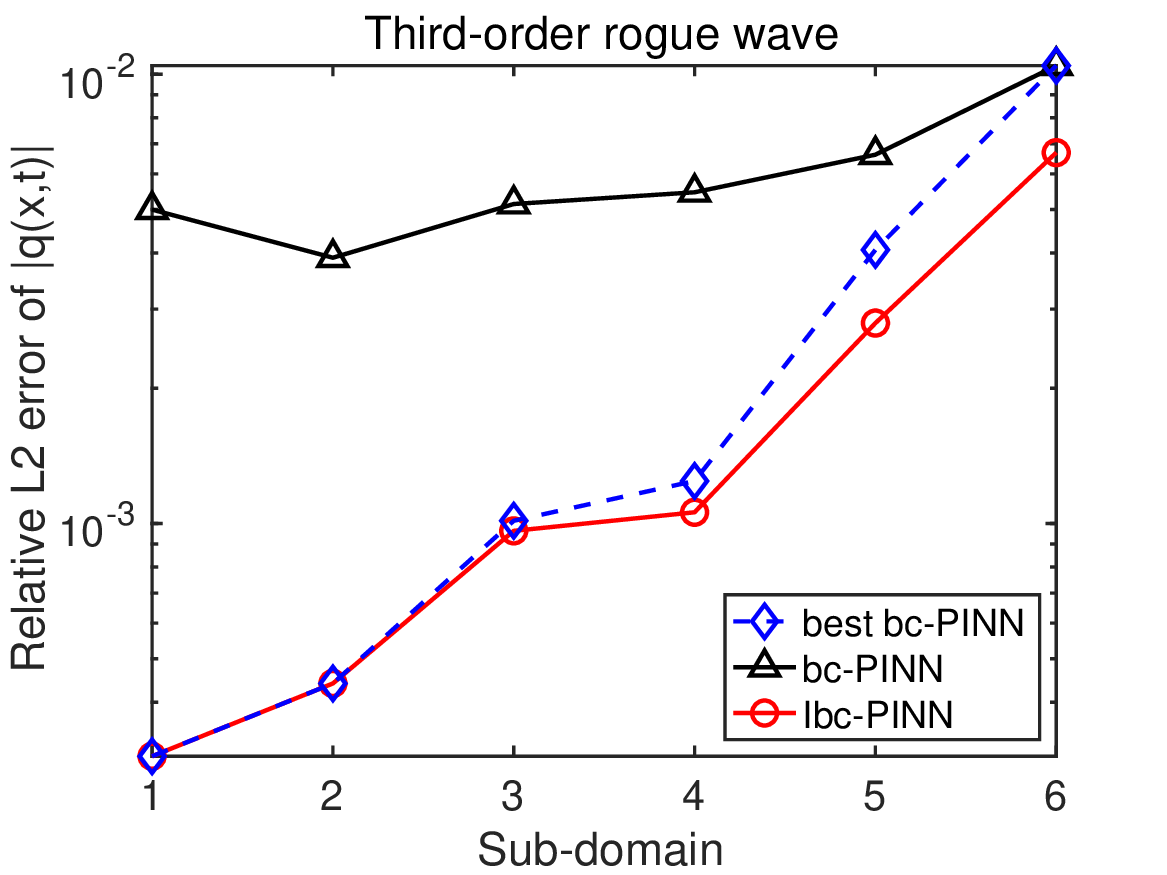}
$b$
\caption{(Color online) Relative $\mathbb{L}_2$ errors in each sub-domain: (a) data-driven second-order rogue wave ($m_1=0, n_1=0$); (b) data-driven third-order rogue wave ($m_1=n_1=n_2=0, m_2=50$).}
\label{fig5-1}
\end{figure}

In order to compare the performance of the two methods more intuitively, we draw a line graph as shown in Fig. \ref{fig5-1}. The black and red lines represent the relative $\mathbb{L}_2$ error of $|q(x,t)|$ achieved by bc-PINN and Ibc-PINN  respectively, while the blue dashed line represents the minimum error of bc-PINN in each sub-domain, namely the diagonal values in Table \ref{table5-1} and Table \ref{table5-3}, annotated on the graph as 'best bc-PINN'. Note that the red line is always below the blue dashed line and furthermore, the difference between the blue dashed line and the red one shows an increasing trend, which indicates that our improvement helps to slow down error accumulation and improve the accuracy of each sub region compared to bc-PINN. In other word, due to the refinement made in the loss term $\operatorname{MSE}_S$, even though bc-PINN adopts a concatenated form of solution, the error propagation speed is also faster than Ibc-PINN. Considering that the examples presented in Fig. \ref{fig5-1} are divided into at most 6 sub regions, if the number of sub-domains divided increases in future research, the slowdown of Ibc-PINN in error accumulation speed may yield greater advantages in accuracy.

\subsection{Error analysis}
\quad

In the last subsection, we demonstrate the necessity of proposing an Ibc-PINN method that can slow down the accumulation of errors by analyzing the errors of two methods in each sub-domain. The previous analysis is conducted from the perspective of sub regions. Here, we analyze the relative $\mathbb{L}_2$ errors and absolute error plots across the entire spatiotemporal region.

The Ibc-PINN method proposed in this study has been improved in two aspects for bc-PINN: one is to modify the loss term $\operatorname{MSE}_S$ and the other is to take the joined form as the final form of the predicted solution, as shown in \eqref{IbcPINNMSES} and \eqref{IbcPINNsolution}. To illustrate that both of these improvements contribute to accuracy, the errors generated by the three methods (bc-PINN, Ibc-PINN(unjoined) and Ibc-PINN) over the entire region are shown in Table \ref{table6-1} and Table \ref{table6-2}. Among them, Ibc-PINN(unjoined) refers to the modification of bc-PINN only in the form of $\operatorname{MSE}_S$ , without using a concatenated version in the final form of the predicted solution.

\begin{table}[htbp]
\caption{Relative $\mathbb{L}_2$ errors of the data-driven rogue wave solutions $|q(x,t)|$ for the NLS and KE equations by bc-PINN, Ibc-PINN(unjoined) and Ibc-PINN.}
\label{table6-1} 
\centering
\begin{tabular}{cc|ccc}
\bottomrule
\multicolumn{2}{c|}{}                                                                                                 & bc-PINN & Ibc-PINN(unjoined) & Ibc-PINN \\ \hline
\multicolumn{1}{c|}{\multirow{2}{*}{\begin{tabular}[c]{@{}c@{}}First-order\\ rogue waves\end{tabular}}}  & $\beta=0$     & 1.062e-03       & 6.776e-04                  & 4.843e-04        \\ \cline{2-2}
\multicolumn{1}{c|}{}                                                                                    & $\beta=\frac{1}{3}$ (KE)   & 4.471e-03       & 3.909e-03                  & 3.789e-03        \\ \cline{1-2}
\multicolumn{1}{c|}{\multirow{2}{*}{\begin{tabular}[c]{@{}c@{}}Second-order\\ rogue waves\end{tabular}}} & $m_1=0$       & 6.368e-03       & 3.902e-03                  & 3.690e-03        \\ \cline{2-2}
\multicolumn{1}{c|}{}                                                                                    & $m_1=10$      & 6.580e-03      & 4.183e-03                 & 3.543e-03       \\ \cline{1-2}
\multicolumn{1}{c|}{\multirow{2}{*}{\begin{tabular}[c]{@{}c@{}}Third-order\\ rogue waves\end{tabular}}}  & $m_1=10,m_2=0$ & 1.932e-02      & 9.368e-03                & 6.953e-03       \\ \cline{2-2}
\multicolumn{1}{c|}{}                                                                                    & $m_1=0,m_2=50$ & 6.419e-03      & 4.020e-03                 & 2.985e-03       \\ \toprule
\end{tabular}
\end{table}

\begin{table}[htbp]
\caption{Relative $\mathbb{L}_2$ errors of the data-driven rogue wave solutions for the AB system by bc-PINN, Ibc-PINN(unjoined) and Ibc-PINN.}
\label{table6-2} 
\centering
\begin{tabular}{cc|ccc}
\bottomrule
\multicolumn{2}{c|}{}                                                                                          & bc-PINN & Ibc-PINN(unjoined) & Ibc-PINN \\ \hline
\multicolumn{1}{c|}{\multirow{2}{*}{\begin{tabular}[c]{@{}c@{}}First-order\\ rogue wave\end{tabular}}}  & $|A(x,t)|$  & 2.083e-01        &  4.612e-03                  & 4.238e-03        \\ \cline{2-2}
\multicolumn{1}{c|}{}                                                                                      & $B(x,t)$ & 5.070e-01        & 1.595e-02                   & 1.442e-02          \\ \cline{1-2}
\multicolumn{1}{c|}{\multirow{2}{*}{\begin{tabular}[c]{@{}c@{}}Second-order\\ rogue wave\end{tabular}}} & $|A(x,t)|$  &3.309e-03         & 2.719e-03                   & 2.671e-03     \\ \cline{2-2}
\multicolumn{1}{c|}{}                                                                                      & $B(x,t)$ &1.212e-02         &  9.881e-03                  &9.891e-03          \\ \toprule
\end{tabular}
\end{table}

According to the results in the tables, it can be seen that Ibc-PINN is optimal in terms of accuracy, while Ibc-PINN(unjoined) is suboptimal, which implies that the improvements in these two aspects can both improve accuracy to a certain extent.

\begin{figure}[htbp]
\centering
\includegraphics[width=5.4cm,height=4cm]{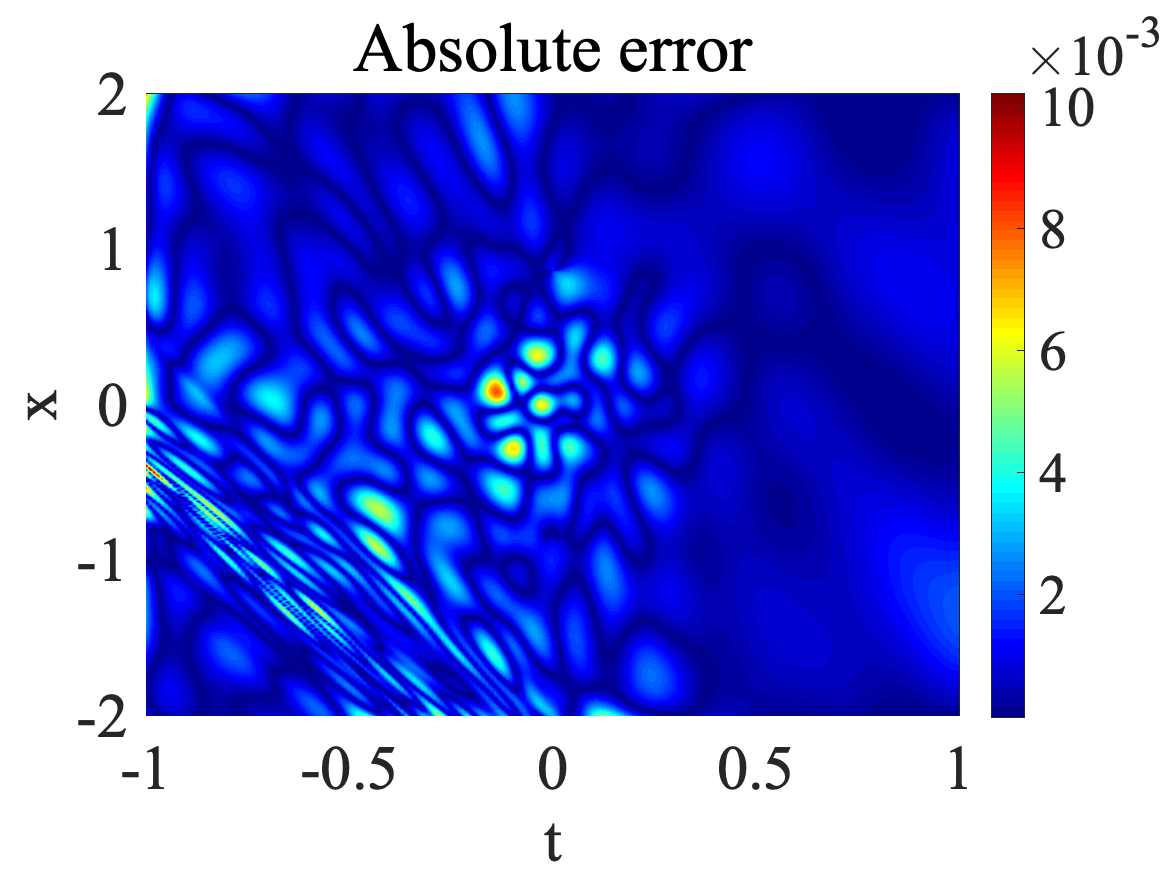}
$a1$
\includegraphics[width=5.4cm,height=4cm]{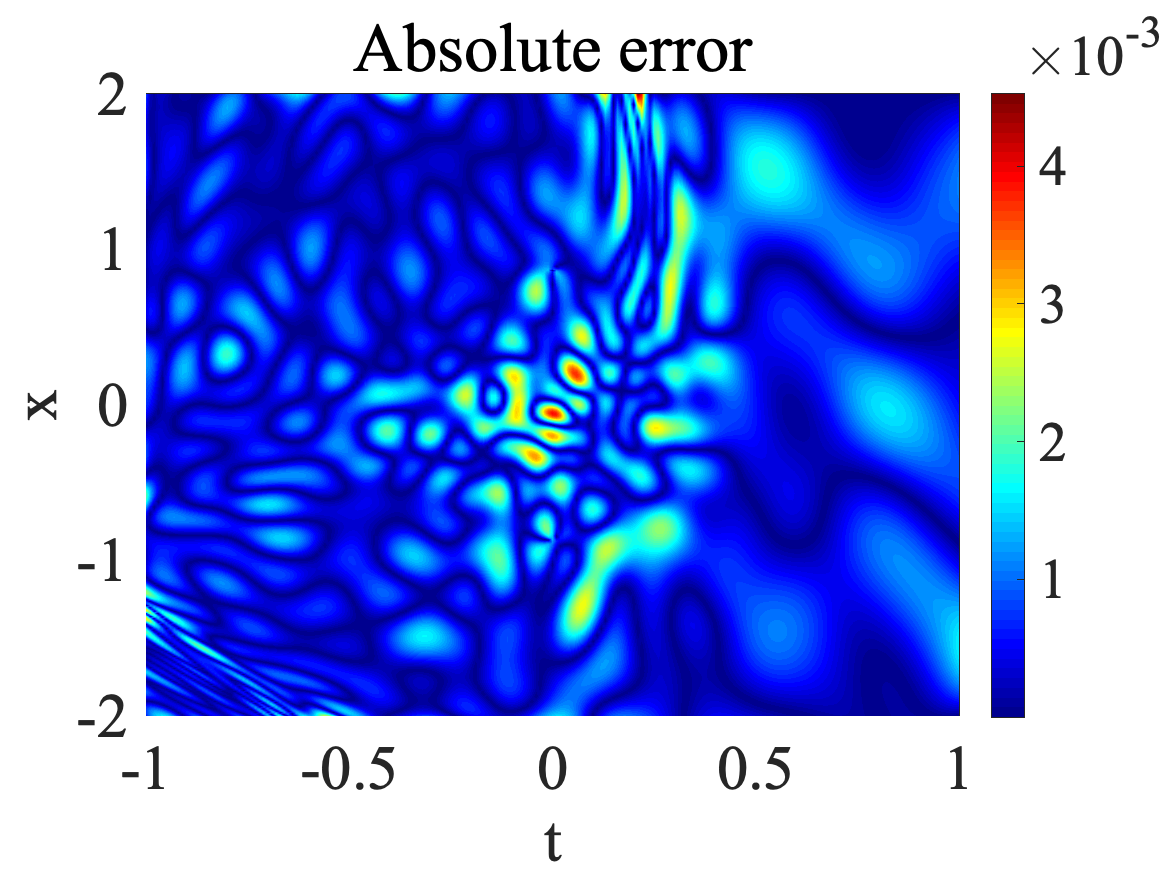}
$a2$
\includegraphics[width=5.4cm,height=4cm]{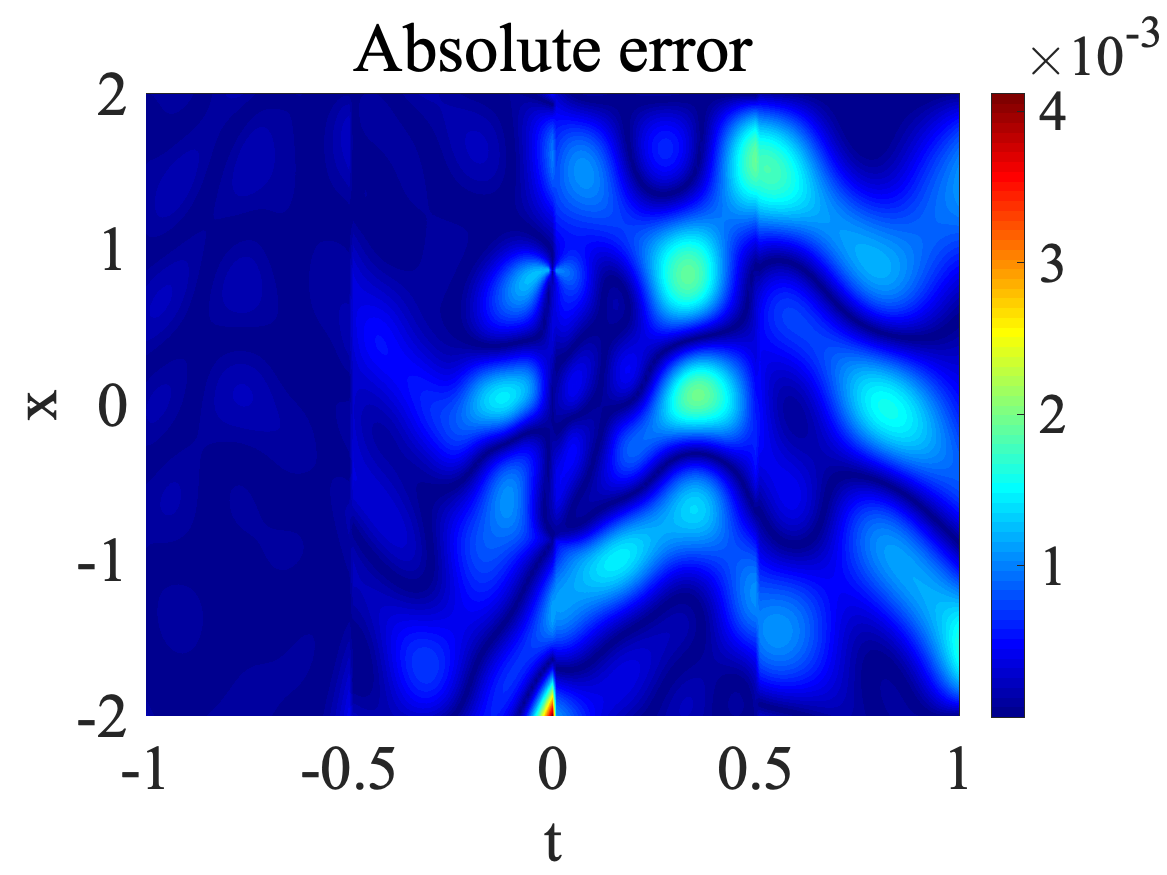}
$a3$\\
\includegraphics[width=5.4cm,height=4cm]{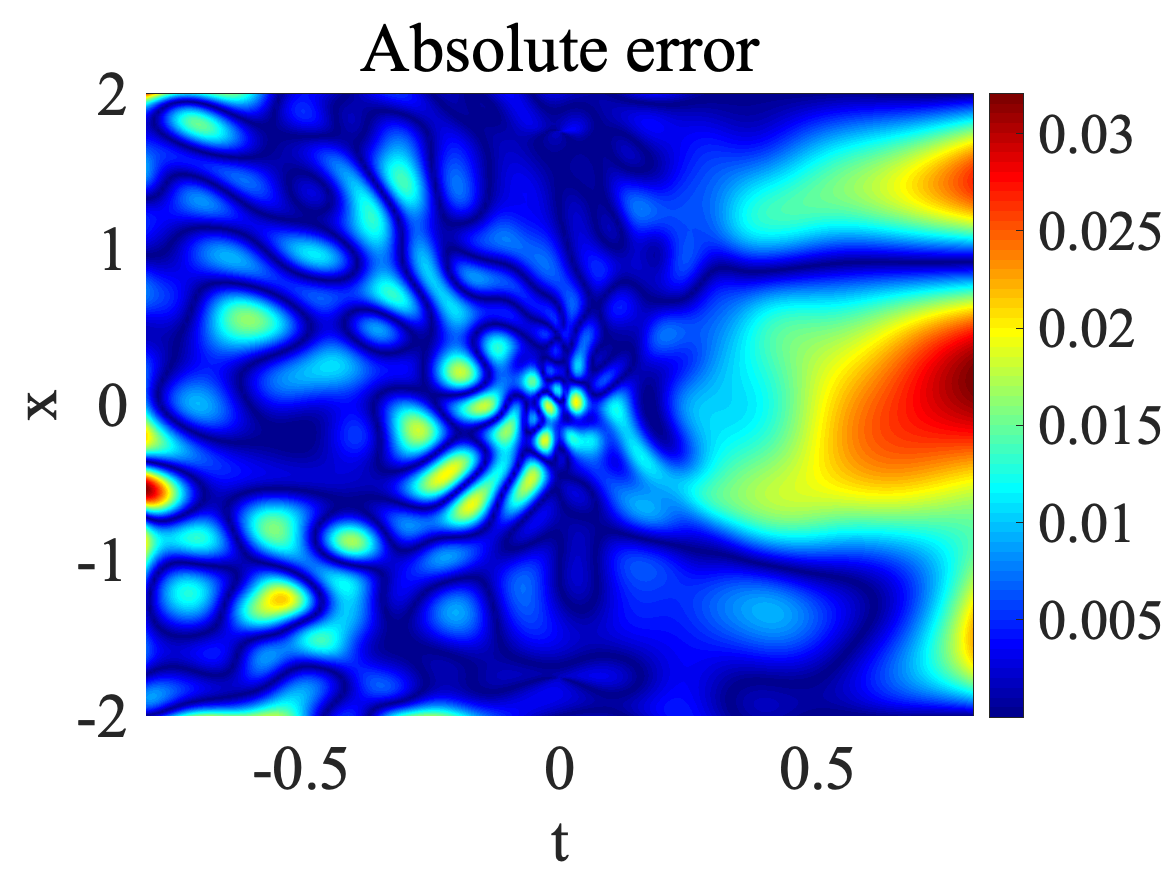}
$b1$
\includegraphics[width=5.4cm,height=4cm]{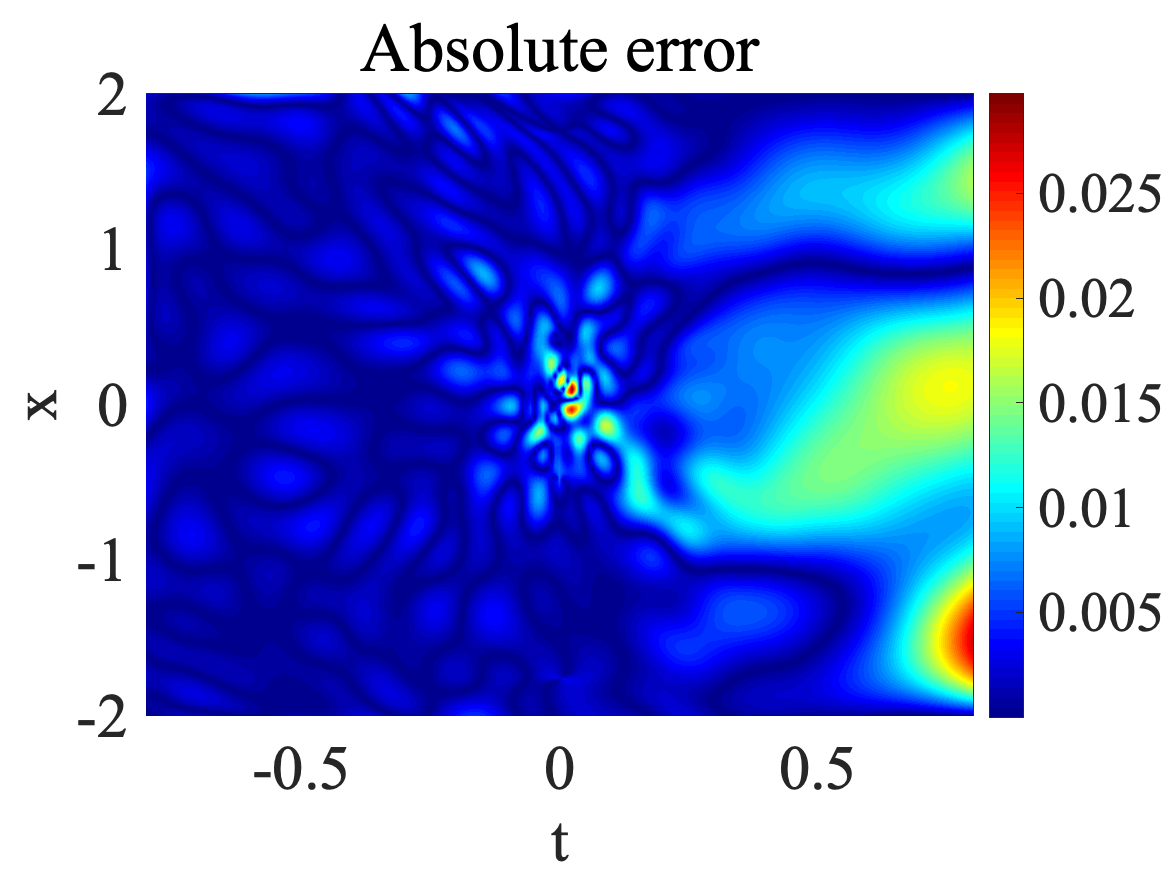}
$b2$
\includegraphics[width=5.4cm,height=4cm]{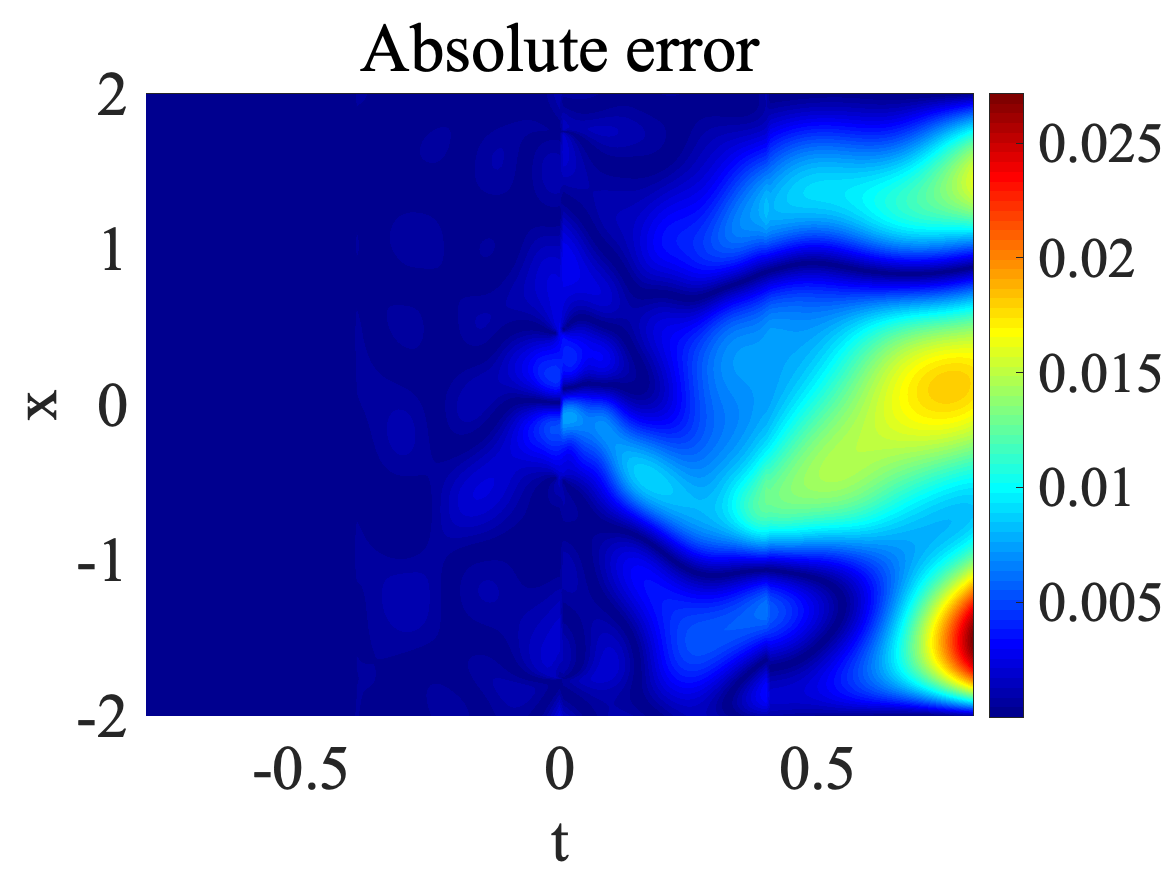}
$b3$
\caption{(Color online) The absolute error density diagrams of data-driven first-order rogue wave ($\beta=0$): (a1) by bc-PINN, (a2) Ibc-PINN(unjoined), and (a3) Ibc-PINN; The absolute error density diagrams of data-driven second-order rogue wave ($m_1=0, n_1=0$): (b1) by bc-PINN, (b2) Ibc-PINN(unjoined), and (b3) Ibc-PINN.}
\label{fig6-1}
\end{figure}

In addition, the absolute error plots of the first-order ($\beta=0$) and second-order rogue waves ($m_1=0, n_1=0$) for the nonlinear Schr\"{o}dinger equation are displayed in Fig. \ref{fig6-1}. The three columns in the figure correspond to the results of the bc-PINN, Ibc-PINN(unjoined), and Ibc-PINN methods, while each row corresponds to a different data-driven rogue wave solution.

We can observe the characteristics of the error distribution of the three methods. Due to the joined form adopted in the final representation of the predicted solutions in Ibc-PINN, there are distinct traces of concatenation in the absolute error graphs of the third column. For the bc-PINN, larger absolute errors primarily occur in the following three typical regions: (1) near the initial time; (2) where the wave amplitude is large or the gradient is significant; (3) near the final time. The phenomenon of significant errors in bc-PINN near the initial time  is contrary to the typical error propagation patterns in time piecewise training methods. For the first sub-domain, accurate initial conditions are given, while the training for subsequent sub-domains is based on pseudo initial values. Theoretically, the accuracy of the first sub region should have been the highest, but the red part indicating a large error value can be observed near the initial time in the first column of absolute error graphs. For Ibc-PINN(unjoined) and Ibc-PINN, the absolute error near the initial time is significantly reduced, especially for Ibc-PINN, which performs well in the first sub-domain with an absolute error of almost zero. It reflects that the violation of error propagation law mentioned above can be corrected by the improvement proposed in this study.

\subsection{The impact of neural network architecture}
\quad

Here, we explore the impact of network architecture on experimental results, specifically investigating whether the improved method still enhances accuracy when changing the number of hidden layers and neurons. Due to the time-consuming nature of time-segmented training approaches, we take the first-order rogue wave solution for the NLS equation as an example to illustrate.

The number of hidden layers changes from 4 to 8 with step size 2 and the number of neurons in each hidden layer changes from 32 to 128 with step size 32. The relative $\mathbb{L}_2$ errors of the bc-PINN and Ibc-PINN methods are presented in Table 11.  It is noteworthy that increasing the number of hidden layers or neurons does not necessarily lead to smaller errors. However, the improved method consistently demonstrate a certain level of accuracy enhancement.

\begin{table}[htbp]
\caption{Relative $\mathbb{L}_2$ errors of the data-driven first-order rogue wave solution $|q(x,t)|$ for the NLS equation by using different number of hidden layers and neurons per layer.}
\label{table7-1} 
\centering
\begin{tabular}{c|cccccccc}
\bottomrule
\multicolumn{1}{c|}{\multirow{2}{*}{\diagbox{Layers}{Neurons}}} & \multicolumn{2}{c|}{32}                 & \multicolumn{2}{c|}{64}                 & \multicolumn{2}{c|}{96}                 & \multicolumn{2}{c}{128} \\ \cline{2-9} 
 & bc-PINN & \multicolumn{1}{c|}{Ibc-PINN} & bc-PINN & \multicolumn{1}{c|}{Ibc-PINN} & bc-PINN & \multicolumn{1}{c|}{Ibc-PINN} & bc-PINN    & Ibc-PINN   \\ \hline
4                       & 8.875e-04        & 5.564e-04                              & 5.889e-04        & 3.224e-04                              & 5.799e-04        & 2.327e-04                              & 1.327e-03           & 1.630e-04           \\
6                       & 8.402e-04        & 3.389e-04                              &5.292e-04         & 1.558e-04                              & 5.854e-04        & 1.874e-04                              & 1.062e-03           & 4.843e-04           \\
8                       &  6.323e-04       & 2.505e-04                              & 6.050e-04        & 3.505e-04                              & 1.398e+00        & 6.662e-04                              & 5.955e-03           & 4.871e-03           \\ \toprule
\end{tabular}
\end{table}

\begin{figure}[htbp]
\centering
\includegraphics[width=8cm,height=6.5cm]{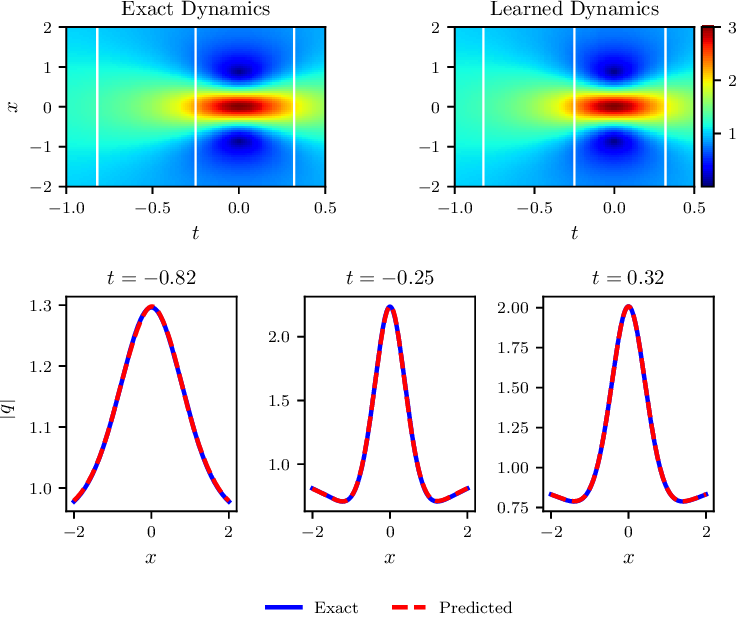}
$a$
\includegraphics[width=8cm,height=6.5cm]{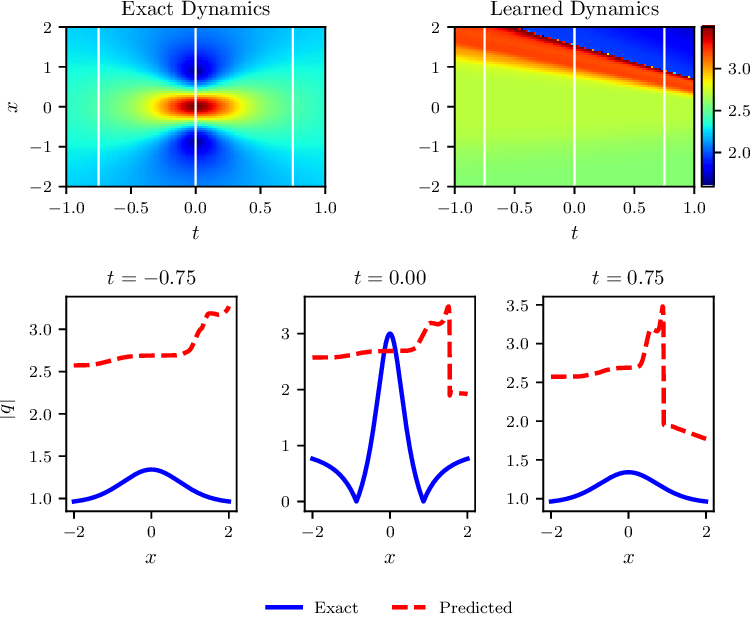}
$b$
\caption{(Color online) Instability of data-driven first-order rogue wave for the NLS equation by bc-PINN: (a) Predicted results of the penultimate stage; (b) Predicted results of the last stage.}
\label{fig7-1}
\end{figure}

\begin{figure}[htbp]
\centering
\includegraphics[width=8cm,height=6.5cm]{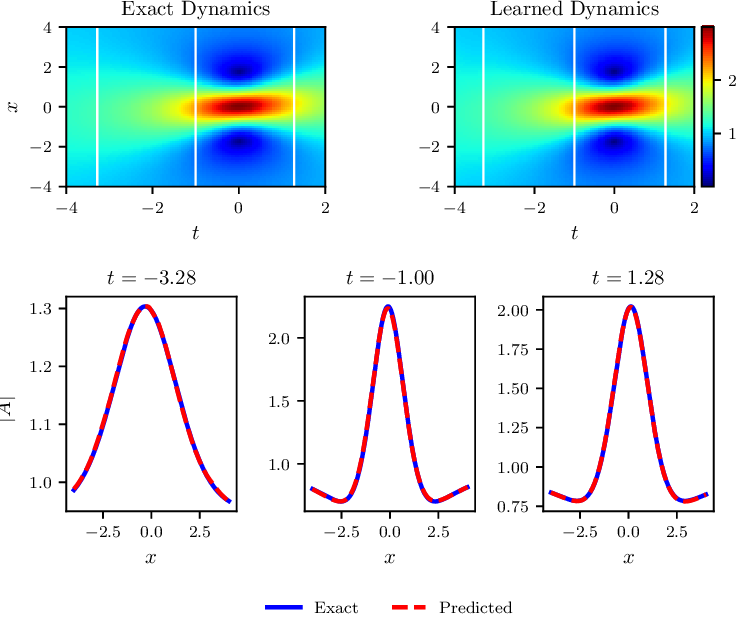}
$a$
\includegraphics[width=8cm,height=6.5cm]{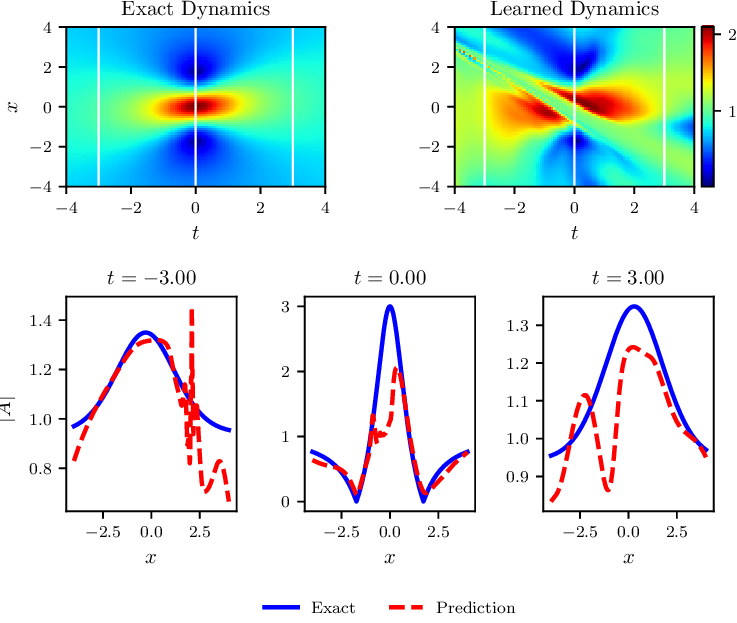}
$b$\\
\includegraphics[width=8cm,height=6.5cm]{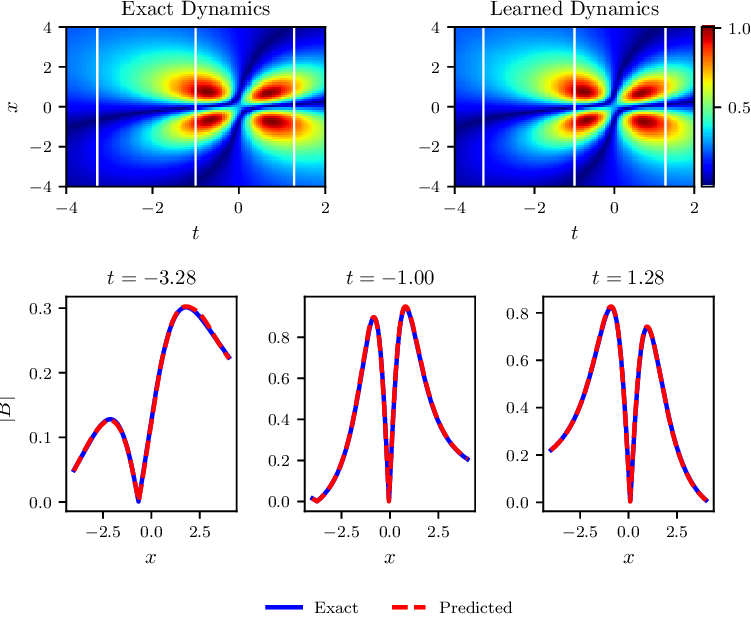}
$c$
\includegraphics[width=8cm,height=6.5cm]{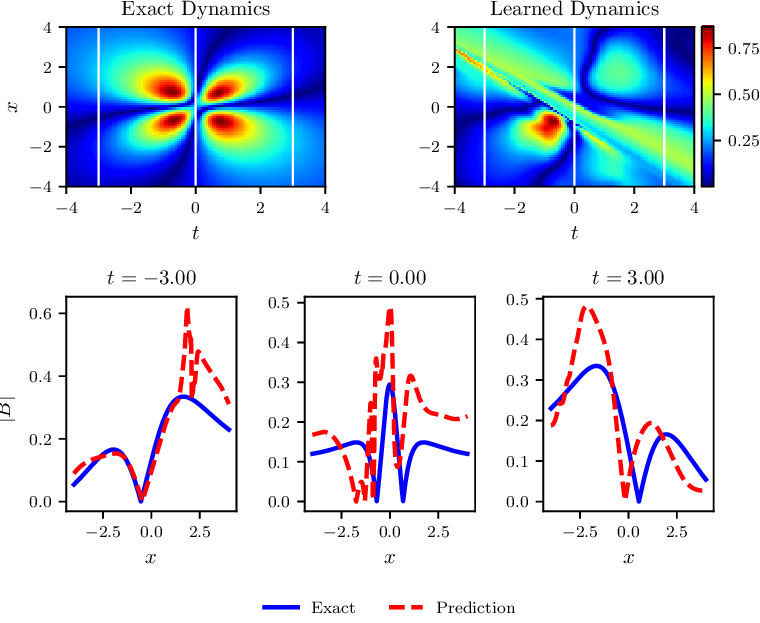}
$d$
\caption{(Color online) Instability of data-driven first-order rogue wave for the AB system by bc-PINN: Predicted results of the penultimate stage: (a) $|A(x,t)|$ and (c) $|B(x,t)|$; Predicted results of the last stage: (b) $|A(x,t)|$ and (d) $|B(x,t)|$.}
\label{fig7-2}
\end{figure}

Additionally, we observed that when the number of hidden layers is 8 and the number of neurons is 96, the bc-PINN method converges to an incorrect solution, resulting in a significant error. This phenomenon is not incidental and is also reflected in the data-driven first-order rogue wave of the AB system, as shown in Table 4. It indicates the instability in the predictive performance of bc-PINN under certain network structure settings, whereas the improved method can effectively alleviate or even avoid such occurrences. More specifically, we further analyzed the specific performance in each sub region based on the results of bc-PINN for these two examples. The predicted results of the penultimate stage and the final stage using the bc-PINN method are shown in the first and second columns of Figs. \ref{fig7-1} - \ref{fig7-2}, respectively. It can be observed that the penultimate subnetwork effectively captures the dynamic behaviors of rogue wave solutions with minor errors, whereas the predictive accuracy of the final subnetwork is significantly compromised. It illustrates that for bc-PINN, once the network training performs poorly in a certain sub-domain, the success achieved in previous regions is nullified and the errors generated by this stage of training instantly pollute the entire region. This is due to the fact that the form of the solution adopted by bc-PINN is entirely dependent on whether the training of the last subnet in the last sub-domain is successful or not. In contrast, for Ibc-PINN, if the training in a specific sub-domain is inadequate, it only affects the accuracy of the predicted solutions from that subregion onwards. This also reveals the necessity for us to propose improvements to the ultimate form of the predicted solution shown in \eqref{IbcPINNsolution}. The underlying reasons for the instability in accuracy of bc-PINN require further investigation, and targeted improvements can be proposed to address this issue in future research.

\section{Conclusion}

The bc-PINN is a sequential method to train physics informed neural networks over successive time segments while satisfying the solution for all previous time segments. Based on the characteristics of error propagation, we have made improvements in two aspects, namely, the loss function and the final form of the predicted solution. Firstly, the loss term for ensuring backward compatibility is modified by selecting the earliest learned solution for each sub-domain as pseudo reference solution. It can reduce the cumulative speed of errors to improve the accuracy of the solution in subsequent training. Secondly, we take the joined form of solutions obtained from individual subnetworks as the final form of the predicted solution, rather than relying solely on the solution learned by the last subnetwork. Its advantage lies in the fact that insufficient training in a specific subdomain only affects the accuracy of the predicted solutions from that subregion onwards. This stands in contrast to bc-PINN, where inadequate training instantly compromises the accuracy of the entire region, rendering prior successful training in other subdomains futile. The improved bc-PINN (Ibc-PINN) is applied to successfully obtain data-driven higher-order rogue wave solutions for the nonlinear Schr\"{o}dinger equation and the AB system. We also explore the impact of neural network architecture on performance and several cases were identified where the accuracy of bc-PINN is significantly compromised, while Ibc-PINN consistently maintains stability. In summary, both improvements contribute to the enhancement of the algorithm's accuracy and stability compared to the original bc-PINN method.

The domain decomposition technique possesses strong flexibility. In each sub-domain, a separate network can be employed, having its own set of parameters, including network width and depth, activation functions, optimization methods. The discussion and research on this part can be carried out in subsequent work. Moreover, different subdomains can be partitioned into varying sizes and adaptive domain decomposition methods can be designed in future study based on the distinct characteristics of solutions in different sub regions. In this paper, we investigate the dynamic characteristics of rogue waves up to the third order, which exhibit geometric structures of triangle and pentagon. In fact, rogue wave patterns have been widely analytically studied but the researches by using deep learning methods are still relatively scarce. Data-driven rogue waves of the fourth order and beyond can be simulated to showcase more intricate geometric structures including heptagon and nonagon in the future research.

\section*{Acknowledgments}
The authors thanks Xin Wang for providing data support. The project is supported by National Natural Science Foundation of China (No. 12175069 and No. 12235007), Science and Technology Commission of Shanghai Municipality (No. 21JC1402500 and No. 22DZ2229014) and Natural Science Foundation of Shanghai (No. 23ZR1418100).

\end{document}